\documentclass{aastex631}

\usepackage{float}
\usepackage{amsmath}
\usepackage{graphicx}
\usepackage{hyperref}

\usepackage[caption=false,justification=centering]{subfig}
\tabletypesize{\scriptsize}
\usepackage{etoolbox}
\makeatletter
\patchcmd\linenumberpar{\@LN@parpgbrk}{\penalty\@LN@parpgpen\relax}{}{}
\makeatother

\received{September 28, 2023}
\revised{December 18, 2023}
\accepted{January 10, 2024}

\submitjournal{PSJ}

\shorttitle{Direct $N$-body simulations of satellite formation around small asteroids}
\shortauthors{Agrusa et al.}
\begin{document}

\title{Direct $N$-body simulations of satellite formation around small asteroids: insights from DART's encounter with the Didymos system}

\author[0000-0002-3544-298X]{Harrison F. Agrusa} 
\affiliation{Universit\'e C\^ote d'Azur, Observatoire de la C\^ote d'Azur, CNRS, Laboratoire Lagrange, Nice, France}
\affiliation{Department of Astronomy, University of Maryland, College Park, MD 20742, USA}

\author[0000-0003-4045-9046]{Yun Zhang}
\affiliation{Department of Climate and Space Sciences and Engineering, University of Michigan, Ann Arbor, MI 48109, USA}

\author[0000-0002-0054-6850]{Derek C. Richardson}
\affiliation{Department of Astronomy, University of Maryland, College Park, MD 20742, USA}

\author[0000-0001-8434-9776]{Petr Pravec}  
\affiliation{Astronomical Institute of the Academy of Sciences of the Czech Republic, Fri\v{c}ova 298,
Ond\v{r}ejov, CZ 25165, Czech Republic}

\author[0000-0003-1226-7960]{Matija \'Cuk}
\affiliation{SETI Institute, 339 N Bernardo Ave, Mountain View, CA 94043, USA}

\author[0000-0002-0884-1993]{Patrick Michel}
\affiliation{Universit\'e C\^ote d'Azur, Observatoire de la C\^ote d'Azur, CNRS, Laboratoire Lagrange, Nice, France}
\affiliation{The University of Tokyo, Department of Systems Innovation, School of Engineering, Tokyo 113-0033 , Japan}

\author[0000-0002-1772-1934]{Ronald-Louis Ballouz}
\affiliation{Johns Hopkins University Applied Physics Laboratory, Laurel, MD 20723, USA}

\author[0000-0002-4952-9007]{Seth A. Jacobson}  
\affiliation{Department of Earth and Environmental Sciences, Michigan State University, East Lansing, MI 48824, USA}

\author[0000-0003-0558-3842]{Daniel J. Scheeres}
\affiliation{Smead Department of Aerospace Engineering Sciences, University of Colorado Boulder, Boulder, CO, USA}

\author[0000-0002-0906-1761]{Kevin Walsh}
\affiliation{Southwest Research Institute, Boulder, CO 80302, USA.}

\author{Olivier Barnouin}
\affiliation{Johns Hopkins University Applied Physics Laboratory, Laurel, MD 20723, USA}

\author{R. Terik Daly}
\affiliation{Johns Hopkins University Applied Physics Laboratory, Laurel, MD 20723, USA}

\author{Eric Palmer}
\affiliation{Planetary Science Institute, Tucson, AZ}

\author{Maurizio Pajola}
\affiliation{INAF-Osservatorio Astronomico di Padova, Padova, Italy}

\author{Alice Lucchetti}
\affiliation{INAF-Osservatorio Astronomico di Padova, Padova, Italy}

\author{Filippo Tusberti}
\affiliation{INAF-Osservatorio Astronomico di Padova, Padova, Italy}

\author[0000-0003-4396-1728]{Joseph V. DeMartini}
\affiliation{Department of Astronomy, University of Maryland, College Park, MD 20742, USA}

\author[0000-0001-7537-4996]{Fabio Ferrari}  
\affiliation{Department of Aerospace Science and Technology, Politecnico di Milano, Milan, 20159, Italy}

\author[0000-0001-8437-1076]{Alex J. Meyer}
\affiliation{Smead Department of Aerospace Engineering Sciences, University of Colorado Boulder, Boulder, CO, USA}

\author[0000-0002-7478-0148]{Sabina D. Raducan} 
\affiliation{Space Research and Planetary Sciences, Physics Institute, University of Bern, Bern, 3012, Switzerland}

\author[0000-0003-3610-5480]{Paul S\'anchez} 
\affiliation{Colorado Center for Astrodynamics Research, University of Colorado Boulder, 3775 Discovery Dr, Boulder, CO 80303, USA}

\begin{abstract}
We explore binary asteroid formation by spin-up and rotational disruption considering the NASA DART mission's encounter with the Didymos-Dimorphos binary, which was the first small binary visited by a spacecraft. Using a suite of $N$-body simulations, we follow the gravitational accumulation of a satellite from meter-sized particles following a mass-shedding event from a rapidly rotating primary. The satellite's formation is chaotic, as it undergoes a series of collisions, mergers, and close gravitational encounters with other moonlets, leading to a wide range of outcomes in terms of the satellite's mass, shape, orbit, and rotation state. We find that a Dimorphos-like satellite can form rapidly, in a matter of days, following a realistic mass-shedding event in which only ${\sim}2-3\%$ of the primary's mass is shed. Satellites can form in synchronous rotation due to their formation near the Roche limit. There is a strong preference for forming prolate (elongated) satellites, although some simulations result in oblate spheroids like Dimorphos. The distribution of simulated secondary shapes is broadly consistent with other binary systems, measured through radar or lightcurves. Unless Dimorphos's shape is an outlier, and considering the observational bias against lightcurve-based determination of secondary elongations for oblate bodies, we suggest there could be a significant population of oblate secondaries. If these satellites initially form with elongated shapes, a yet-unidentified pathway is needed to explain how they become oblate. Finally, we show that this chaotic formation pathway occasionally forms asteroid pairs and stable triples, including co-orbital satellites and satellites in mean motion resonances. 
\end{abstract}
\keywords{Asteroid dynamics (2210) — Asteroid satellites (2207) — Asteroid rotation (2211) — Near-Earth objects (1092) — Small Solar System bodies (1469)}


% \begin{center}\flag{Note: we include movies of several simulations which are available at the following link. Except for the movie corresponding to Fig. 1, these movies are rendered in a rotating frame with a period of 10 h. \href{https://zenodo.org/record/8387044}{https://zenodo.org/record/8387044}}
% \end{center}
\section{Introduction} \label{sec:intro}

Binaries make up ${\sim}15\%$ of the near-Earth asteroid (NEA) population \citep{Margot2002,Pravec2006}. This fraction increases to ${\sim}65\%$ for fast-rotators greater than  ${\sim}300$ m in diameter \citep{Pravec2006}. Given the relatively short median dynamical lifetimes of NEAs of about 10 Myr \citep{Gladman2000}, this high binary fraction implies an efficient formation mechanism that can maintain a steady-state population or formation before exiting the main asteroid belt. Among NEA binaries, the primary component is typically small, less than ${\sim}5\text{ km}$ and with a moderately sized secondary, usually less than ${\sim}60\%$ of the primary's diameter. Furthermore, the two bodies are typically on close orbits, separated by less than ${\sim}10$ primary radii. Typically, the primary is rapidly rotating and has a near-spherical shape with an equatorial bulge \citep{Pravec2006,Pravec2007, Benner2015}. In addition, the secondaries of these systems frequently have an elongated shape and are typically in synchronous rotation when the secondary is on a close, near-circular orbit \citep{Pravec2016}. The specific angular momentum of these systems is often close to the angular momentum of a sphere having the same total mass rotating at its critical disruption limit, indicating that the creation mechanism of these binaries must be related to some sort of fission or mass-shedding process \citep{Pravec2007, Pravec2010}. 

The Yarkovsky–O'Keefe–Radzievskii–Paddack (YORP) effect is a process in which the absorption and re-emission of solar radiation imparts a small torque on irregularly-shaped bodies \citep{Rubincam2000}. The strength of the YORP effect is highly dependent on the size and shape of the body and its heliocentric distance, and is thought to be the dominant mechanism that spins-up small asteroids leading to rotational disruption and the formation of a binary \citep{Vokrouhlicky2002,Bottke2002}. For a review of YORP and the formation of small binary asteroids, see \cite{Walsh2015}.

The recent Double Asteroid Redirection Test (DART) mission was the first spacecraft to visit (albeit briefly) a small  binary asteroid \citep{Rivkin2021}. The primary, Didymos, has a flattened top-shape with a volume-equivalent diameter of ${\sim}760 $ \text{ m} and a short spin period of 2.26 h \citep{Naidu2020,Daly2023}. The satellite, Dimorphos, has an approximately circular orbit with a semimajor axis of ${\sim}3$ primary radii \citep{Naidu2022,Scheirich2022}. Dimorphos was thought to be in synchronous rotation prior to DART's kinetic impact \citep{Richardson2022}, although this could not be determined directly. Both bodies have surfaces consistent with being gravitational aggregates or ``rubble piles'' \citep{Barnouin2023_submitted}, although any direct measurements of their interiors will have to wait for the arrival of the European Space Agency's Hera mission in 2027 \citep{Michel2022}. Dimorphos has a volume-equivalent diameter of ${\sim}150$ m, corresponding to a secondary-to-primary size ratio of ${\sim}0.2$. If Didymos and Dimorphos have equal bulk densities (which not known), then they would have a mass ratio of ${\sim}0.01$. Dimorphos appears to have a remarkably oblate shape, in which its semi-major axis, $a$, and semi-intermediate axis, $b$, are nearly identical in length and roughly ${\sim}1.5$ times longer than its semi-minor axis, $c$ (i.e, principal rotation axis) \citep{Daly2023}. Additionally, the presence of rocks perched on boulders with slopes no higher than ${\sim}35^{\circ}$ suggests that Dimorphos's surface may be cohesionless with a friction angle of ${\sim}35^\circ$ \citep{Barnouin2023_submitted}. A ${\sim}35^\circ$ friction angle is also derived from an independent method based on the angularity of boulders on Dimorphos's surface \citep{Robin2023_submitted}. Although there could be some strength at depth, some recent models of DART's impact are consistent with a cohesionless interior \citep{Raducan2023_submitted}.

The shape of Dimorphos came as a surprise, as previously observed systems (either via radar or lightcurve) have found that the secondary tends to have a more prolate, or elongated, shape in which $a>b>c$, where $a$, $b$, and $c$ are the body's respective semimajor, semi-intermediate, and semi-minor axes. Based on Dimorphos's physical extents (which are well-fit to a uniform ellipsoid), a preliminary estimate puts Dimorphos's elongation at $a/b=1.02\pm0.03$ \citep{Daly2023} which has since been updated to $a/b=1.06\pm0.03$ \citep{Daly2023_submitted}. Nevertheless, no previously published studies employing lightcurve- or radar-based shape determination of similar binary systems have found a satellite with $a/b\lesssim1.1$ \citep{Ostro2006,Naidu2015b,Becker2015,Pravec2016}. The principal motivation for this study was to understand how Dimorphos might have formed with its present shape, but as we will show later, the results are broadly applicable to all small binaries formed by rotational disruption.

In Sections \ref{subsec:review} and \ref{subsec:spinup}, we summarize previous work on binary formation and YORP spin-up. Then we introduce \textsc{pkdgrav} in Section \ref{sec:methods} and show a full end-to-end example simulation demonstrating the formation of a binary system via a single YORP-driven mass-shedding event in Section \ref{subsec:mass_shedding_example}. Then, based on this simulation and constraints derived by DART, we introduce the simulation set-up and initial conditions used for this study in Section \ref{subsec:sim_setup}. The results from the numerical simulations are shown in Section \ref{sec:results} followed by conclusions in Section \ref{sec:conclusions}.

\subsection{Previous work}\label{subsec:review}

There is general agreement that many small binary asteroids are likely created via YORP-induced spin-up in which a satellite forms when the primary exceeds its critical spin limit. However, there is still some debate about precisely how the binary system forms, which we address here.

\cite{Walsh2008b} modeled the formation of a binary using a rubble-pile asteroid model consisting of thousands of spherical particles in which angular momentum is slowly added to the system as a proxy for YORP-induced spin-up. Through the spin-up process, the primary reshapes, forming an equatorial bulge, and the secondary is gradually built in orbit via gravitational accumulation of material shed from the primary's equator. This idea was revisited in \cite{Walsh2012} with a broader simulation suite at higher resolution, finding that that the resulting top-shaped primary and secondary properties are broadly consistent with the observed population. 

However, this model suffers from a few issues. First, it has been shown that the magnitude \textit{and} direction of the YORP torque is highly-sensitive to small changes in the primary's shape, meaning that each landslides, mass shedding, and natural impacts could change the strength and direction of the YORP effect. This could make YORP spin-up effectively a random walk process and may significantly increase the amount of time required to build a secondary in orbit \citep{Statler2009, Cotto-Figueroa2015, Zhou2022}. This effect may significantly decrease the efficiency of such a formation process, making a scenario in which the secondary forms from a single rotational disruption event more attractive\footnote{{Recent work shows that all twelve of the available and reliable YORP detections show that that the asteroid's spin rate is \textit{increasing} in time \citep{Durech2023}. This could due to YORP having an underlying preference to increase the spin rates of asteroids rather than decrease them \citep{Golubov2012}. If there really is an underlying preference for spin-up rather than spin-down, then the idea of stochastic YORP does not present a significant issue to binary formation models.}}. On a related note, \cite{Jacobson2011a} point out that a gradual process of satellite formation suffers from the fact that single particles can escape the system before the satellite is fully formed. The argument is that escape can occur before the next shedding event because the system has positive free energy, and the single (or multiple) shed particles would escape before the next YORP cycle. Second, in the \cite{Walsh2008b,Walsh2012} works, the primary was initialized in a hexagonal close-packed (HCP) arrangement and particle contacts were handled using the Hard-sphere Discrete Element Method (HSDEM), which is ill-suited to simulate particles undergoing persistent contact with multiple other particles \citep[see][]{Murdoch2015,Sanchez2012,Sanchez2016}. Although this model captures the general process of binary formation, the HCP packing and HSDEM contact physics may affect the precise details of the mass shedding and satellite formation.

\cite{Jacobson2011a} presented an alternative theory for binary formation in which the secondary arises from a single large rotational disruption event, dubbed ``rotational fission'' \citep{Scheeres2007b}. There seems to be some confusion about these two ideas in the literature, and we attempt to point out here that the ``gravitational accumulation'' idea of \cite{Walsh2008b} and the ``rotational fission'' idea of \cite{Jacobson2011a} are similar, yet fundamentally distinct. \cite{Jacobson2011a} present a model that follows the spin-orbit coupled dynamical evolution of a binary system following a fission event. Their model assumes the initial rubble pile consists of two ellipsoidal components (that will later become the primary and secondary). Angular momentum is slowly added to the single body and at fast spin rates, the long axes of the two constituent ellipsoids become aligned while still at rest on one another, owing to this being the only stable equilibrium configuration for two ellipsoids \citep{Scheeres2007b}. When the critical spin limit is exceeded, the bodies then ``fission'' and begin to orbit each other. \cite{Jacobson2011a} integrate the fully-coupled spin and orbital evolution of the binary system, accounting for their nonspherical gravitational potentials along with a treatment for tidal evolution. For mass ratios $<0.2$, they argue that the post-fission system has positive free energy (i.e., the sum of the bodies' kinetic energies and the mutual potential), meaning that there are one of two possible outcomes: the secondary must either escape the system or it must fission itself to form (temporarily) a triple system. These \textit{secondary} fissions occur when the secondary is gravitationally torqued by the primary such that it exceeds its stability limit and is then split into two separate ellipsoids (whose shapes are generated randomly). In most of their simulations the third body re-impacts the primary, but it can also escape the system or re-impact the secondary. This ongoing fission process, along with tidal evolution, will eventually lead to a state with negative free energy, ultimately allowing the binary system to be stable.

This model has the advantage that it only requires a single rotational disruption event, thus avoiding the ``stochastic YORP'' problem. This model broadly reproduces the characteristics of the NEA population, including the existence of asteroid pairs. A follow-up study that includes a more sophisticated asteroid population model shows that the asteroid fission theory can reproduce many of the characteristics of binary asteroids \citep{Jacobson2016}. Recently, the rotational fission model has been extended to include 3D dynamics, and many of the conclusions still largely hold true \citep{Boldrin2016,Davis2020b,Ho2022}.

The \cite{Jacobson2011a} suffers from one potential weakness; in order to allow for secondary fission events, which are necessary to achieve stability, they invoke that the secondary itself is a rubble pile. This is of course a sensible assumption, however, it presents a major issue; at the very moment of the initial fission event, when the secondary (which is itself a rubble pile!) detaches from the primary, it \textit{must} tidally disrupt, as it is lying within the primary's Roche limit \citep{Holsapple2006,Sharma2009} rather than continue to evolve as a coherent body.  This dilemma can be avoided in cases where the secondary's bulk density is much higher than the primary's such that the Roche limit lies within the primary itself, or if the secondary has a small amount of cohesion to prevent a tidal disruption \citep[e.g.][]{Holsapple2008, Sanchez2016, Tardivel2018}, or if the secondary has a bilobate shape in which the neck region could fail before the rest of the body \citep{Hirabayashi2019a}, or some combination of all three. However, neither of these three contingencies seem viable for Dimorphos to form purely by fission if it is a rubble pile with little-to-no cohesion and a friction angle of ${\sim}35^\circ$ \citep{Barnouin2023_submitted, Raducan2023_submitted}. If such a body were to rotationally fission from Didymos's equator, it would immediately undergo tidal disruption \citep[e.g.][]{Agrusa2022a}.

In this paper, we present an alternate formation path that incorporates the ideas of both the \cite{Walsh2008b} and \cite{Jacobson2011a} models. In our model, the secondary forms via gravitational reaccumulation, similar to the \cite{Walsh2008b} theory, however, all the required mass is shed in a single impulsive event, like \cite{Jacobson2011a}. We will demonstrate that this simple model of accumulation from debris shed from a single rotational disruption event can lead to stable binary systems and secondaries having a wide range of shapes.

Recently, \cite{Madeira2023} considered the gravitational accumulation of Dimorphos from debris produced by Didymos using a 1D ring-satellite model (HYDRORINGS) \citep{Charnoz2010, Charnoz2011, Salmon2010}. In this model, material migrates outwards via viscous spreading until it reaches the Roche limit, at which point it is immediately converted to a moonlet. Assuming the initial debris ring is narrowly confined at the primary's surface, they find that it takes ${\sim}1$ yr for the ring to spread to the Roche limit, after which Dimorphos would accrete rapidly, reaching ${\sim}90\%$ of its current mass within days. This process requires ${\sim}25\%$ of Didymos's mass to be put into orbit to form a satellite with ${\sim}1\%$ of Didymos's mass. Once Dimorphos forms, they estimate that the ring would take thousands of years to disappear. In terms of timescales and the required amount of mass, the predictions of this model differ significantly than what we present in this work, most likely due to different assumptions in the initial conditions and necessary simplifications of the 1D model. A significant body of literature suggests that a rubble pile undergoing rotational failure and mass shedding, due to its nonzero friction and/or cohesion, will shed material onto much wider initial orbits \citep[e.g.][]{Hirabayashi2015a, Sanchez2016, Sanchez2020,Zhang2018, Hyodo2022} rather than within a narrowly confined region, including several studies focused on Didymos itself \citep[e.g.,][]{Yu2018, Zhang2017, Zhang2021}. This means that satellites can start forming much quicker, without having to wait for a ring to spread to the Roche limit. Also, the \cite{Madeira2023} study assumes that all ring particles are 1 m in diameter, whereas Dimorphos's contains boulders at least as large as 16 m \citep{Pajola2023_submitted}. The presence of larger boulders would speed up the accumulation process, due to their higher mass and larger collision cross section. Finally, this model neglects gravity between moonlets, assuming that they undergo perfect mergers whenever they enter their mutual Hill spheres. In this work, we will show that this is often not the case, and that the merger process of moonlets is highly chaotic, leading to a wide range of outcomes.

\subsection{YORP spin-up and mass shedding}\label{subsec:spinup}

In this study, the considered binary formation scenario is based on the idea that the progenitor body sheds substantial surface materials in a relatively short timescale. However, this mass shedding failure behavior is not unconditional.  Prior theoretical work, relying on static stress analyses, suggests that a homogeneous ellipsoidal body would first structurally fail internally at its center instead of at the surface during spin up \citep{Holsapple2010, Hirabayashi2015b}. This internal failure would lead to internal deformation, suppressing surface landslides and mass shedding.  Modeling of a rubble-pile body's spin-up using the soft-sphere discrete element method (SSDEM) supports these static analyses, showing that homogeneous cohesionless bodies could indeed fail through internal deformation \citep{Sanchez2012, Hirabayashi2015a} and homogeneous cohesive bodies could even fail through global tensile disruption \citep{Sanchez2016, Zhang2018}.  

For mass shedding to occur on the surface, the body's interior must be mechanically stronger and/or denser than the exterior layer. Research by \cite{Sanchez2018rotational} using SSDEM simulations on a spherical rubble pile with an internal core occupying 70\% of its radius demonstrates that the interior needs to be 3 to 4 times stronger than the shell to prevent internal failure.  Similarly, \cite{Ferrari2022a} use a polyhedron discrete element method to show that a rigid core with a volume fraction exceeding 25\% can also facilitate mass shedding, as shown for the case of Didymos. By employing higher particle resolution (${\sim}10^5$ particles) and a highly polydisperse particle size distribution, the SSDEM spin-up simulations of \cite{Zhang2017, Zhang2021} reveal that the large particle size difference can reduce the internal porosity and increase the mobility of surface regolith in a rubble-pile. The resultant small scale heterogeneity could trigger surface failure via mass shedding, even if mechanical interactions at the particle level are uniform throughout the body. A similar effect is produced by the interactions between non-spherical particles with irregular shape, where geometrical interlocking mechanism adds mechanical strength to the inner structure \citep{Ferrari2020a}. In all cases, maintaining some internal shear strength and a low surface cohesion are crucial to initiate surface mass shedding \citep{Sanchez2020, Zhang2022}.

Beyond considering the internal structure, the evolution of a rubble-pile body after reaching its spin limit is also heavily influenced by numerous other factors, such as shape and surface morphology \citep{Hirabayashi2019a, Hirabayashi2020, Zhang2022}.  This complexity implies that the surface failure conditions could vary across different bodies, which determines why certain asteroids evolved into binary systems while others didn't, and could play some role in the apparent deficit of small binaries and pairs among primitive asteroid types relative to silicate-rich bodies \citep{Minker2023}.  In Section \ref{subsec:mass_shedding_example}, to validate the assumed binary formation scenario for Didymos, we will present an SSDEM simulation example to showcase the feasibility of Didymos' mass shedding failure behavior under reasonable conditions.

We finally note that the mass shedding doesn't have to occur solely as a result of YORP spin up. Once YORP spins up a body near its critical limit, a natural impact could trigger the mass shedding event, for example. In fact, this study is somewhat agnostic to the exact process causing the mass shedding, we merely require that the mass shedding occur all at once such that enough material is put into orbit to allow a rapid reaccretion process.

\section{Methodology}\label{sec:methods}
We use \textsc{pkdgrav}, a gravitational $N$-body tree code to model the accumulation of a rubble-pile satellite \citep{Richardson2000,Stadel2001}. Particle contacts are handled using the soft-sphere discrete element method (SSDEM), in which a Hooke's Law spring provides a normal force between overlapping particles \citep{Schwartz2012}. Parameters such as the spring constant and coefficients of static, twisting, and rolling friction and cohesion can be adjusted to represent desired material properties. We refer the reader to \cite{Zhang2017} for a detailed description of the model. In short,  $k_N$ and $k_T$ are the two spring constants that control the particle's stiffness in the normal and tangential directions respectively, $\epsilon_N$ and $\epsilon_T$ are the coefficients of restitution in the normal and tangential direction for controlling energy dissipation,  $\mu_S$, $\mu_R$, and $\mu_T$ are the coefficients of static, rolling, and twisting friction. Finally, the shape parameter $\beta$ is used to approximate the non-spherical nature of real particles by statistically defining their contact area, enabling the calculation of the associated rotational resistance.  

The normal spring constant ($k_N$) and timestep are selected such that typical particle overlaps don't exceed ${\sim}1\%$ of a particles radius and that ${\sim}30$ timesteps take place over the course of a collision \citep{Schwartz2012}. This ensures that particle contacts are resolved properly and that particles do not undergo a nonphysical level of interpenetration. Based on our typical particle mass, sizes, and collision speeds this corresponds to a $k_N\sim4\times10^6 \text{ kg s}^{-2}$ and a punishingly small timestep of ${\sim}0.15$ s. Following common practice, we set $k_T=\frac{2}{7}k_N$ \citep{Walton1995, Schwartz2012}. In all simulations presented here, we set the restitution coefficients to $\epsilon_N=\epsilon_T=0.55$, a typical value for rocky material \citep{Chau2002}. $\mu_R$ and $\mu_T$ are set to 1.05 and 1.3, respectively, to represent the rough surfaces of medium hardness rocks \citep{Jiang2015}, leaving $\mu_S$ and $\beta$ as the only two parameters that are varied to achieve different friction angles. These values and their corresponding friction angle are provided in Table \ref{tab:params}.

\subsection{Satellite formation via mass shedding}\label{subsec:mass_shedding_example}
Here, we show an example simulation to test Didymos's structural failure behaviors at its spin limit.  Didymos is modeled as a granular aggregate composed of 87,635 spheres with radii ranging from about 4 to 16 m.  The particle size distribution follows a cumulative power-law distribution with an exponent of $-2.5$, aligning with the boulder size frequency distribution with similar radii found on Dimorphos, as observed in the images taken by the camera onboard the DART spacecraft \citep{Pajola2023_submitted}. The granular aggregate was configured based on the pre-impact Didymos shape model \citep{Barnouin2023_submitted} to ensure accurate representation of Didymos. To allow the initiation of surface mass shedding and maintain the stability of Didymos at its current spin period of 2.26 h, the shape parameter $\beta$ is adopted to be 0.8, and $\mu_S$ is taken to be 1.0, representing a material internal friction angle of $\sim$40$^\circ$ \citep{Zhang2022}.  This friction angle is within the typical range of compacted dry sand, i.e., 33$^\circ$--43$^\circ$ \citep{Bareither2008}. To resolve the quasi-static mechanical contact between particles, we adopted a smaller timestep of $\sim$0.02 s and a larger $k_N{\sim}8\times10^7$ kg s$^{-2}$.  The bulk density of the body is set to 2.7 g cm$^{-3}$, consistent with Didymos' latest bulk density estimate constrained by the updated shape model and the binary separation, i.e., $2.760\pm.130$ g cm$^{-3}$ \citep{Naidu2023_submitted}\footnote{Due to uncertainties in the size and volume of Dimorphos, which has some degeneracy in the body separation, there is significant uncertainty in Didymos's bulk density. The formal uncertainty of ${\pm}.130$ g cm$^{-3}$ is likely an underestimate and a realistic uncertainty might be larger.}. Our numerical investigation found that cohesion is no longer required to maintain the bulk structural stability of Didymos at its current spin state for a bulk density of $\gtrsim$2.7 g cm$^{-3}$, and, therefore, the interparticle cohesive strength was set to zero\footnote{We also performed simulations using a friction angle of $35^\circ$ ($\mu_S=0.6$, $\beta=0.5$), and found that a bulk cohesive strength of about 8 Pa is needed for the structural stability. The mass-shedding structural failure behavior at faster spin is similar to the case presented here.}. All other parameters are the same as those introduced in the previous section.

Didymos is quasi-statically spun-up (as a proxy for YORP) until structural failure is detected{, which we define by the body's longest dimension changing by more than 1\% relative to the starting shape}. Then the spin-up procedure is halted and the granular system evolves purely under its own self gravity.  The results show that the primary structurally failed at a spin period of ${\sim}2.2596$ h, where it then shed surface particles from mid-to-low latitudes, putting ${\sim}3\%$ of its total mass onto low inclination orbits. Much of this material rapidly clumps into moonlets, and a Dimorphos-mass satellite is formed within days following a series of moonlet mergers. As a result of the conservation of angular momentum, the primary's spin period drops to ${\sim}2.5$ h by the end of the simulation due to mass shedding and reshaping. {Material that falls back onto Didymos preferentially lands on the equator, which contributes to forming Didymos's equatorial ridge, similar to the process demonstrated by \cite{Hyodo2022}. The present-day shape of Didymos is the subject of other ongoing and future studies \citep[][Zhang et al., in prep]{Barnouin2023_submitted}.}

\begin{figure}[H]
\includegraphics[width=0.5\textwidth, trim={2.5cm 6cm 2.5cm 6cm},clip]{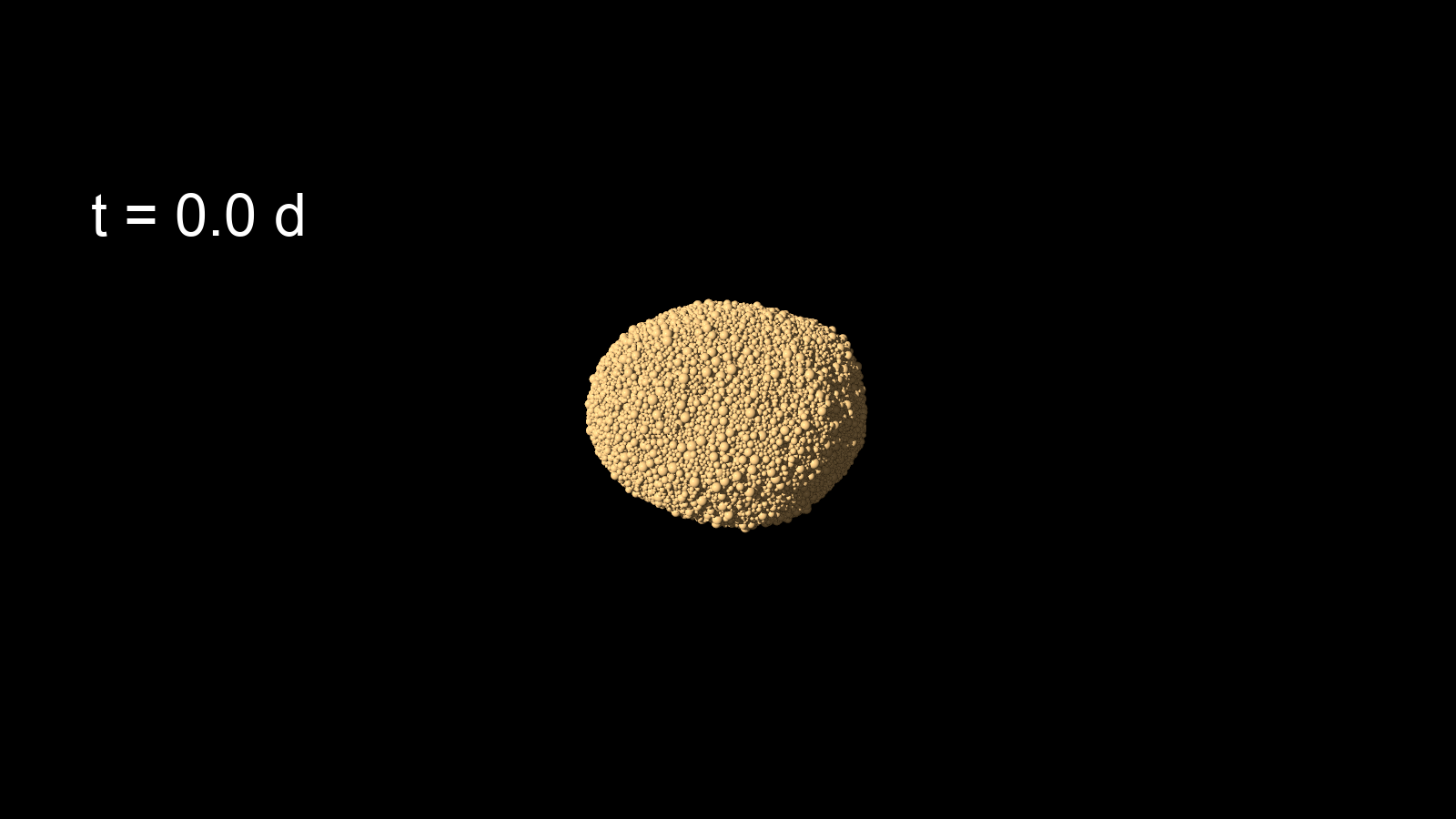}
\includegraphics[width=0.5\textwidth, trim={2.5cm 6cm 2.5cm 6cm},clip]{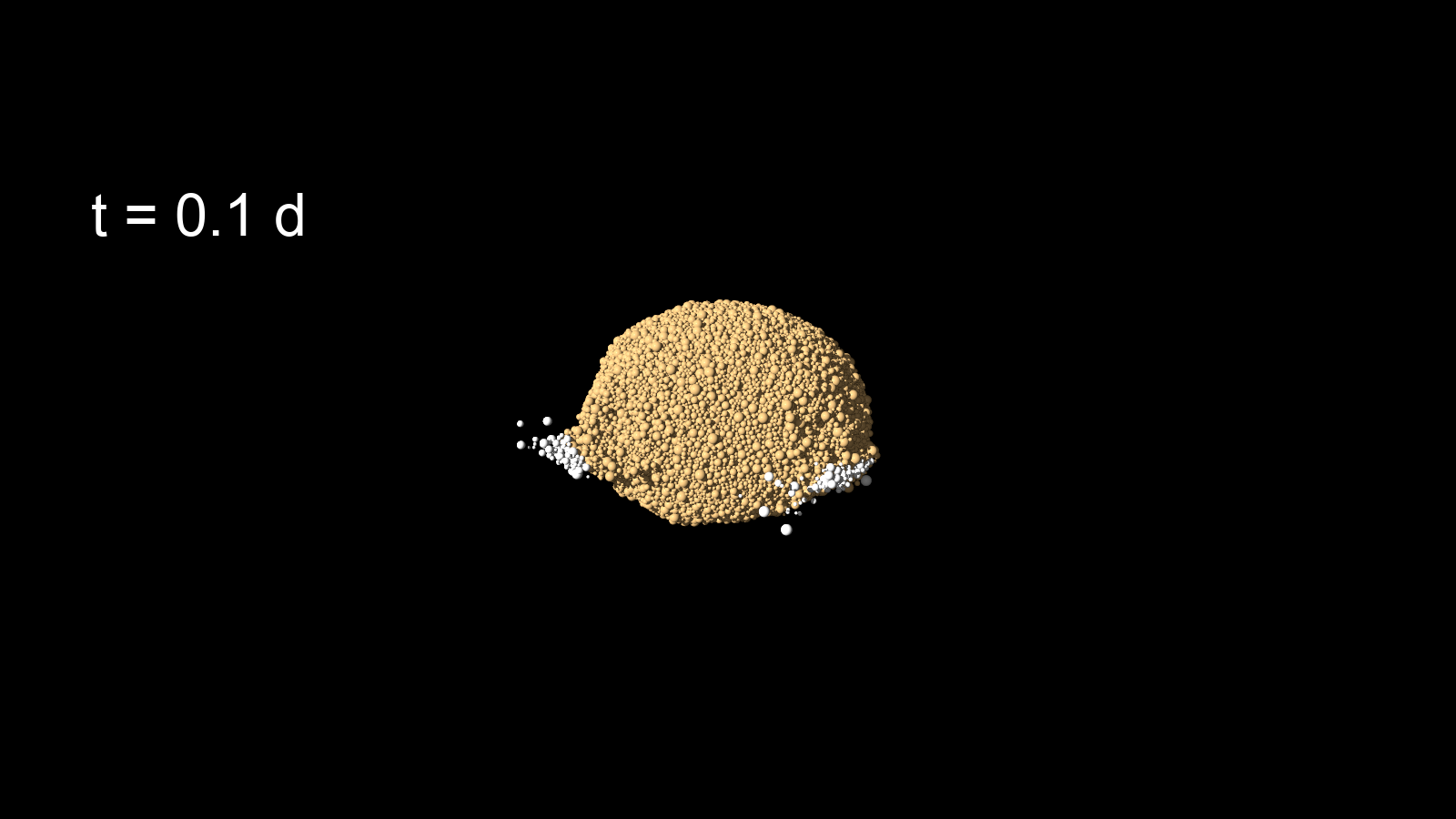}
\includegraphics[width=0.5\textwidth, trim={2.5cm 6cm 2.5cm 6cm},clip]{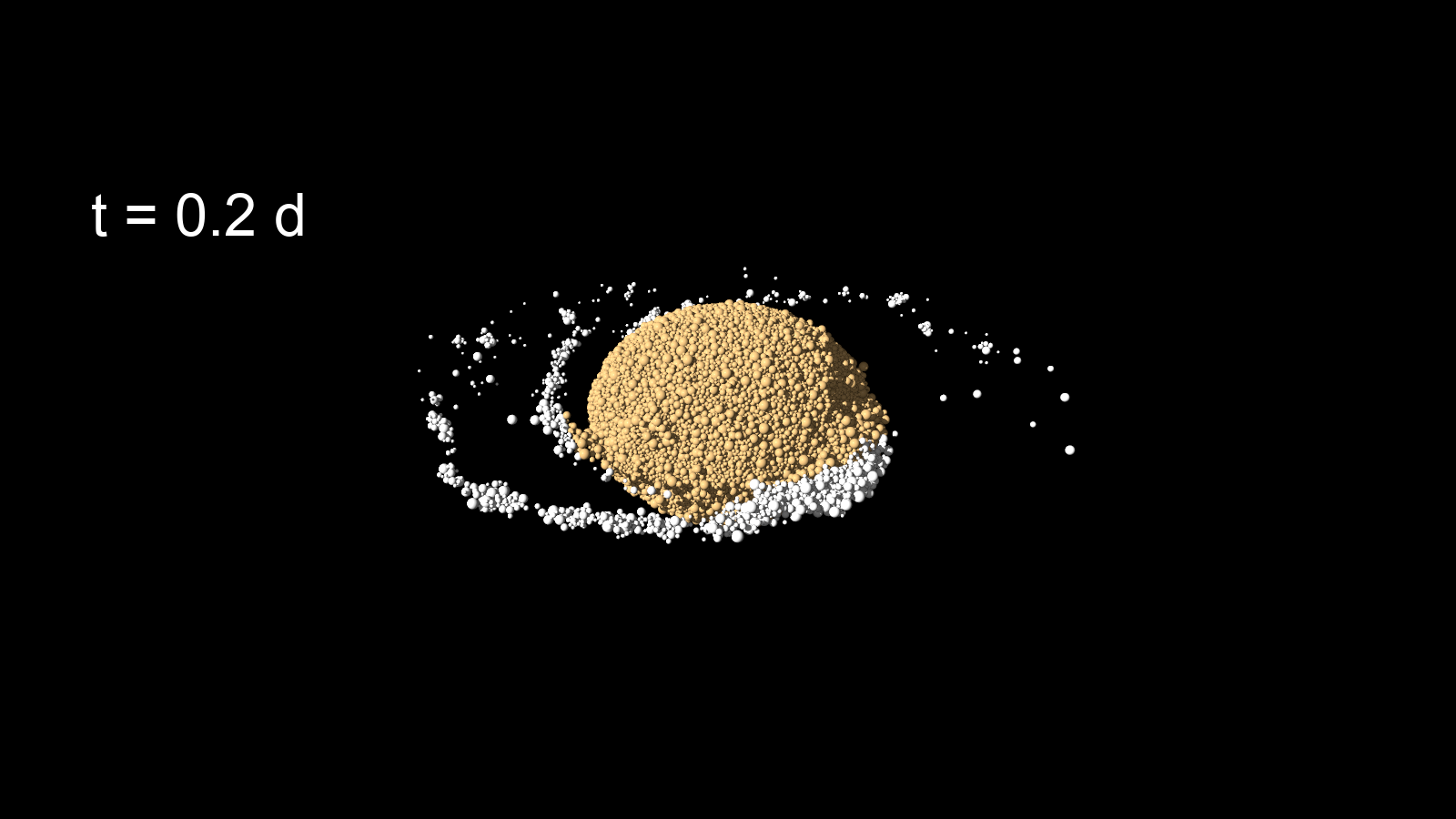}
\includegraphics[width=0.5\textwidth, trim={2.5cm 6cm 2.5cm 6cm},clip]{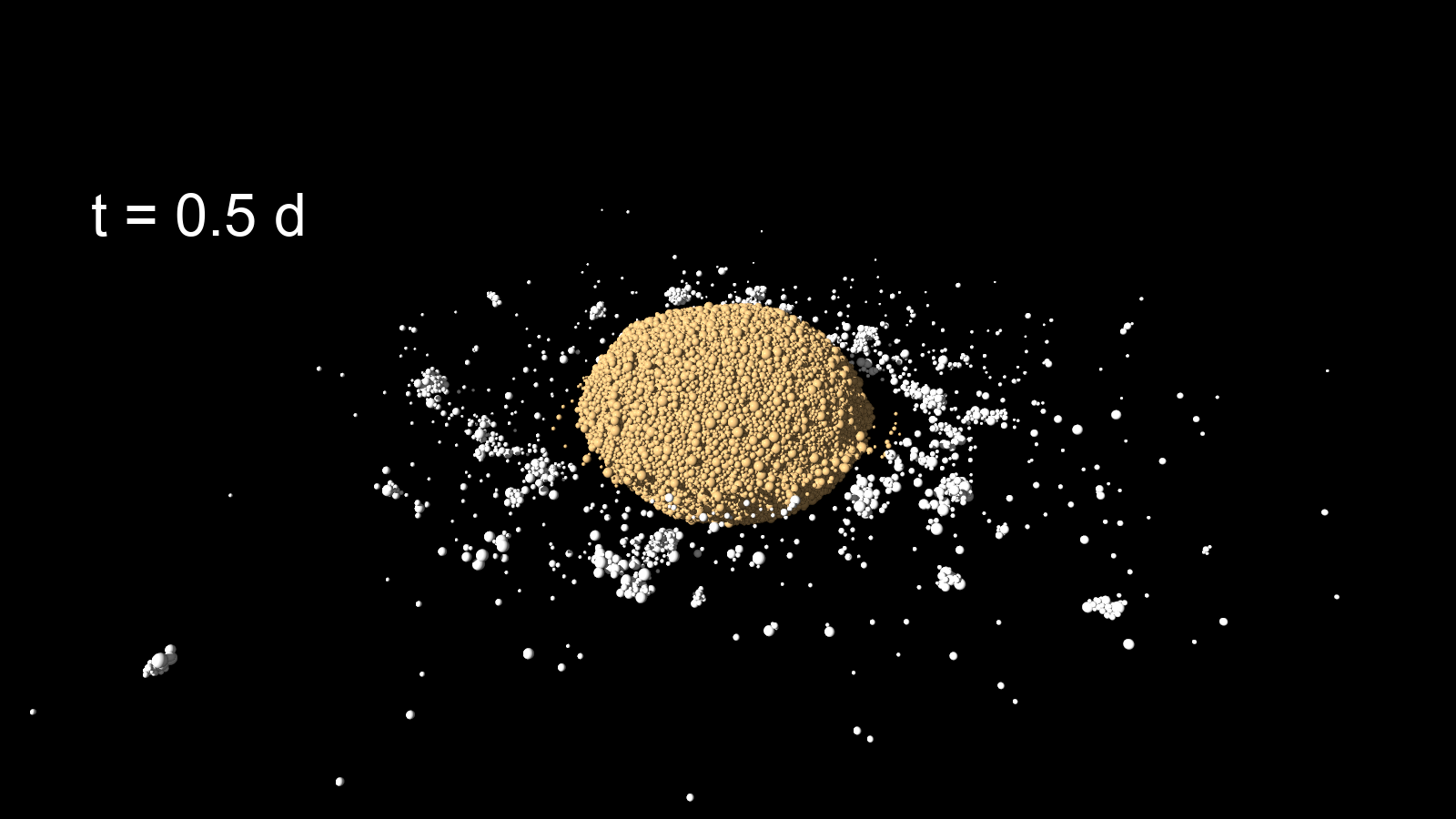}
\includegraphics[width=0.5\textwidth, trim={2.5cm 6cm 2.5cm 6cm},clip]{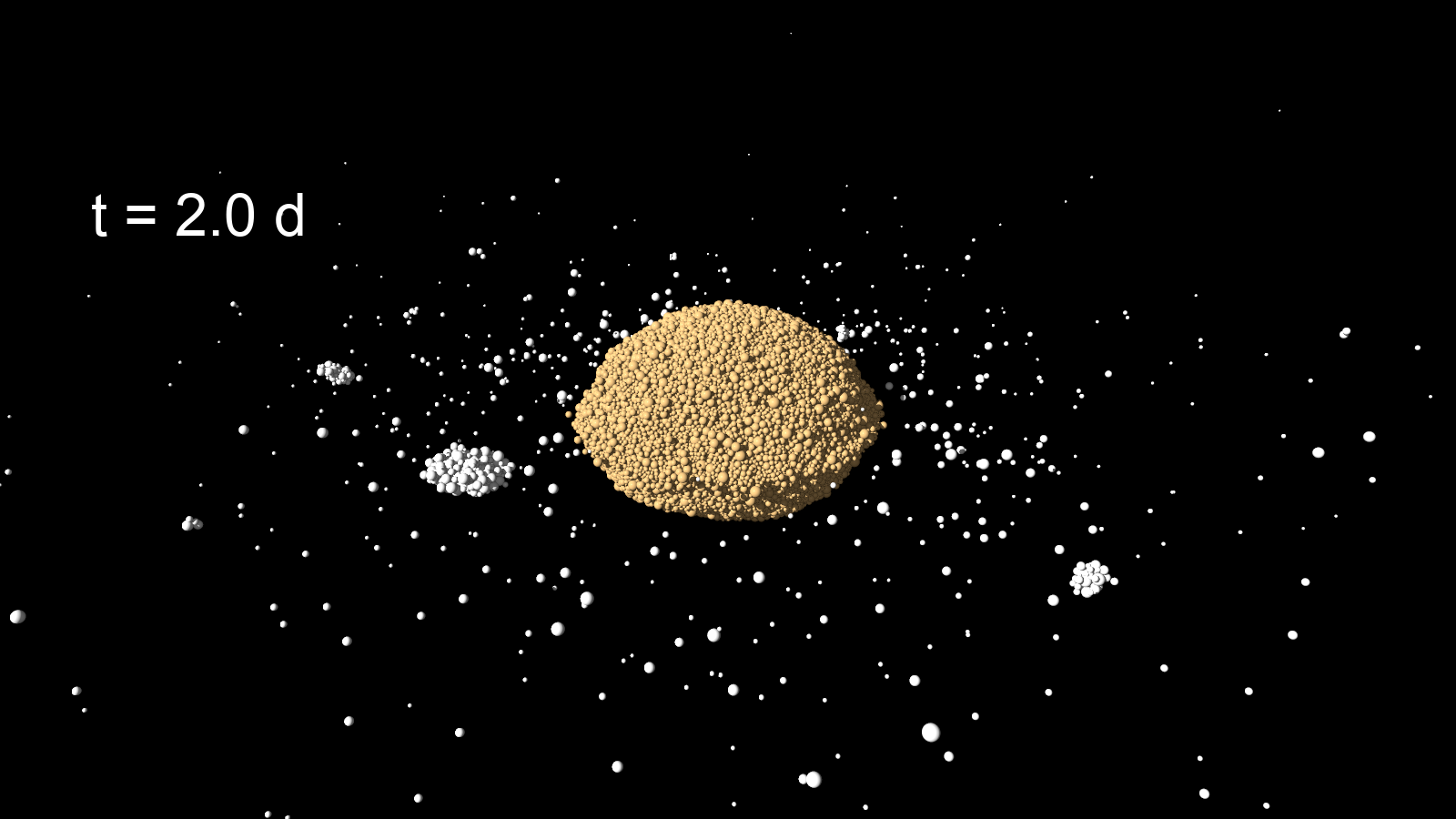}
\includegraphics[width=0.5\textwidth, trim={2.5cm 6cm 2.5cm 6cm},clip]{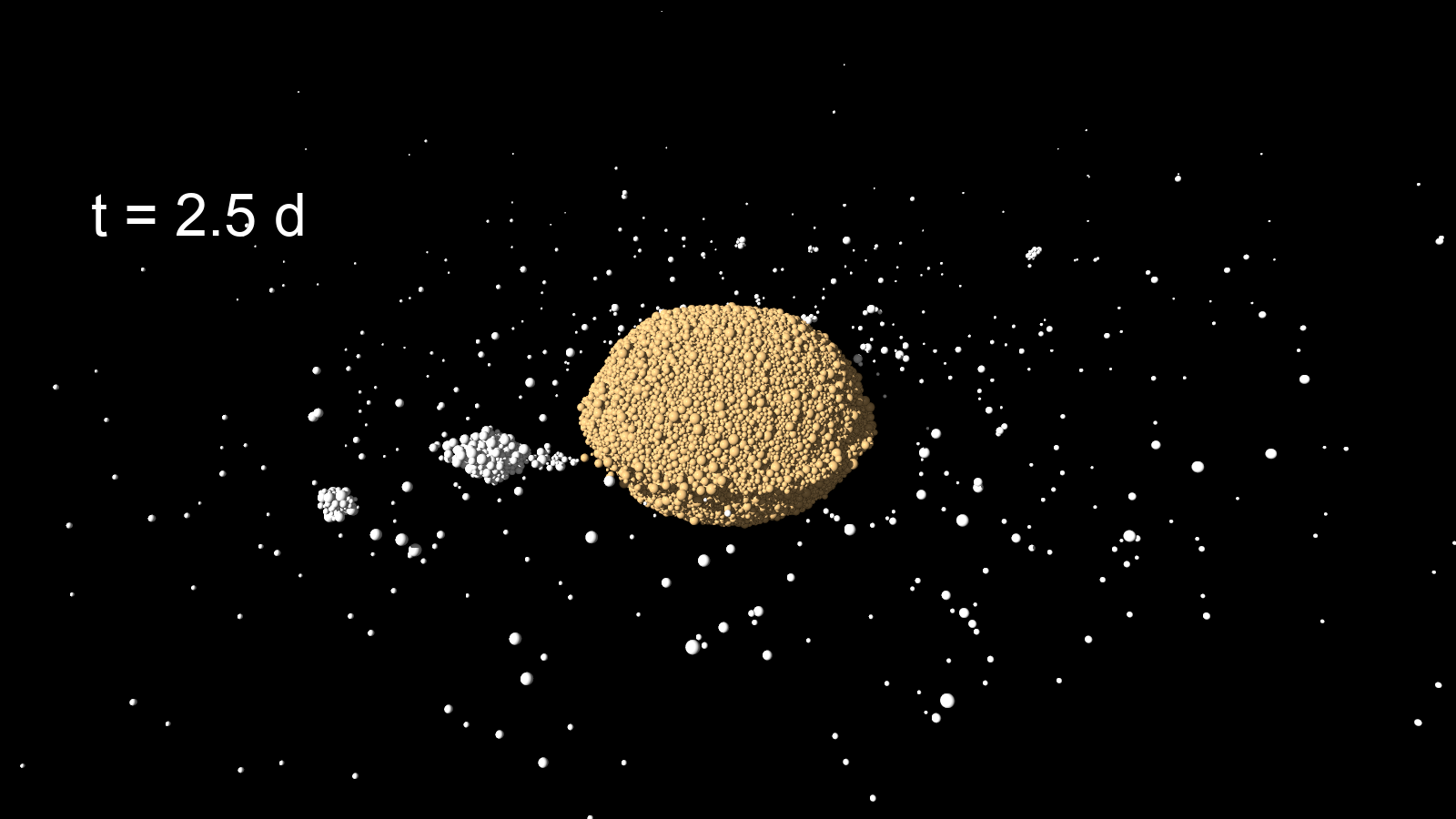}
\includegraphics[width=0.5\textwidth, trim={2.5cm 6cm 2.5cm 6cm},clip]{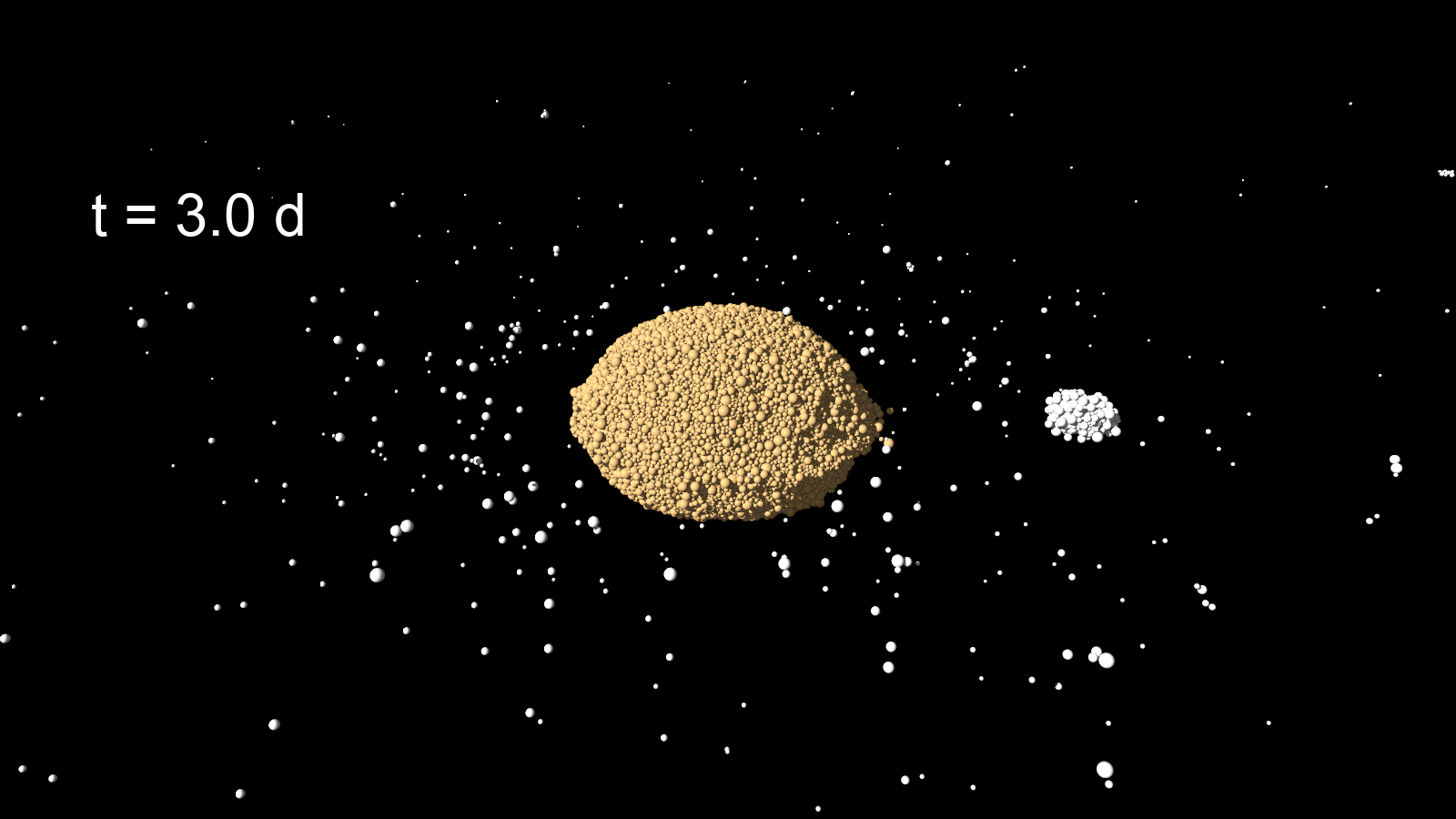}
\includegraphics[width=0.5\textwidth, trim={2.5cm 6cm 2.5cm 6cm},clip]{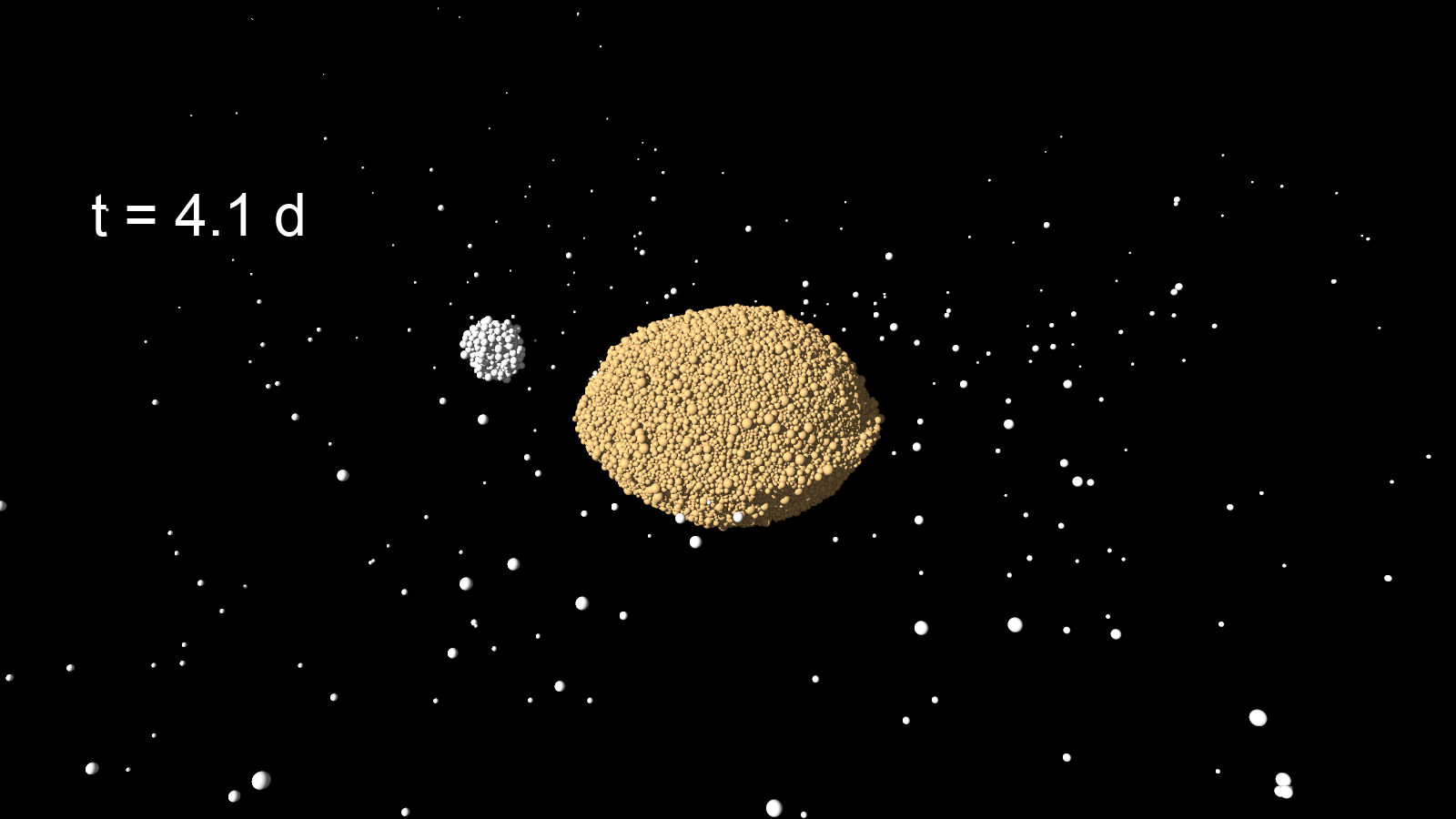}
\caption{An example of a satellite forming after mass shedding from a Didymos-shaped primary. Particles are colored in white being shed from the primary. The initial spiral arms are caused by mass shedding occurring at localized regions rather than across the entire body. By ${\sim}0.5$ d, this asymmetry largely smooths out in azimuth before moonlets begin forming. A movie of this simulation is provided.}
\label{fig:mosaic_yun}
\end{figure}

\begin{figure}[H]
\centering
\includegraphics[width=0.75\textwidth]{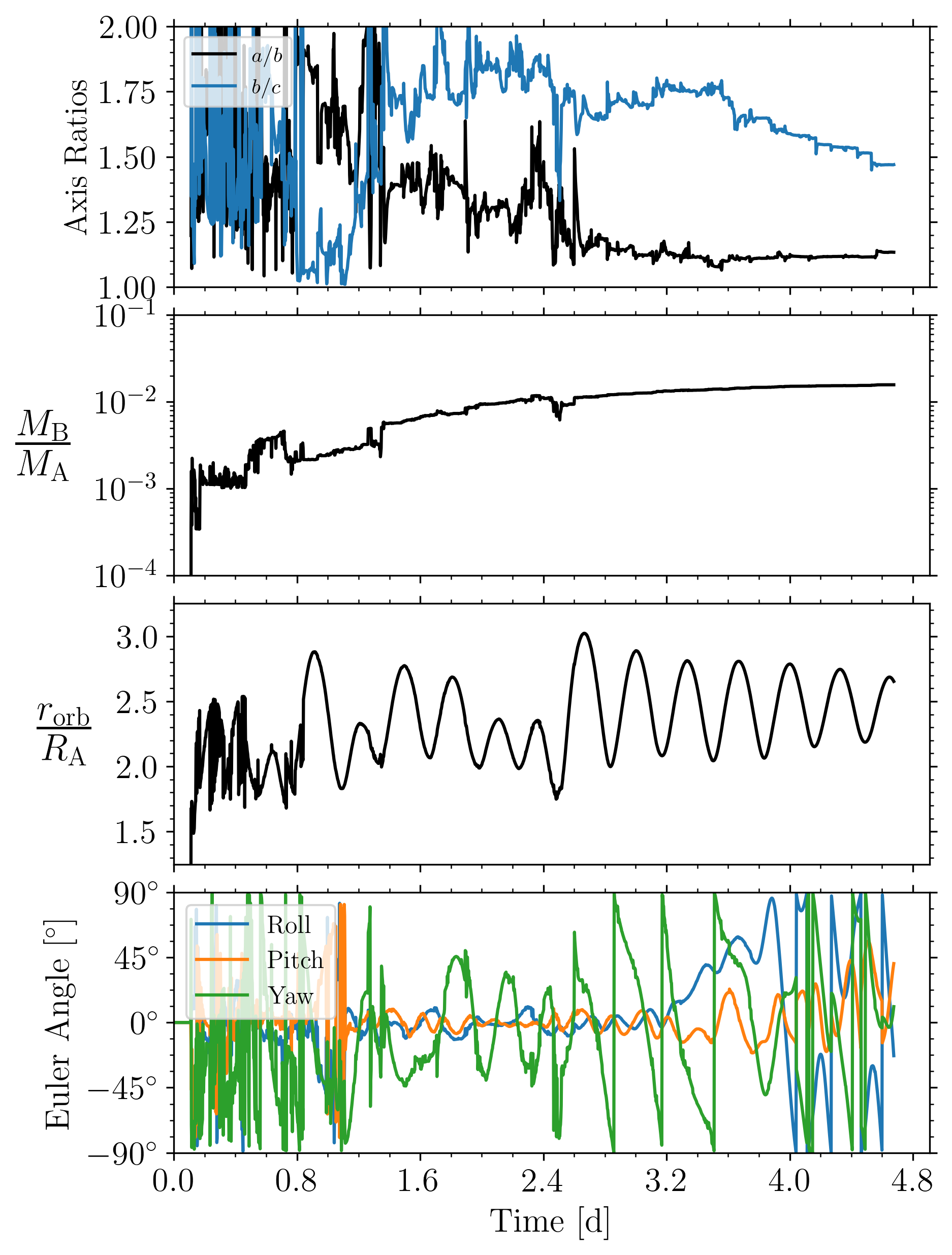}
\caption{A time series plot showing the shape ($a/b$ and $b/c$, based on the satellite's DEEVE), mass ratio ($M_\text{B}/M_\text{A}$), body separation ($r_\text{orb}/R_\text{A}$), and the attitude of the satellite formed via mass shedding. Within only several days, a Dimorphos-mass satellite is formed. The satellite is initially tidally locked to the primary (starting at ${\sim}1.3$ d), but then undergoes a partial tidal disruption followed by a merger with a moonlet, causing it to break from synchronous rotation (at ${\sim}2.5$ d). The satellite has an oblate shape, similar to Dimorphos. However, the satellite's shape, mass, orbit and rotational state will continue to change as it continues to accrete more material. }
\label{fig:timeSeries_yun}
\end{figure}

Snapshots of this simulation are shown in Fig.\ \ref{fig:mosaic_yun} along with a time-series plot of the satellite's shape, mass, orbit, and attitude in Fig.\ \ref{fig:timeSeries_yun}.  In this simulation, we form an approximately Dimorphos-mass satellite in a matter of only a few days. The satellite is also relatively oblate compared to the measured shapes of other binary systems, having axis ratios of $a/b{\sim}1.15$ and $b/c{\sim}1.5$, based on the satellites dynamically equivalent equal-volume ellipsoid (DEEVE)\footnote{For a body of mass $m$ and principal moments of inertia $A$, $B$, $C$, its corresponding dynamically equivalent equal-volume ellipsoid axis lengths $a$, $b$, $c$ are given by the following relations: $A = \frac{m}{5}(b^2 + c^2), B = \frac{m}{5}(a^2 + c^2), C = \frac{m}{5}(a^2 + b^2)$.}. The satellite is initially in synchronous rotation with a libration amplitude of ${\sim}45^\circ$. However, this tidally-locked state is broken when the satellite has a close encounter with a smaller moonlet, sending it inwards where it undergoes a partial tidal disruption followed by an immediate merger with an another moonlet. This sequence of events breaks the satellite's synchronous rotation state and happens to lead to a relatively oblate shape. We encourage the reader to view the provided movie of this simulation.

This simulation robustly demonstrates that a Dimorphos-like satellite can form rapidly from a single rotational disruption event. In addition, the satellite's series of close encounters and collisions with other moonlets demonstrates that this process is highly chaotic. An infinitesimal change to the initial conditions of this simulation could lead to a very different outcome in terms of the satellite's physical and dynamical properties. However, due to the computational expense of these high resolution, full spin-up-to-satellite-formation simulations, they are impractical for longer simulations as well as studying outcomes statistically, which is the focus of the rest of this study. {Before proceeding to the rest of this study, we note that the simulation presented in this subsection is a new result in its own right, although its primary purpose is to motivate the initial conditions used in the rest of this study. This simulation also highlights some key differences between our study and recently published papers on binary formation, which we address here.}

{\paragraph{Comparison with \cite{Sugiura2021} and \cite{Hyodo2022}}
Recently, \cite{Sugiura2021} studied the shape evolution of a rubble pile under YORP spinup using smoothed-particle hydrodynamics (SPH) simulations. In contrast to DEM simulations where particles represent discrete objects (i.e., rocks, boulders, etc.), SPH is used to simulate continuum mechanics where the particles sample local quantities such as density, internal energy, and pressure. \cite{Sugiura2021} find that spin-up can result in a ``top-shaped'' body for friction angles exceeding $70^\circ$. However, it is challenging to compare this study to the result obtained due to differences in  initial conditions, material models, and spin-up procedure. First, the work of \cite{Sugiura2021} starts from spherical shapes, whereas this simulation starts from a Didymos-like shape. Second, these simulations consider friction angles exceeding those of terrestrial and lunar granular material \citep[typically $\sim35{-}45^\circ$, e.g.,][]{Bareither2008,Beakawi2018,Mitchell1972} as well as friction angle estimates of recently visited asteroid surfaces \citep[also on the order of ${\sim}35^\circ$, e.g.,][]{Fujiwara2006,Watanabe2019,Barnouin2019,Barnouin2023_submitted,Robin2023_submitted}. This is because \cite{Sugiura2021} consider a single ``effective friction angle'' that accounts for both friction and cohesion. Cohesion is defined as the shear strength at zero pressure, where the shear strength of a granular material can be written as $Y=\tan(\phi)p + c$, where $\phi$ is the friction angle, $p$ is the confining pressure, and $c$ is the cohesion. \cite{Sugiura2021} instead adopt a material mod4el of the form $Y=\tan(\phi)p$ where $\phi$ is now an ``effective friction angle'' that encompasses the effects due to cohesion, which justifies their choice to explore values of $\phi$ that are significantly higher than real granular materials. With this material model, we can see that as the confining pressure goes to zero (i.e., near the surface), the shear strength goes to zero, which effectively results in the material having zero cohesion. Therefore, we suspect that the failure mechanisms observed by \cite{Sugiura2021} are a result of the rubble pile implicitly having a cohesionless upper layer with a stronger internal core. Ryugu and Bennu seem to have sub-Pa levels of cohesion at their surfaces and potentially some strength at depth, so this implicit assumption by \cite{Sugiura2021} isn't unreasonable \citep{Arakawa2020,Jutzi2022,Walsh2022,Barnouin2019}. However, recent work by \cite{Zhang2022} demonstrates that rubble pile failure mechanisms are highly sensitive to small changes in cohesion at a fixed friction angle, and to small changes in friction at constant cohesion, indicating that friction and cohesion cannot be combined into a single parameter and should be treated separately.}

{Finally, the spin-up procedure used here differs from \cite{Sugiura2021} in a critical way. Both methods apply a similar angular acceleration to the rubble piles (${\sim}10^{-10} \text{ rad s}^{-2}$), which ensures that the artificially induced Euler acceleration is always negligible compared to the centrifugal acceleration. However, \cite{Sugiura2021} spin up the rubble pile until 1$\%$ of its mass is put \textit{into orbit}, whereas this study halts the spin-up process at the moment of failure to ensure a realistic mass shedding process. Although the total angular momentum added during the mass-shedding process by \cite{Sugiura2021} is small, this can make a critical difference for a rubble pile on the edge of stability. As a result, \cite{Sugiura2021} typically find that ${\sim}10\%$ of the primary's mass ends up going into orbit following the spin-up process. \cite{Hyodo2022} then track the formation of moons after such a mass-shedding event, where they find similar formation timescales to what is presented in this study. However, due to their simulations starting with $10-20\%$ of the primary's mass in a debris disk, their simulations result in a system of satellites with a combined mass of ${\sim}5\%$ of the primary's mass. Although many binary asteroids have similarly high mass ratios (on the order of $5-10\%$), this model does not produce binary systems with smaller satellites, such as the Didymos system, where the mass ratio is ${\sim}1\%$ \citep{Pravec2016,Pravec2019}.}

{
\paragraph{Comparison with \cite{Madeira2023}}
In comparison to the simplified 1D model of \cite{Madeira2023}, there are several key differences to highlight based on this simulation. Of course, this model has the advantage of modeling the system for a much longer timescales that what is possible with a direct $N$-body approach. In addition, their model can provide useful insight into the formation of binaries. However, the result presented here (as well as the result of \cite{Hyodo2022}) demonstrates that the mass-shedding and accretion timescales are relatively short, meaning that it may not be necessary to simplify the problem to one dimension. In addition, the insight of the simplified model of \cite{Madeira2023} is only useful to the extent to which it reflects the true nature of the problem. \cite{Madeira2023} suppose an idealized disk in which only a single satellite accretes from the Roche limit at a time. On the contrary, this simulation demonstrates that following a realistic YORP-driven mass-shedding event, multiple moonlets can form simultaneously before undergoing a chaotic series of close encounters and mergers until one satellite remains, which cannot be captured by a 1D model. It is not clear that mass shedding would actually lead to a disk matching the assumptions of \cite{Madeira2023}.}

{Recently, \cite{Madeira2024} extended their study, claiming that their model explains the oblate shape of Dimorphos, the contact binary shape Dinkinesh's recently discovered satellite, Selam, and the prolate shapes of other binary asteroids, where the key difference between these three outcomes solely comes down to the mass-shedding timescale. Some caution should be applied here until the model can address some key issues. First, the mass-shedding timescales explored by \cite{Madeira2024} are chosen arbitrarily and need some physical justification, as there is currently no connection between the initial conditions of their model and YORP processes or the geophysical properties of the primary body. In addition, these 1D disk-moon models do not directly account for the gravitational interactions and subsequent mergers between moonlets. Rather, mergers are assumed to occur anytime two moonlets enter one another's Hill sphere. The shape outcome of these mergers is not demonstrated either; rather a post-merger shape is assumed based on simulations of mergers of Saturn's inner moons \citep{Leleu2018}. However, the geophysics and dynamics of mergers leading to the formation of Saturn's inner moons are very different than in the case of two merging satellites of an S-type asteroid. For example, the moons Pan, Atlas, and Prometheus have densities below $1 \text{ g cm}^{-3}$ and orbit within or near Saturn's Roche limit while Dimorphos is thought to have a bulk density on the order of ${\sim}2.4\text{ g cm}^{-3}$ and any mergers leading up to its formation are thought to occur beyond the Roche limit, according to \cite{Madeira2023}. Any model that can explain the shapes of Dimorphos, Selam, other binary satellites needs to self-consistently address the spin-up, mass-shedding, accretion, and (potentially) merger processes. For example, the shape of a post-merger satellite will depend on parameters such as its density, friction angle, and cohesion. At the same time, these parameters are intimately connected with the primary's failure mechanisms and the mass shedding process, which will then determine the initial conditions of any orbiting debris. This is a challenging problem, and the work presented here only attempts to address a small part of this process.}

\subsection{Simulation set up}\label{subsec:sim_setup}
The simulation shown in Figs.\ \ref{fig:mosaic_yun} and \ref{fig:timeSeries_yun} are computational expensive requiring ${\sim}10^5$ particles, are limited to short integration times and are impractical for determining statistical outcomes. Therefore, we use the above result, along with known properties of the Didymos system, to inform a simplified initial condition for a large suite of simulations. This allows us to determine the properties of the resulting satellite from a statistical point of view. It is important to note that the primary's shape, density, internal structure, and material properties will all play some role in determining the initial conditions of the shed material \citep[e.g.,][]{Sanchez2020, Walsh2012, Zhang2022, McMahon2020}. Therefore, the simplified initial conditions presented here are intentionally generic and may not be perfectly representative of a fully realistic scenario. However, we will show this generic initial condition produces many of the properties of both Dimorphos and other binary systems. 

\begin{figure}[H]
\centering
\includegraphics[height=0.4\textwidth]{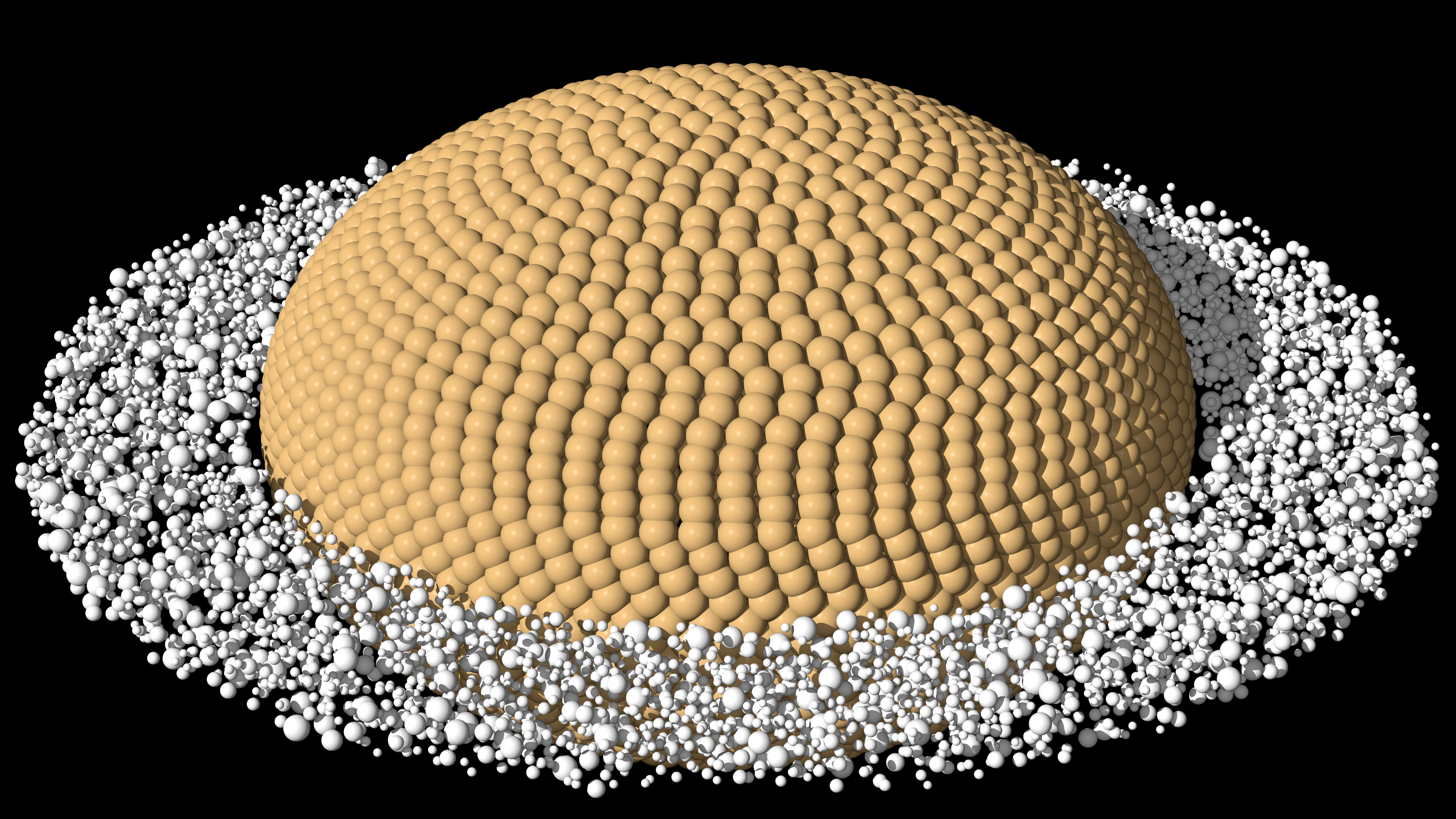}
\includegraphics[height=0.4\textwidth]{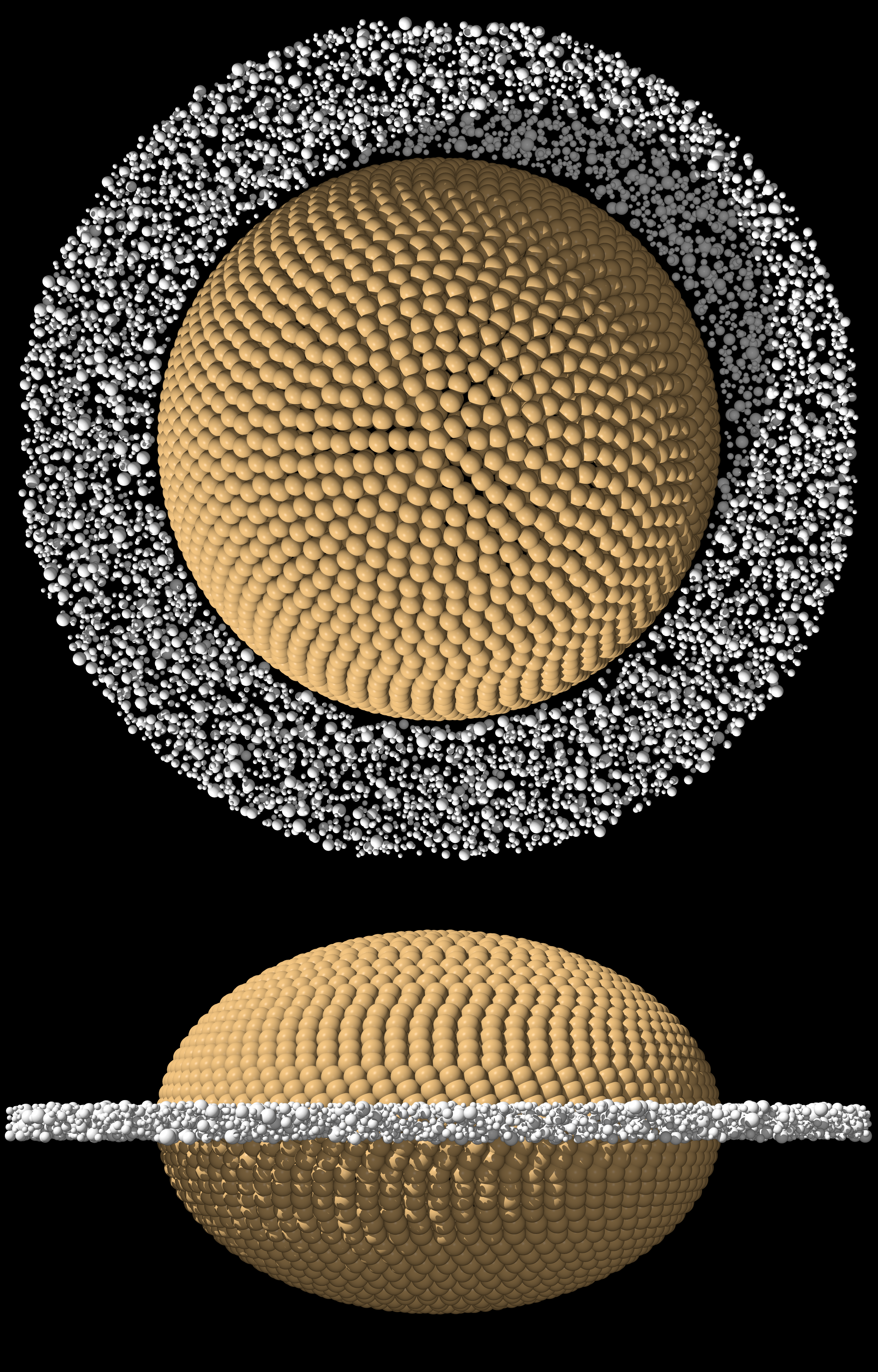}
\caption{A view of the initial conditions from the disk simulations. The primary (gold colored particles) behaves as a single rigid body.}
\label{fig:init_example}
\end{figure}

The primary was treated as a uniform oblate spheroid with semiaxes $a=b>c$, with an equatorial radius of $a=b=400$ m and $a/c=1.5$, and a bulk density of $2.4 \text{ g cm}^{-3}$, similar to the best estimate for Didymos's shape and density at the time this study began \citep{Daly2023}. In order to reduce the total number of particles in the simulation, the primary is modeled as a hollow shell of particles locked together into a rigid aggregate. In order for the primary to have moments of inertia as if it were a solid body (which is important for both gravity and collisions), we include a single point-mass particle at the center, and then adjust the mass of the central particle and the shell particles to achieve moments of inertia of an equivalent uniform oblate spheroid\footnote{For a homogeneous spheroid, the moments of inertia can be written in terms of its mass $M$ and its semi-axes $a,b,c$: $A=\frac{M}{5}(b^2 + c^2), B=\frac{M}{5}(a^2+c^2), C=\frac{M}{5}(a^2+b^2)$.}. In other words, the point-mass and surrounding spheroidal shell collectively behave as if they are a single, solid body with the correct moments of inertia. As shown in Fig.\ \ref{fig:init_example}, the radius of the shell particles are over-inflated to prevent small particles from falling through the cracks. This approach enables a physically realistic solid-body primary without requiring an excessive number of particles in the interior. The spin period of the primary is initially set to 2.5 h in all simulations, approximately matching the rotation period of the primary from Section \ref{subsec:mass_shedding_example} following its rotational disruption.

 The orbiting material is generated following a power-law size-frequency distribution (SFD), with a power-law index of -3 and particle radii ranging between 3 and 10 m. This SFD is informed by (but does not match exactly) the observed boulder SFD on Dimorphos \citep{Pajola2022, Pajola2023_LPSC}, as well as the boulder SFD seen on other rubble pile asteroids \citep[e.g.,][]{Dellagiustina2019, Burke2021, Michikami2019, Michikami2021}. Including the smallest boulders observed on Dimorphos is impractical due to computational constraints; however, the smallest particles in the simulation are 6 m in diameter, approximately the size of the two largest boulders \textit{Bodhran Saxum} (${\sim}6.1$ m) and \textit{Atabaque Saxum} (${\sim}6.5$ m) near the DART impact site \citep{Daly2023}. Therefore, we consider this particle resolution sufficient for modeling the formation and shape of a Dimorphos-like satellite, as the particle sizes in these simulations approach the actual sizes of individual large boulders on Dimorphos's surface. Cohesion between boulders is ignored in these simulations, as cohesion forces on rubble piles likely arise from Van der Waals forces between fine-grained material less than ${\lesssim}$10 $\mu$m \citep[e.g.,][]{Sanchez2014}. In a mass shedding scenario, much of this fine-grained material (if present) would be released when the mass is first shed, and when moonlets undergo collisions and tidal disruptions. For a Dimorphos-like system, solar radiation pressure would rapidly remove these fine grains \citep[e.g.,][]{Ferrari2022b}. Therefore, we might expect the cohesion of the secondary to be no higher than the primary's cohesion (if present). This may explain why Didymos's surface requires a small amount of cohesion to maintain its surface stability while Dimorphos does not \citep{Barnouin2023_submitted}.

A disk of orbiting debris is placed in a circular region extending out to 1.5 times the equatorial radius of the primary, which approximately corresponds to the effective Roche limit for a body with a friction angle of $35^{\circ}$ \citep{Holsapple2006}. The disk also has a finite thickness of 40 m (roughly two diameters of the largest disk particle), to allow enough space to initialize the required number of particles. The example simulation presented in Section \ref{subsec:spinup} demonstrates that mass shedding does not lead to a stable  disk but instead material clumps almost immediately to form moonlets. To approximate this effect here, particles are purposefully not given any initial velocity dispersion to ensure a disk that is gravitationally unstable (i.e., a Toomre $Q<1$) so material will begin clumping immediately. All disk particles have a density of $3.5 \text{ g cm}^{-3}$, to match the grain density of L and LL chondrites \citep{Flynn2018}, which are the best meteoritic analogues for Didymos \citep{deLeon2006,Dunn2013}.

Each disk particle is placed on a circular orbit, accounting for the mass of the primary and its oblateness, as well as the mass of all other enclosed disk particles. Each disk particle's mean motion (i.e., its initial orbital angular velocity), can be written as a function of its radial distance $r_i$,
\begin{equation*}
    n_i^2 = \frac{GM_\text{A}}{r_i^3}\bigg(1 + \frac{3}{2}J_2 \bigg(\frac{R^\text{eq}_\text{A}}{r}\bigg)^2\bigg)  + \frac{M_\text{disk}(<r_i)}{r_i^3},
\end{equation*}
where $M_A$, $R^{eq}_A$, and $J_2$ are the primary's respective mass, equatorial radius, and oblateness, $M(<r_i)$ is the mass of all the disk particles enclosed within particle $i$'s position, and $G$ is the gravitational constant. {Owing to the finite thickness of the disk, particles are initialized with inclinations on the order of $2-3^\circ$. Particles are generated in a symmetric disk to simplify the process of generating initial conditions, which is not a realistic starting condition. However, we emphasize that, owing to the disk being gravitationally unstable, collisions and close encounters quickly excite the eccentricities and inclinations of the disk particles, which we deomnstrate below rapidly leads to particle orbits representative of the more realistic starting conditions following a mass-shedding event, such as that shown in Fig.\ \ref{fig:mosaic_yun}.}

Any particles that reach a distance of 40 km ($100R^{eq}_A$) are automatically deleted from the simulation. This boundary was set purely as a precaution to prevent a single particle from being ejected from the system, which could significantly slow down \textsc{pkdgrav}'s tree due to the single extremely distant particle, and corresponds to ${\sim}2/3$ of the system's Hill sphere if it were located at 1 AU. We found that a small fraction of particles end up reaching this boundary and are deleted. Typically only 1-2\% of the disk's initial mass is ejected and we verified that the vast majority of the particles that hit this boundary and were deleted were either on escape trajectories ($e_\text{orb}>1$), or had an apoapse distance well outside the Hill sphere and would have not returned to the binary system. Therefore, deleting these particles has a negligible effect on the binary's formation. In a some cases, a large aggregate is ejected from the system forming an asteroid pair and is discussed in Section \ref{subsec:pairs}.

A core motivation of this study was to demonstrate that a Dimorphos-sized satellite could plausibly form via a single mass shedding event. Therefore, most simulations simply vary the initial mass in the disk (i.e., number of particles), however we also vary the friction angle of the material between $29^\circ$ and $40^\circ$. For each set of parameters, we randomize the initial locations of the particles to understand the chaotic nature of the satellites formation. For each disk mass ($M_\text{disk}$) and friction angle ($\phi$), we run a given number of `clone' simulations, which are listed in Fig.\ \ref{tab:params}. In total, we run 120 simulations. 96 of the simulations have $\phi=35^\circ$ and have an initial disk mass ranging between $0.02M_{A}$ and $0.04M_{A}$. These 96 simulations constitute the bulk of the results shown in the following sections. With the remaining 24 simulations, $M_\text{disk}$ is kept fixed at $0.03M_{A}$ and friction angles of $29^\circ$, $32^\circ$, and $40^\circ$ are tested. All simulations were limited to ${\sim}100$ d ($5.8\times10^7$ steps) due to the high computational cost imposed by the small timestep. Each simulation is assigned a number between 1 and 120, and the result of each simulation is tabulated in Table \ref{tab:sim_results} in the Appendix. 

\begin{deluxetable*}{ccccc}
\tablenum{1}
\tablecaption{Number of simulations for each combination of $M_\text{disk}$ and $\phi$, along with the static friction coefficient ($\mu_S$) and shape parameter ($\beta$) used to achieve the given friction angle ($\phi$).\label{tab:params}}
\tablewidth{0pt}
\tablehead{
\colhead{$M_\text{disk}/M_\text{A}$} & \colhead{$\phi$} & \colhead{$N_\text{sims}$} & \colhead{$\mu_S$} & \colhead{$\beta$}}
\startdata
0.02 & $35^\circ$  & 32 & 0.6 & 0.5\\
0.03 & $35^\circ$  & 32 & 0.6 & 0.5\\
0.04 & $35^\circ$  & 32 & 0.6 & 0.5\\
0.03 & $29^\circ$  & 8 & 0.2 & 0.3\\
0.03 & $32^\circ$  & 8 & 0.4 & 0.4\\
0.03 & $40^\circ$  & 8 & 1.0 & 0.8\\
\enddata
\tablecomments{In all simulations, the primary consists of 2,371 particles, while the number of debris particles varies between ${\sim}4,200$ and ${\sim}8,400$, depending on the total mass in the disk. See Table \ref{tab:sim_results} to see a full listing of each simulation and its result.}
\end{deluxetable*}

\section{Results}\label{sec:results}
\begin{figure}[H]
\includegraphics[width=0.5\textwidth, trim={2.5cm 6cm 2.5cm 6cm},clip]{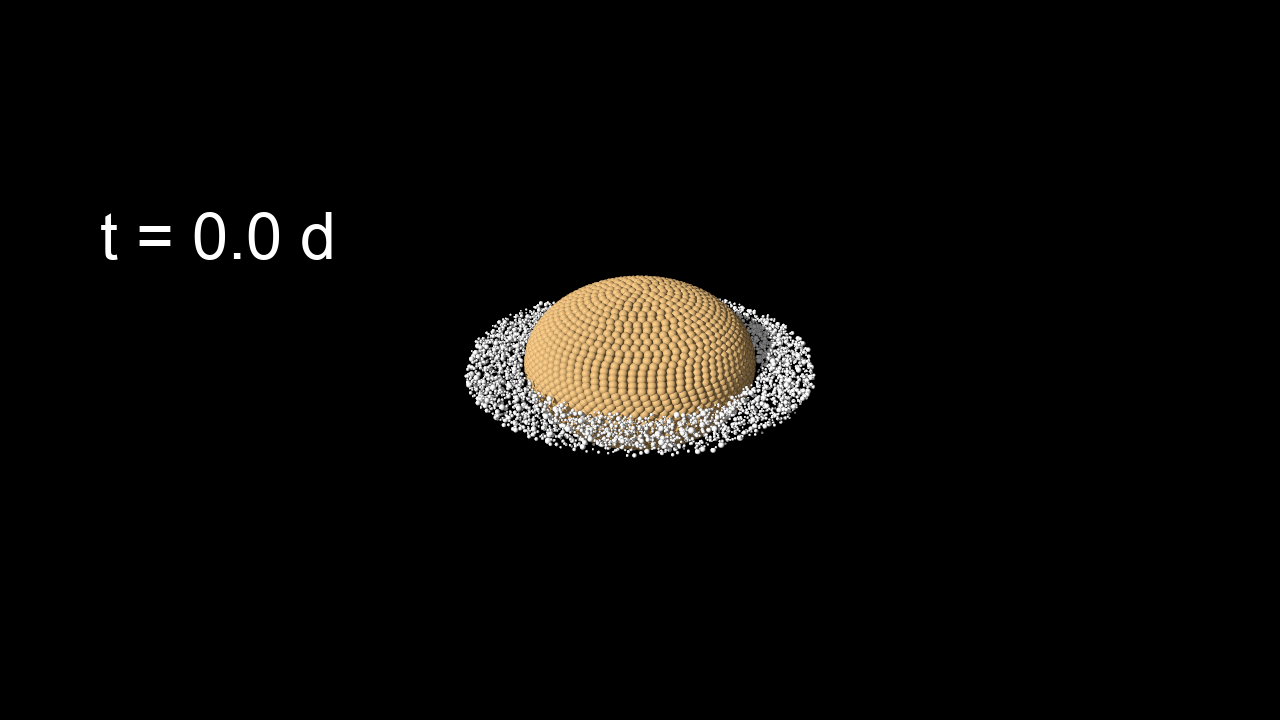}
\includegraphics[width=0.5\textwidth, trim={2.5cm 6cm 2.5cm 6cm},clip]{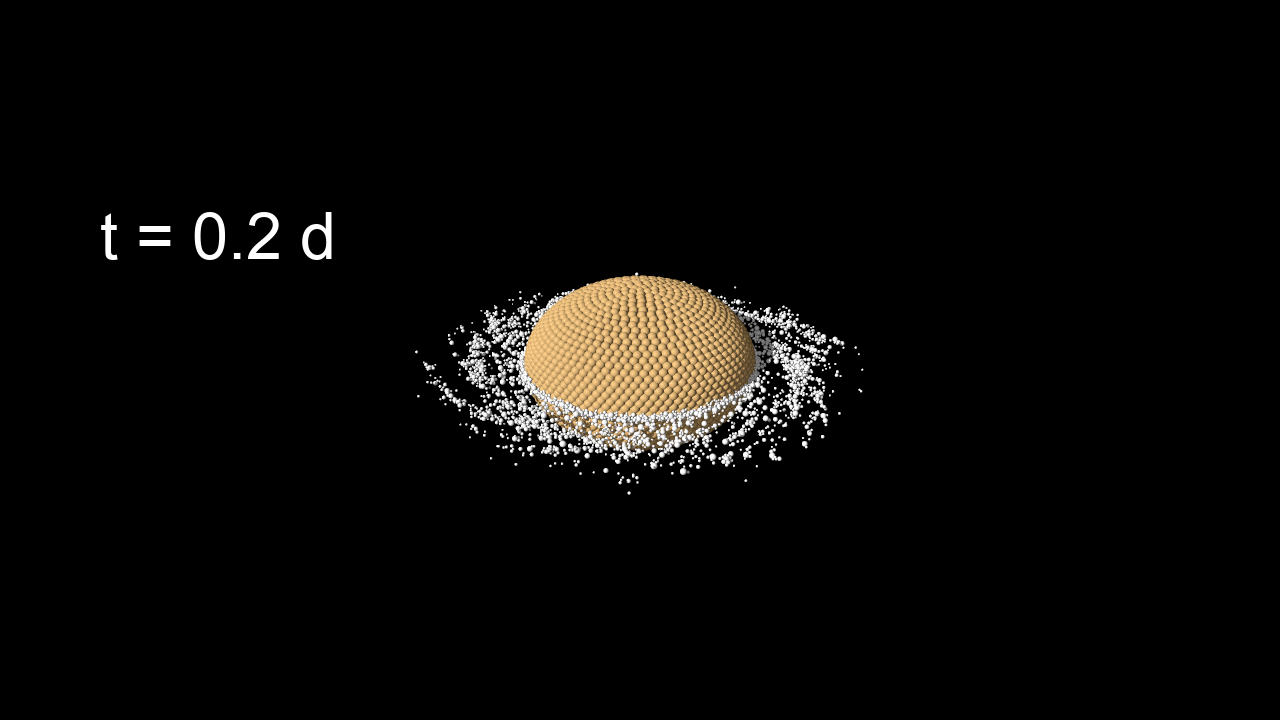}
\includegraphics[width=0.5\textwidth, trim={2.5cm 6cm 2.5cm 6cm},clip]{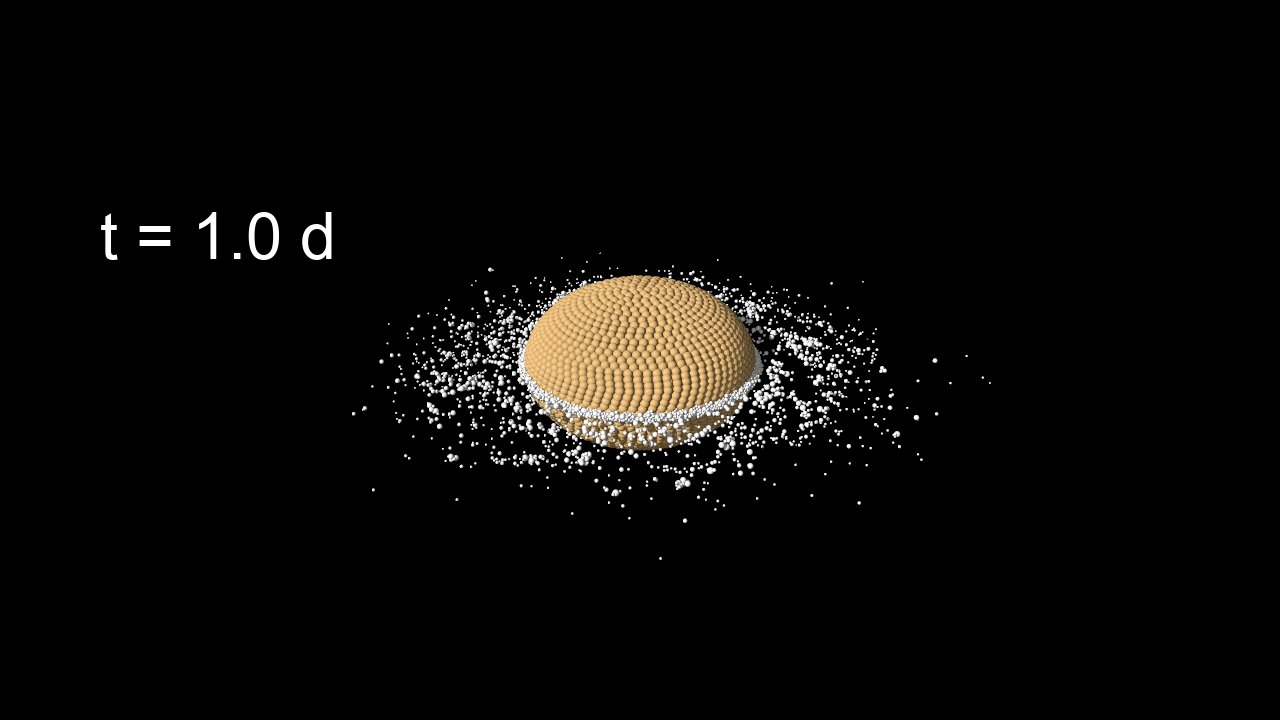}
\includegraphics[width=0.5\textwidth, trim={2.5cm 6cm 2.5cm 6cm},clip]{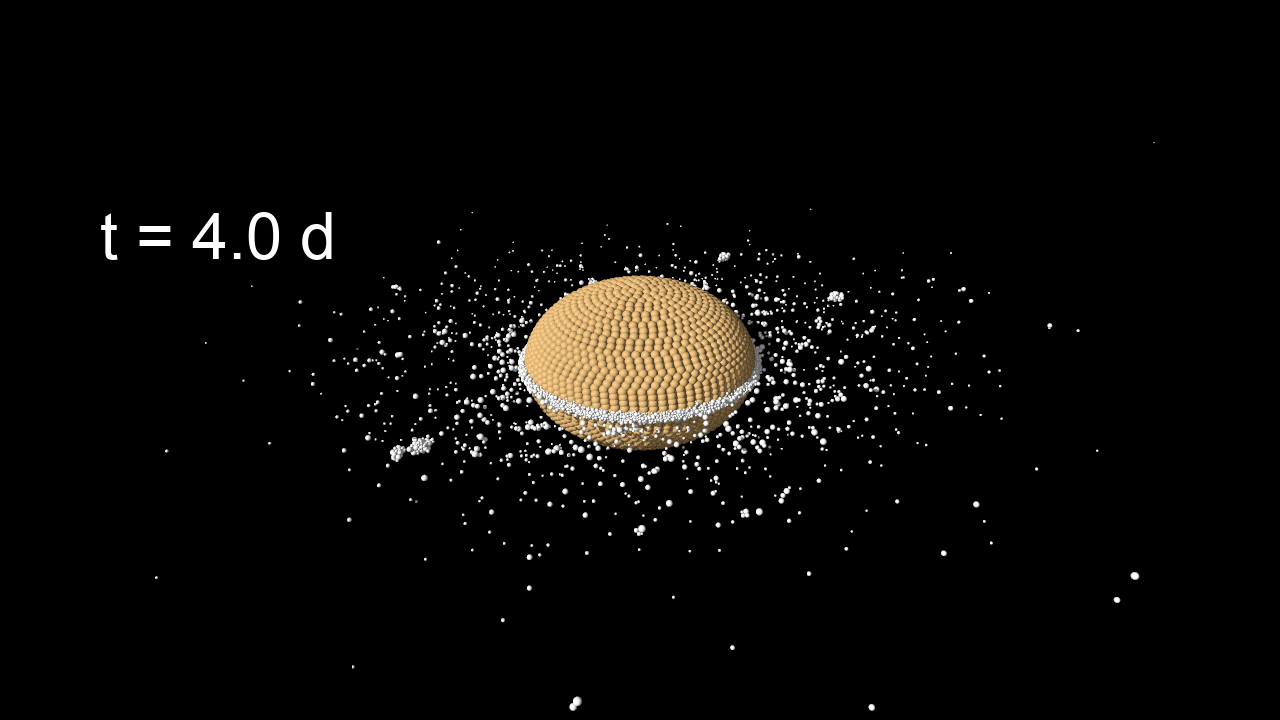}
\includegraphics[width=0.5\textwidth, trim={2.5cm 6cm 2.5cm 6cm},clip]{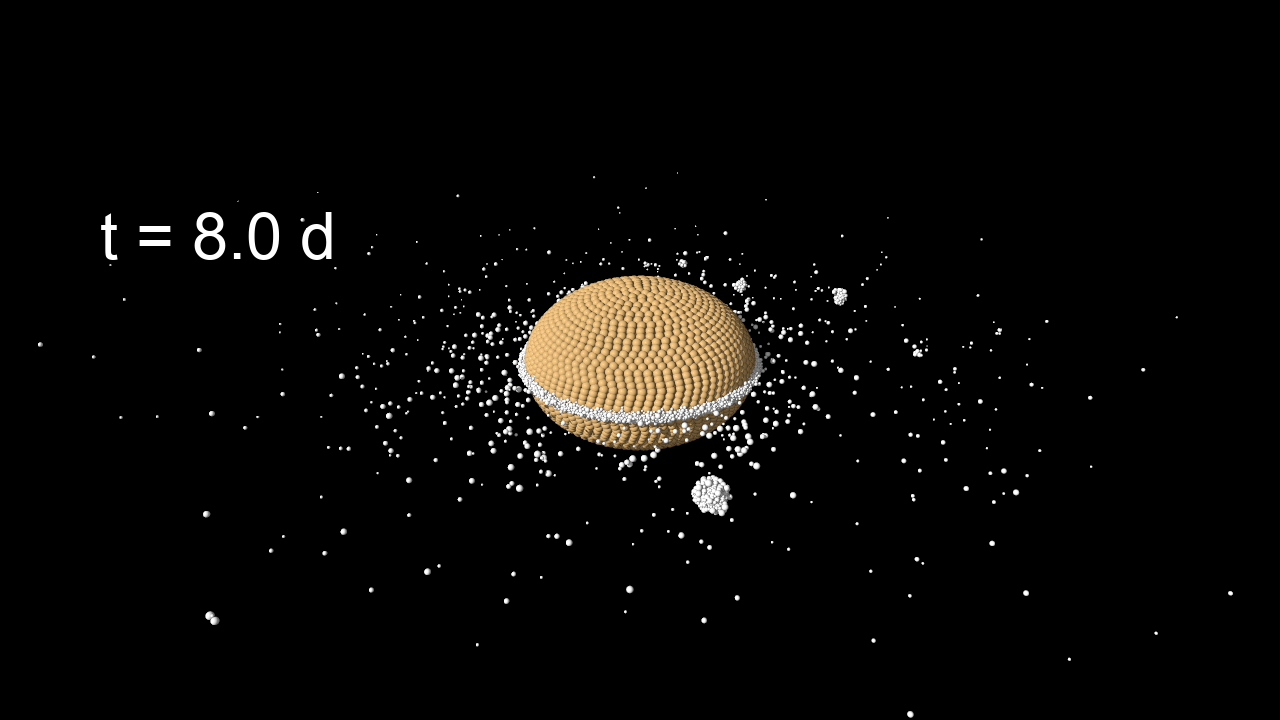}
\includegraphics[width=0.5\textwidth, trim={2.5cm 6cm 2.5cm 6cm},clip]{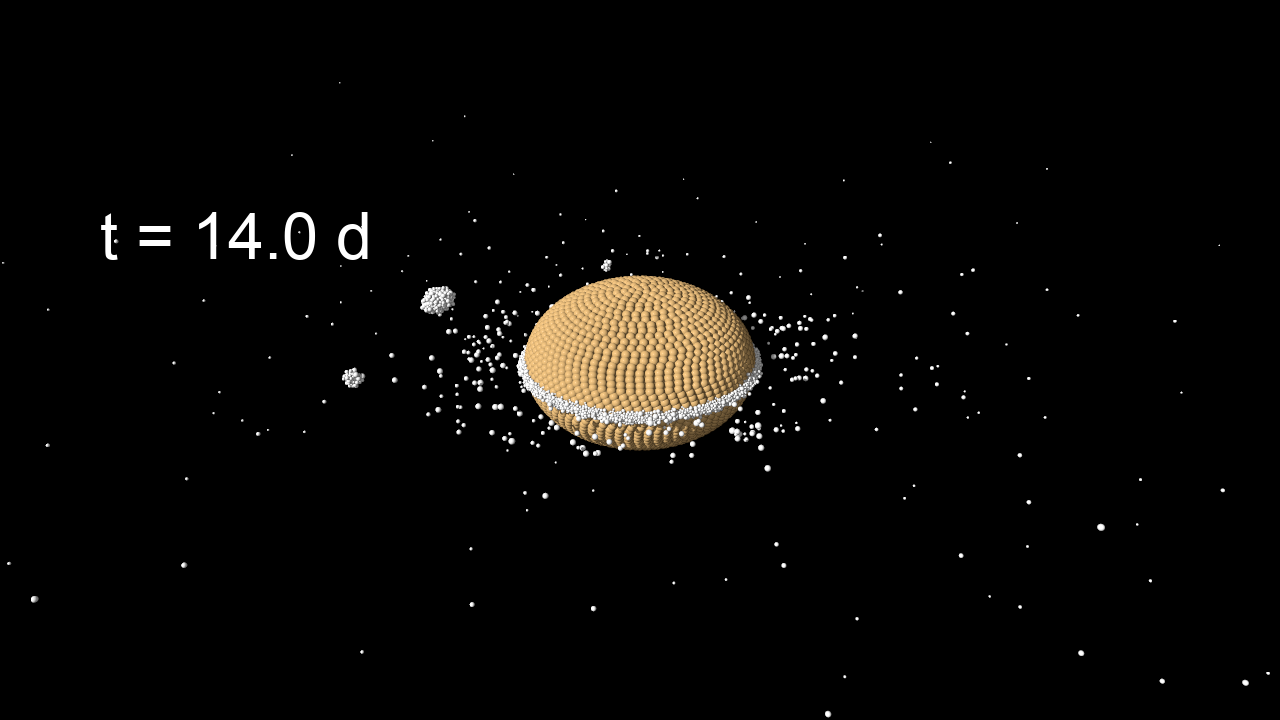}
\includegraphics[width=0.5\textwidth, trim={2.5cm 6cm 2.5cm 6cm},clip]{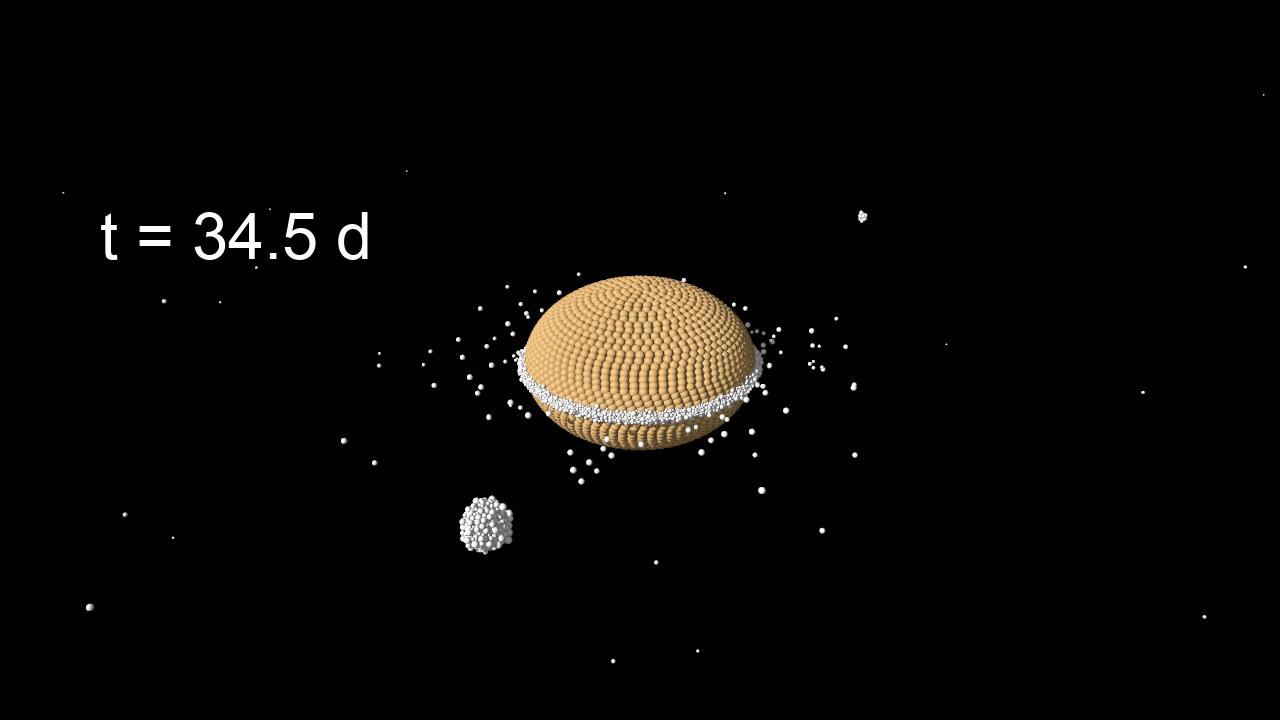}
\includegraphics[width=0.5\textwidth, trim={2.5cm 6cm 2.5cm 6cm},clip]{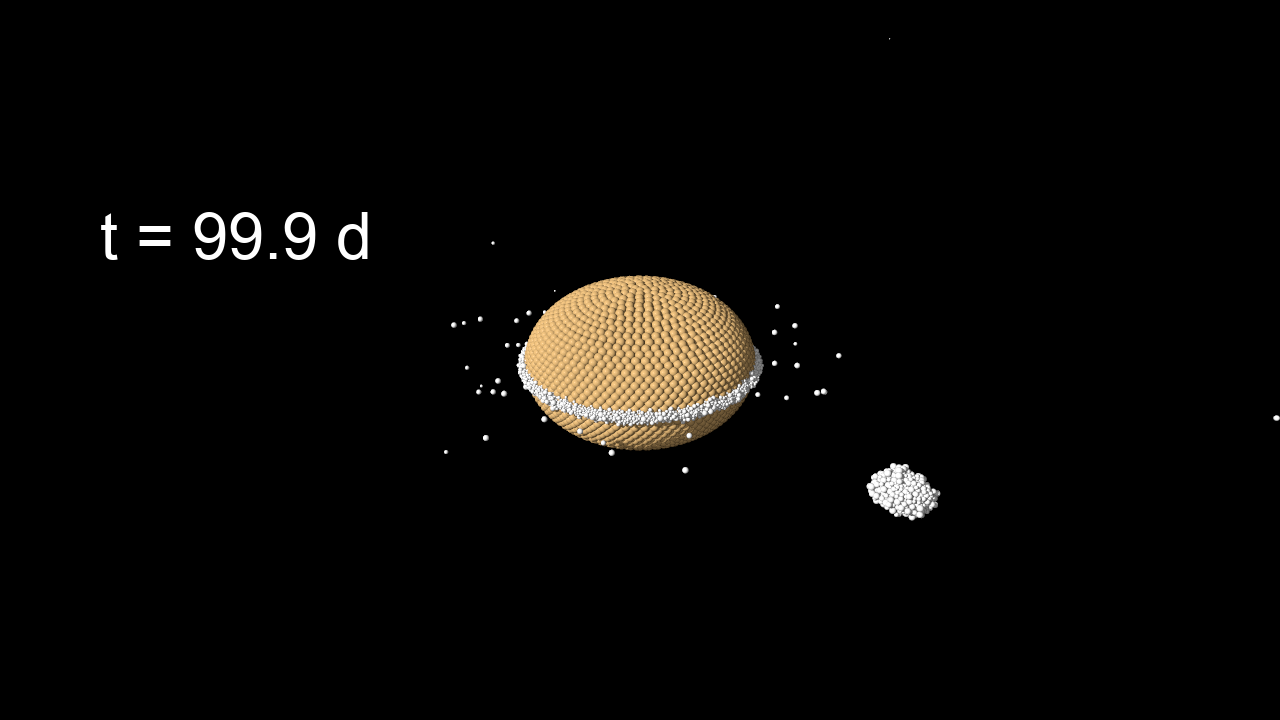}

\caption{An example of a simulation where the secondary is formed in synchronous rotation. This is Disk 008 in Table \ref{tab:sim_results}, and has an initial disk mass of $M_\text{disk} = 0.02M_\text{A}$ and a friction angle of $\phi=35^\circ$. A movie of this simulation is available.}
\label{fig:typical_example_mosaic}
\end{figure}

{\subsection{An example case}\label{subsec:example}}
First, we show a representative simulation to demonstrate the model. In Fig.\ \ref{fig:typical_example_mosaic} we show 8 snapshots of a simulation in which the $M_\text{disk}=0.02M_\text{A}$ and $\phi=35^\circ$. Almost immediately, particles begin clumping and forming short-lived spiral arms due to Keplerian shear, as a result of the disk being gravitationally unstable \citep[e.g.][]{Kokubo2000}. A couple days later, several moonlets form and undergo a chaotic series of close encounters and mergers until a single large satellite remains. Interestingly, the satellite is born in a tidally-locked configuration with the primary. The immediate tidal locking of the secondary is likely due to several factors, including its low eccentricity and formation near the Roche limit. In addition, all the mergers in this simulation occur rather gently without significantly perturbing the secondary's spin state. As we will see in the next several subsections, the chaotic nature of the satellite's formation leads to a wide range out outcomes in terms of the satellite's physical and dynamical properties. Therefore, this simulation is not necessarily typical of all cases.

% phi35/mDisk0.02/rMin3.0/disk_07
% [trim={left bottom right top},clip]

\begin{figure}[H]
\centering
\includegraphics[width=0.5\textwidth]{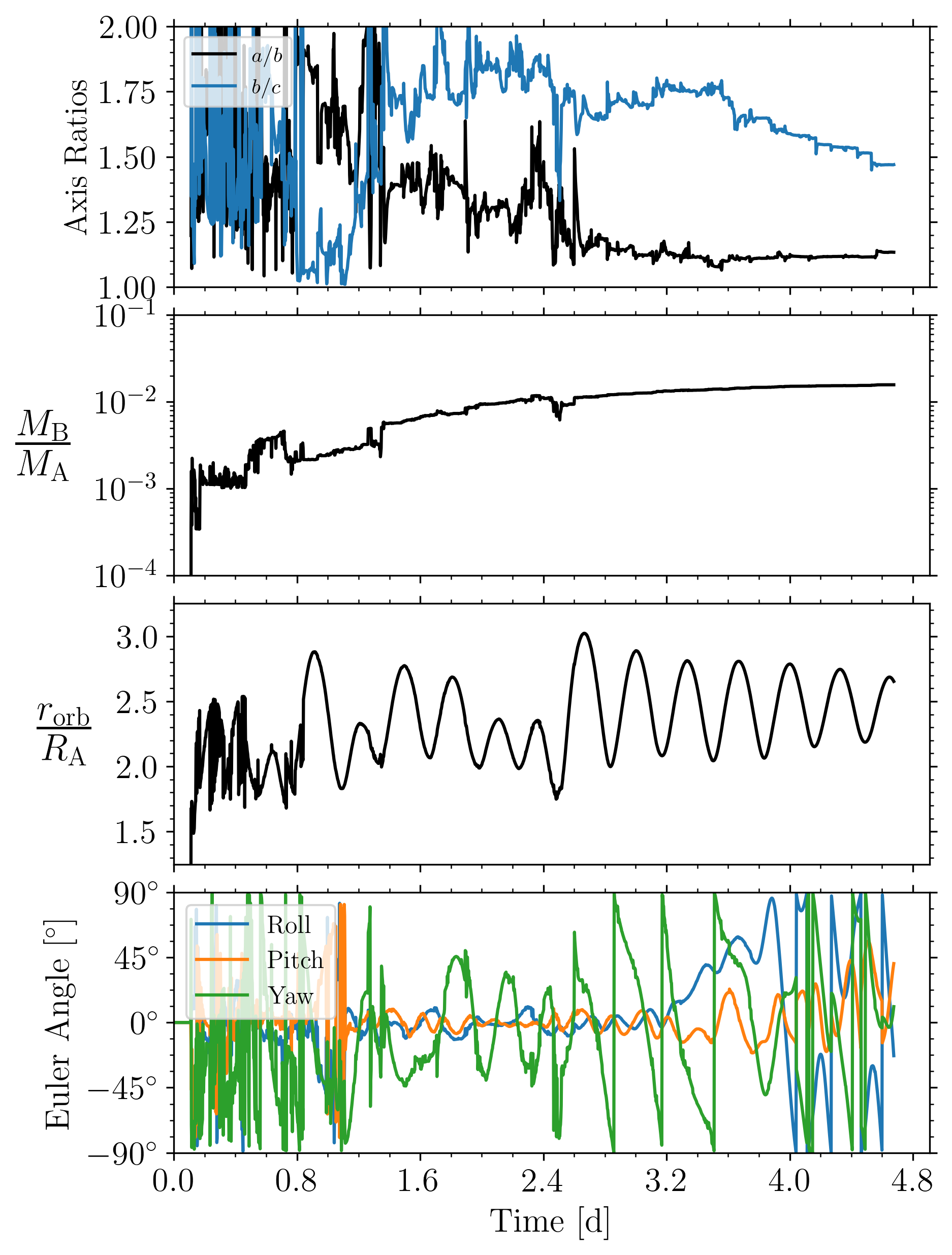}
\caption{Time-series plots of the example simulation from Fig.\ \ref{fig:typical_example_mosaic} (Disk 008). Starting from the top, we plot the satellites DEEVE axis ratios, mass ratio, orbital distance, and Euler angles, which describe its orientation in a frame rotating with the orbit. In only a handful of days, the satellite reaches a near-circular orbit at ${\sim}3R_{A}$ with a mass just under $0.01M_{A}$. The rotation state of the satellite is excited, but tidally locked, given by all three Euler angles being large but still well under $90^\circ$.}
\label{fig:timeSeries_lockedCase}
\end{figure}

{
In order to compare whether the simplified initial condition of the disk is a reasonable representation of the more realistic initial conditions following mass shedding, as demonstrated in Section \ref{subsec:mass_shedding_example}, we compare the distributions of semimajor axis, eccentricity, and inclinations of orbiting debris at early times. Fig.\ \ref{fig:disk_elements} compares the orbital elements of orbiting material between the more realistic mass-shedding example from Fig.\ \ref{fig:mosaic_yun} and the simplified case shown in Fig.\ \ref{fig:typical_example_mosaic}. Owing to the gravitationally unstable disk and the role of collisions, the orbits of disk particles are rapidly excited to wider orbits, with higher eccentricities and inclinations, despite initially starting with circular, nearly co-planar orbits. At least qualitatively, the distribution of orbital elements several days into the simulation reasonably reflects the orbits of post-mass-shedding debris (i.e., Fig.\ \ref{fig:mosaic_yun}), although the more realistic initial conditions have a slightly wider distribution in both inclination and eccentricity.}
{
\begin{figure}[H]
\centering
\includegraphics[width=0.5\textwidth]{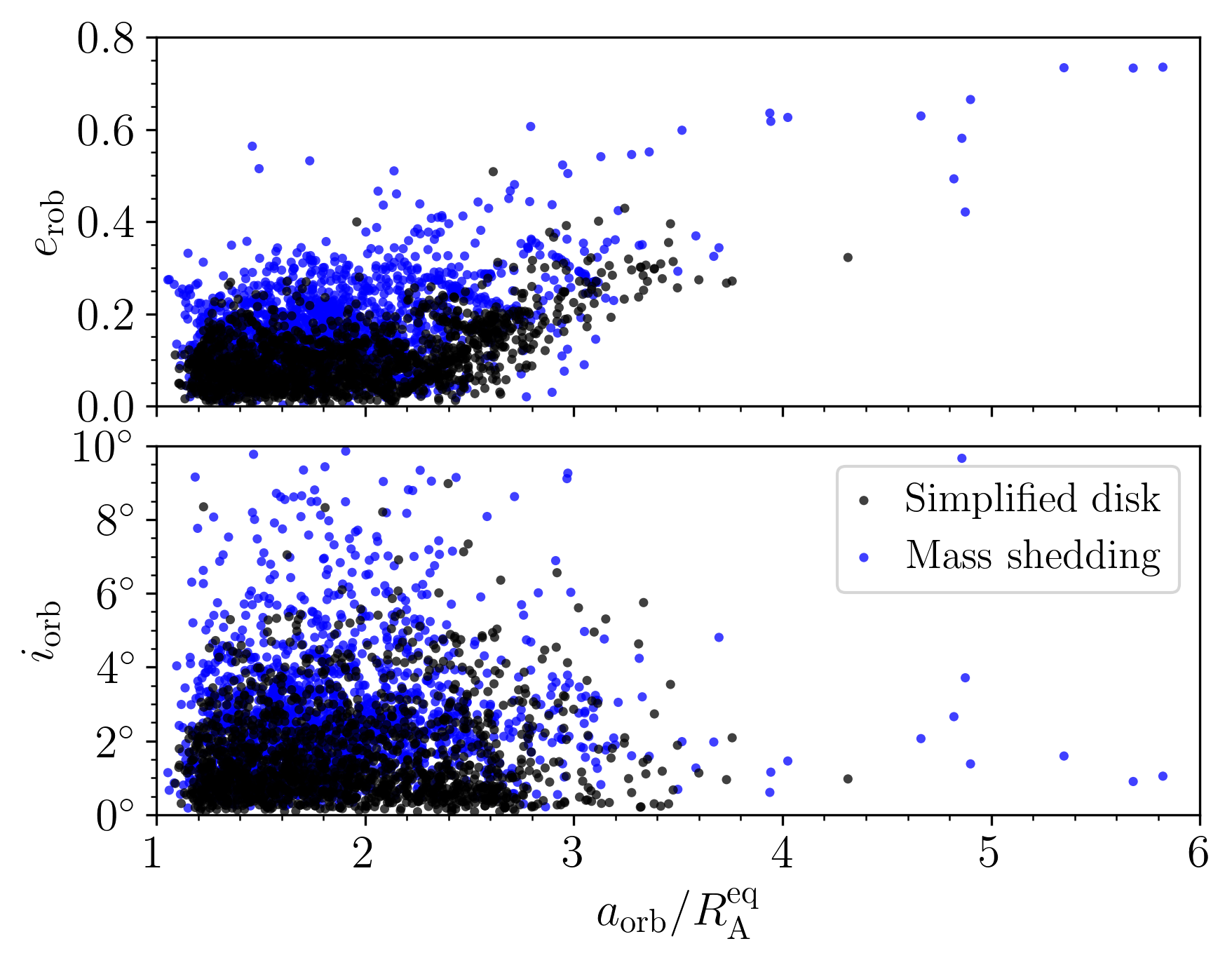}
\caption{{A comparison of orbiting particles between the more realistic, full spin-up simulation shown in Fig.\ \ref{fig:mosaic_yun} and the simplified initial conditions from the simulation shown in Fig.\ \ref{fig:typical_example_mosaic}. The orbital elements are plotted when moonlets first begin forming. For the mass-shedding simulation, this corresponds to ${\sim}0.5$ d after mass is initially shed, while for the simplified disk, this occurs around ${\sim}4$ d into the simulation. Although the distribution of orbital elements between the two cases does not match perfectly, due to the gravitationally unstable initial conditions of the simplified disk, we arrive at a similar range in semimajor axis, eccentricities, and inclinations seen in the more realistic simulation.}}
\label{fig:disk_elements}
\end{figure}
}

\subsection{Satellite mass and density}
\begin{figure}[H]%
\centering
\includegraphics[width=0.75\textwidth]{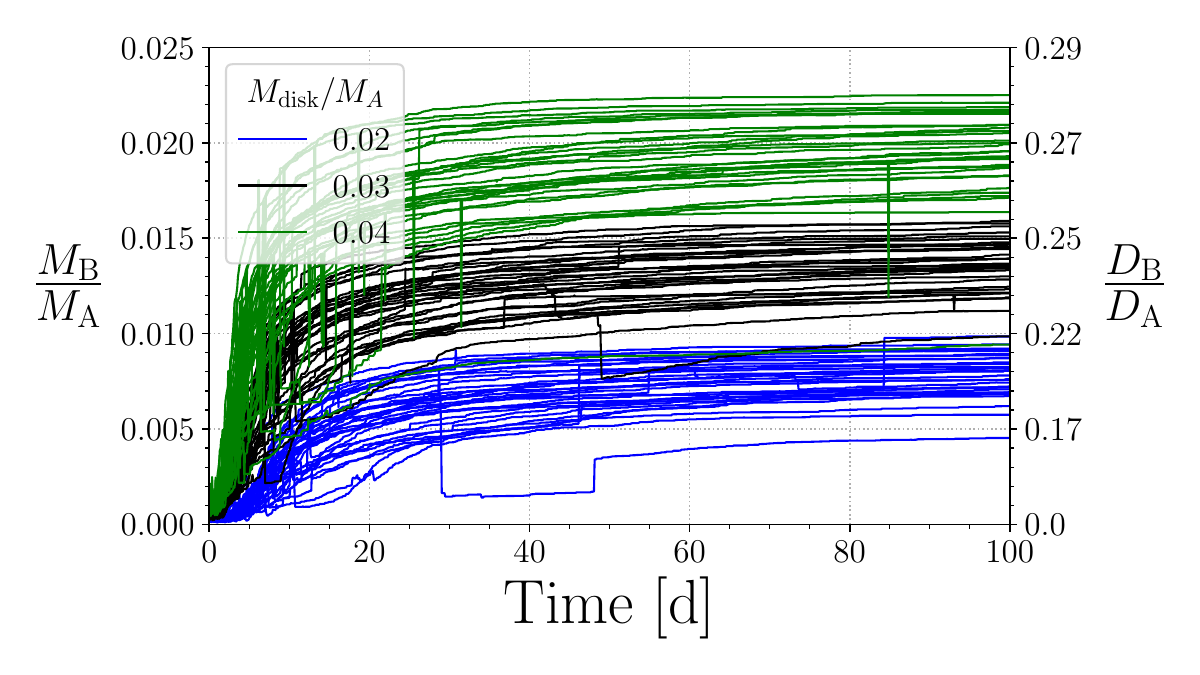}
\caption{The secondary-to-primary mass ratio over time, for all simulations with $\phi=35^\circ$. The accretion of the satellite is highly efficient, occurring in only a matter of days. The second y-axis shows the secondary-to-primary size ratio, assuming that the two bodies have the same bulk density (which is approximately true in these simulations). The spread in final mass among disks with the same initial mass is due to the chaotic formation history of each system. Note: most discontinuities in this plot are real and are due to mergers or tidal disruptions of the satellites. However, there are a couple discontinuities that result from the nearest-neighbor search algorithm misidentifying the largest orbiting fragment.}\label{fig:massRatios}
\end{figure}

In Fig.\ \ref{fig:massRatios}, we plot the secondary-to-primary mass ratio for all simulations with $\phi=35^\circ$ in order to understand how the mass of the initial disk determines the resulting satellite mass. Assuming the primary and secondary have the same bulk density (which is approximately true in these simulations), we also provide the size ratio $\frac{D_\text{B}}{D_\text{A}} \approx \big(\frac{M_\text{B}}{M_\text{A}}\big)^{1/3}$ on the second y-axis. In the 96 simulations shown here, the satellite tends to reach its final mass in the first few tens of days apart from a select few special cases that undergo late mergers or disruptions. For context, the Dimorphos-to-Didymos size ratio is ${\sim}0.2$ which corresponds to a mass ratio ${\sim}0.01$ if the bodies have equal densities \citep{Daly2023}. Therefore, we find that an initial disk mass of only ${\sim}0.02-0.03M_\text{A}$ is capable of producing a Dimorphos-mass satellite in only a matter of \textit{days}. Although their study focuses on a regime where significantly more mass and angular momentum is put into orbit, we find that our formation timescale is broadly consistent with that found by \citep{Hyodo2022}.  Our result is orders of magnitude lower (both in time and mass) than the calculation of \cite{Madeira2023} which requires 25\% of Didymos's mass to be shed into a ring that will then take years to form a Dimorphos-mass satellite. This disagreement likely stems from the different approach to the problem, namely in the initial conditions. Our model starts with a disk of particles initialized on much wider orbits, given the existing body of literature on spin-up and mass shedding \citep[e.g.,][]{Yu2019, Zhang2018, Zhang2022, Hyodo2022} and based on the spin-up example provided in Section \ref{subsec:mass_shedding_example}, whereas the \cite{Madeira2023} model supposes that the orbiting debris starts in a narrowly confined region at the surface of the primary and slowly spreads outwards, which substantially increases the timescale for the satellite's formation.

\begin{figure}[H]
    \centering
    \includegraphics[width=0.5\textwidth]{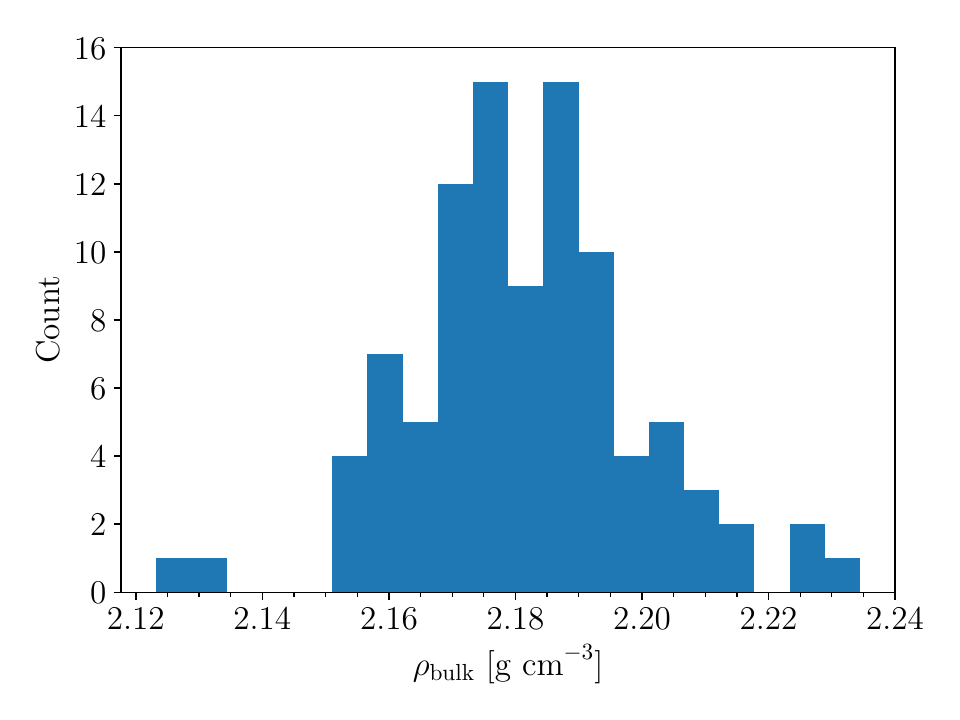}
    \caption{The bulk density of all satellites with $\phi=35^\circ$ at the end of the simulations. The density of all individual spheres is 3.5 g cm$^{-3}$, so these rubble piles have packing fractions of ${\sim}0.62$. Realistically, it is possible for the ``real'' bulk density to be slightly higher than what is measured here, as small particles which are below the resolution limit of these simulations would fill in some void space.}
    \label{fig:rho_bulk}
\end{figure}

We plot a histogram of the bulk density of each satellite at the end of the simulation in Fig.\ \ref{fig:rho_bulk}. The volume of the rubble pile is computed by the concave hull (or ``alpha shape''; \citep{Edelsbrunner1995}) of the set of points defined by the edges of the outermost spheres which make up the satellite. This method of calculating bulk density provides high accuracy by providing a ``tighter fit'' to the true shape of the rubble pile than other volume estimates such as the volume of the convex hull or the dynamically equivalent ellipsoid. Since some of the large void spaces could be filled with smaller boulders that are below the resolution of the simulations, the measured bulk density here is only notional and should be thought of as a soft lower limit. 

\subsection{Satellite orbit and rotational state}
\begin{figure}[H]%
\centering
\includegraphics[width=0.6\textwidth]{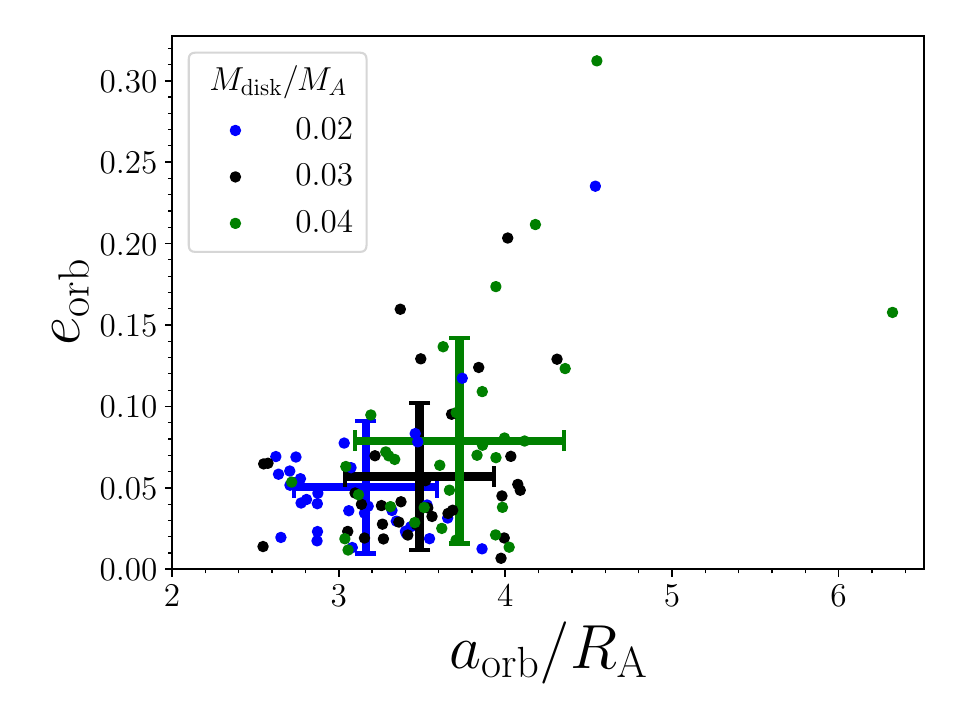}
\caption{The semimajor axis (in terms of the primary's radius) and eccentricity of the newly formed satellite, over all simulations with $\phi=35^\circ$. The colors indicate the starting disk mass, and the three points with error bars show the mean and $1\sigma$ standard deviation for the three different disk masses. Although the variance is quite large, a larger disk mass typically leads to a higher eccentricity and larger semimajor axis, due to the disk having a higher angular momentum and more collisions that can drive up the largest satellite's eccentricity.}
\label{fig:finalOrbits}
\end{figure}

The majority of the simulations end with a single satellite on a near-circular orbit having a semimajor axis between ${\sim}2.5$ and ${\sim}4$ primary radii. Fig.\ \ref{fig:finalOrbits} shows the final semimajor axis and eccentricity for the satellite in the 96 simulations where $\phi=35^\circ$. Generally, we find that more massive disks tend to produce a satellite on a wider, more eccentric orbit, simply as a result of the disk having a larger initial angular momentum. This is demonstrated in Fig.\ \ref{fig:finalOrbits} where we show the mean $a_\text{orb}$ and $e_\text{orb}$ along with their standard deviations for each of the 3 different disk mass cases.

\begin{figure}[H]
    \centering
    \includegraphics[width=0.6\textwidth]{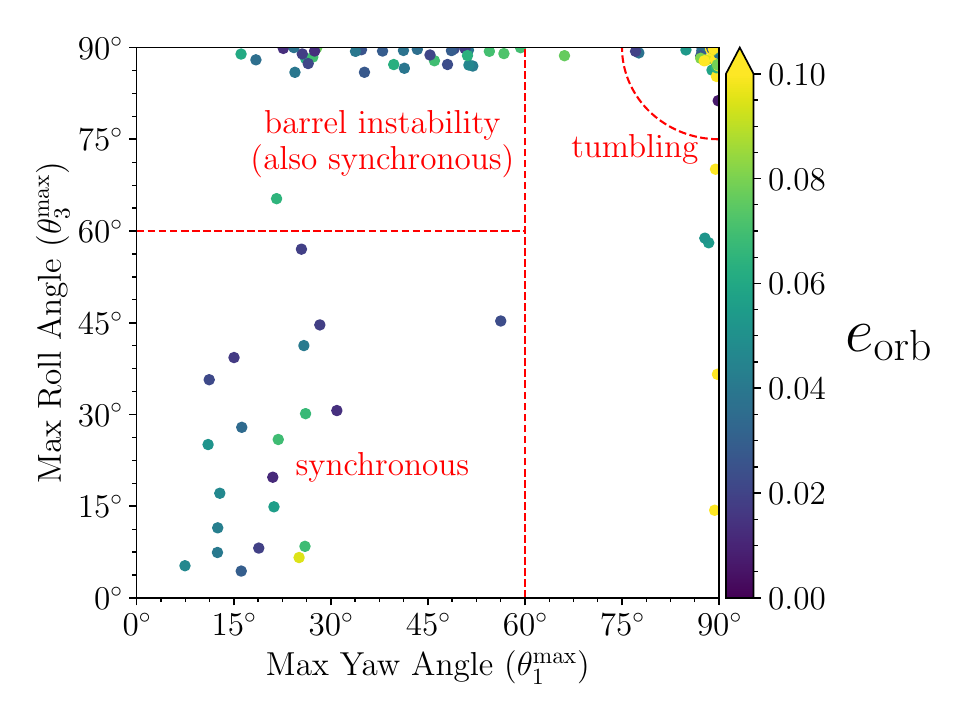}
    \caption{The maximum roll and yaw angles for each satellite, based on the final 10 days of the simulation. There are three distinct post-formation rotation states for the satellites, which are annotated on the plot to guide the reader's eye. We consider the satellite to be in synchronous rotation if its long axis remains aligned towards the secondary (i.e., yaw  $\lesssim60^\circ$). The ``barrel instability'' is a subset of the synchronous rotators where the secondary's long-axis stays largely aligned with the primary, yet it is able to roll about its long axis like a barrel. Finally, a significant fraction of satellites form in a tumbling state where all three Euler angles (roll, pitch, and yaw) can reach $90^\circ$. }
    \label{fig:max_roll_libration}
\end{figure}

For each satellite, we determine its rotation state based on the 1-2-3 Euler angle set, consisting of the satellites roll, pitch, and yaw angles in a frame rotating with its orbit (see \cite{Agrusa2021}). When the roll and pitch angles are small, the yaw angle can be thought of as the classic planar libration angle, i.e., the angle between the secondary's long axis and the line-of-centers. However, many of the satellites have undergone several mergers or close gravitational encounters, and are in excited, non-planar rotation states, necessitating the use of the Euler angle convention. In Fig.\ \ref{fig:max_roll_libration}, we plot the maximum roll and yaw angle for each satellite over the final 10 days of the simulation. We see three distinct regions in this plot. If the yaw angle stays below ${\lesssim}60^\circ$, we consider the satellite to be in synchronous rotation, since its long axis stays pointed in the direction of the primary.  Out of the synchronous rotators, a minority are in ``pure'' (albeit highly excited) synchronous rotation where the roll and pitch angles are also ${\lesssim}60^\circ$. Most of the synchronous rotators, however, are in a rolling state about their long axis. In this so-called ``barrel instability'', the satellite remains on-average tidally locked to the primary although it continues to roll about its long axis \citep{Cuk2021}. This non-principal axis rotation state within the 1:1 spin-orbit resonance could be long-lived as this mode would dissipate inefficiently by tides. Since binary YORP (BYORP) requires a synchronous secondary, this would also substantially weaken the BYORP effect or prevent it entirely \citep{Cuk2005, Quillen2022a}. Finally, we also see many satellites in an end-over-end tumbling state, where their roll and yaw angles both reach $90^\circ$. Generally, the tumblers have a higher eccentricity than the synchronous rotators as indicated in the plot, however, the onset of chaotic tumbling also depends on the satellite's inertia ratios \citep{Wisdom1984, Agrusa2021} as well as its collision history.

Of course, it is no surprise that these satellites are never perfectly tidally-locked, as these are short-term simulations in which the satellite has undergone many collisions and mergers so its rotation state is naturally excited. However, synchronization is the fastest-evolving tidal effect \citep{Goldreich2009}, and we would therefore expect the satellite's free libration to damp on relatively short timescales, potentially within hundreds of years since they are already in synchronous rotation \citep{Meyer2023a}. This may provide a simple explanation for why we observe so many synchronous satellites \citep{Pravec2016}, without needing to invoke the more complicated dynamical processes of rotational fission to achieve this equilibrium \citep{Jacobson2011a}. Instead, if the satellite forms via accumulation of shed material, then it has a reasonable chance to form in or near a tidally-locked state.

We are not aware of any studies that demonstrate a binary asteroid can immediately form in a synchronous rotation state following rotational disruption, although this seems like a relatively natural outcome. Formation with synchronous rotation has been found in other studies on gravitational accumulation near the Roche limit, such as the accretion of moonlets from Saturn's rings \citep{Karjalainen2004} and from circumplanetary disks \citep{Hyodo2015}.

\subsection{Satellite shape}
While there is significant literature on binary asteroid formation, there are no studies to our knowledge that directly model the expected shapes of the satellite. Many studies \textit{consider} the shape of the secondary, but do not model that shape being formed directly \citep[e.g.][]{Jacobson2011a,Davis2020b}. 

\begin{figure}[H]
\centering
\includegraphics[width=0.5\textwidth]{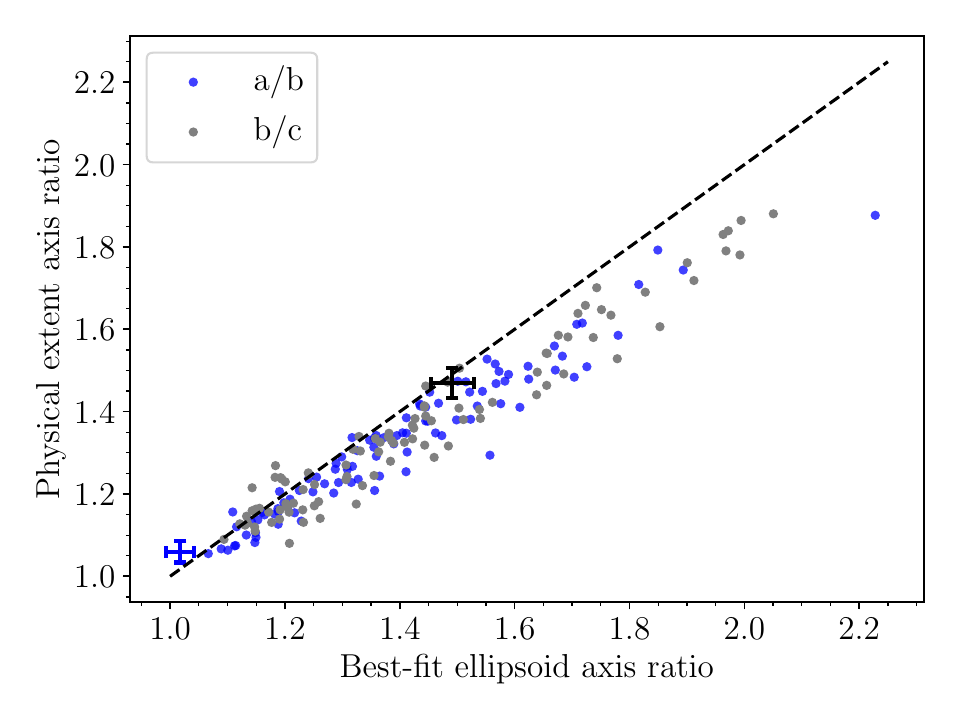}
\caption{The DEEVE axis ratios of each simulated satellite compared to the body's axis ratios defined by its physical extents. Points lying under the diagonal line have a DEEVE axis ratio (either $a/b$ or $b/c$) that is \textit{greater} than its physical extents and vice versa. Points lying on the line indicate that the DEEVE shape is a good approximation of the physical shape. For reference, we include the axis ratios of Dimorphos defined by its physical extents and best-fit ellipsoid, demonstrating that Dimorphos's shape is exceptionally well-fit to an ellipsoid. Dimorphos's $a/b$ and $b/c$ axis lengths, including $1\sigma$ uncertainties, are indicated in blue and black, respectively \citep{Daly2023_submitted}
}\label{fig:deeve_vs_extents}
\end{figure}

It is important to be clear with the definition of the satellite's shape. In this study, we define the shape of the satellite by its three principal axis $a, b$ and $c$, which correspond to the body's three principal moments $A, B$, and $C$. We measure its axis lengths in two different ways. The first is simply the physical extents of the body along these axes. The second measure is the axis lengths of the dynamically equivalent equal-volume ellipsoid (DEEVE). This is a uniform density ellipsoid having the same mass and moments of inertia of the rubble pile. If the body has an approximately ellipsoidal shape, then these two measures of its shape will match closely. Measuring the body's axis ratios by its DEEVE can be useful as they don't fluctuate significantly as a result of the motion of a single particle on the surface, which is common as the satellite is forming. Therefore, we use the DEEVE axis ratios in any time series plots as they are much less noisy. However, the physical extent of the body tends to better represent the ``true'' shape of the body and is most analogous to real life observations. In our simulations, these two measures tend to differ significantly for highly irregular shapes (high $a/b$ and/or $b/c$). This is demonstrated in Fig. \ref{fig:deeve_vs_extents}, where the DEEVE axis ratio tends to be quite larger than the physical extent axis ratio for large axis ratios.

\begin{figure}
\centering 
\subfloat[DEEVE\label{fig:subfig:deeveRatios}]{%
    \includegraphics[width=0.49\linewidth]{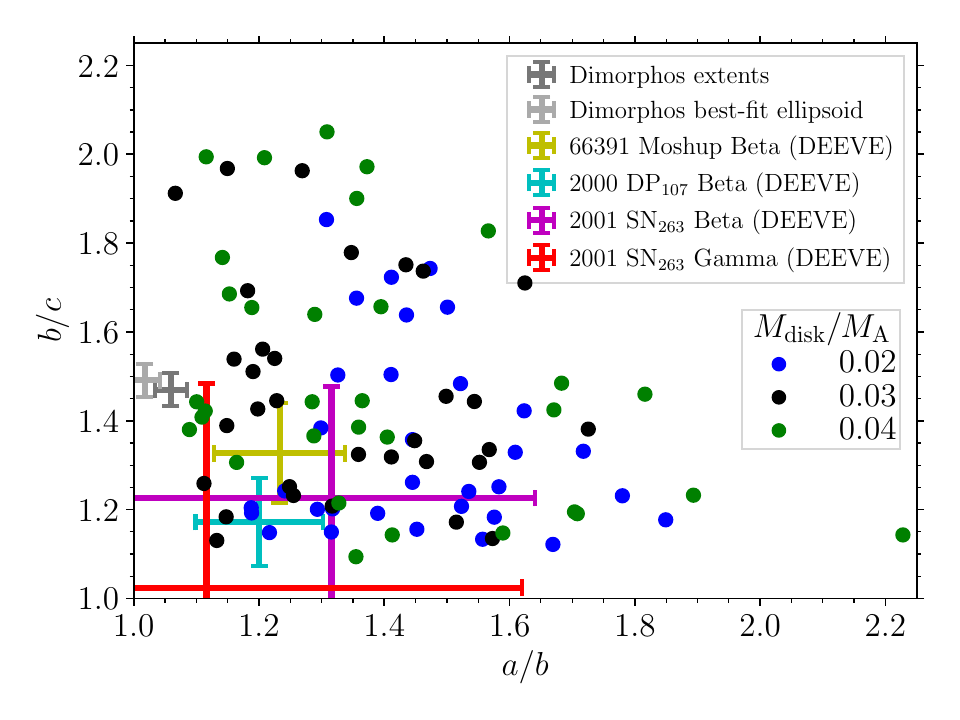}%
    }%
\subfloat[Physical extents\label{fig:subfig:extentRatios}]{%
    \includegraphics[width=0.49\linewidth]{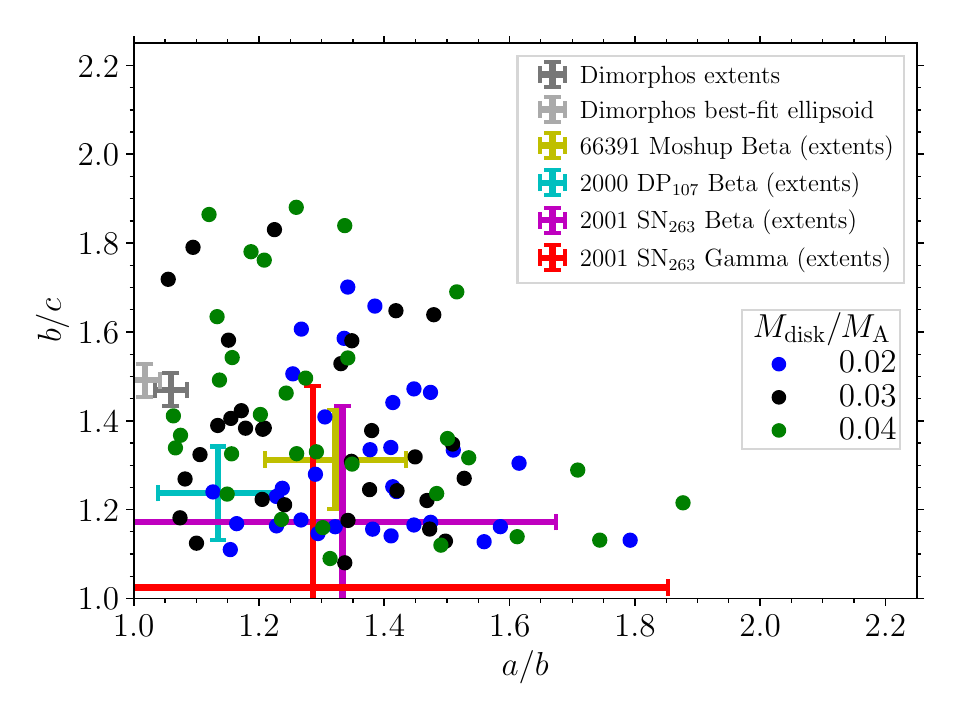}%
    }
\caption{The satellite's shape, parameterized by its (a) DEEVE or (b) physical axis ratios $a/b$ and $b/c$ for the 96 simulations with $\phi=35^\circ$. The colored dots indicate the simulation results with varying initial disk masses. The shape and $1\sigma$ uncertainty of Dimorphos, along with the radar-derived shapes of the satellites of Moshup (1999 KW$_4$), DP$_{107}$, and SN$_{263}$ are shown for comparison \citep{Daly2023_submitted,Ostro2006,Naidu2015b,Becker2015}.\label{fig:finalAxisRatios_vs_radar}}
\end{figure}

Due to the discrepancy between the DEEVE-derived and extend-derived shapes highlighted above, we compare both quantities to known secondary shapes for thoroughness. In Fig.\ \ref{fig:finalAxisRatios_vs_radar} we show both the DEEVE- and extent-derived semi-axis ratios $a/b$ and $b/c$ at the end of each simulation with $\phi=35^\circ$. We also include the shapes of the satellites of 66391 Moshup (formerly 1999 KW$_4$), 2000 DP$_{107}$, and 2001 SN$_{263}$, which are the only publicly available radar-derived shapes for the satellites of other binary asteroids \citep{Ostro2006,Naidu2015b,Becker2015}, as well as the axis ratios of Dimorphos \citep{Daly2023, Daly2023_submitted}. The updated shape of Dimorphos provided by \citep{Daly2023_submitted} differs slightly from the initial assessment in \citep{Daly2023}, but the difference is small enough that we only plot the latest values to avoid confusion. For each real asteroid system, we include $1\sigma$ uncertainties in $a/b$ and $b/c$, assuming that the reported uncertainties in $a, b$ and $c$ from each respective paper are uncorrelated, which may not be true. Therefore, the uncertainties should only be used to guide the reader's eye. The satellites of Moshup and DP$_{107}$ are the best comparisons with Dimorphos, as they are both S-type binaries, whereas SN$_{263}$ is a C-type triple (with large uncertainties in the satellite shapes). Fig.\ \ref{fig:finalAxisRatios_vs_radar}  demonstrates that, generally speaking, these simulations produce satellites that are \textit{more elongated} (high $a/b$) and \textit{more flattened} (high $b/c$ or $a/c$) than the radar-derived secondaries. However, there are many cases that produce shapes similar to radar-observed secondaries, given their uncertainties. Although no satellites are produced within the $1\sigma$ uncertainty region of Dimorphos's shape, several simulations do come close. Dimorphos's $b/c$ ratio lies approximately in the middle of the simulated $b/c$ range. Although there is a preference for elongated satellites, several simulations result in a low elongation ($a/b\lesssim1.1$), like Dimorphos. These simulations demonstrate that more elongated shapes are preferred, although immediately forming a Dimorphos-like shape by mass shedding is not implausible. Due to the satellite's accretion near the Roche limit, this strong preference for more elongated shapes comes as no surprise and has been seen in analogous studies \citep[e.g.][]{Porco2007,Hyodo2015}. It is important to note that the simulations here are run for only 100 days and there could be longer term processes that will modify the satellite's shapes (impacts, landslides, etc.), so it is important to interpret any comparison between the simulated and observed shapes with caution.

\begin{figure}[H]
        \centering
        \includegraphics[width=0.5\textwidth]{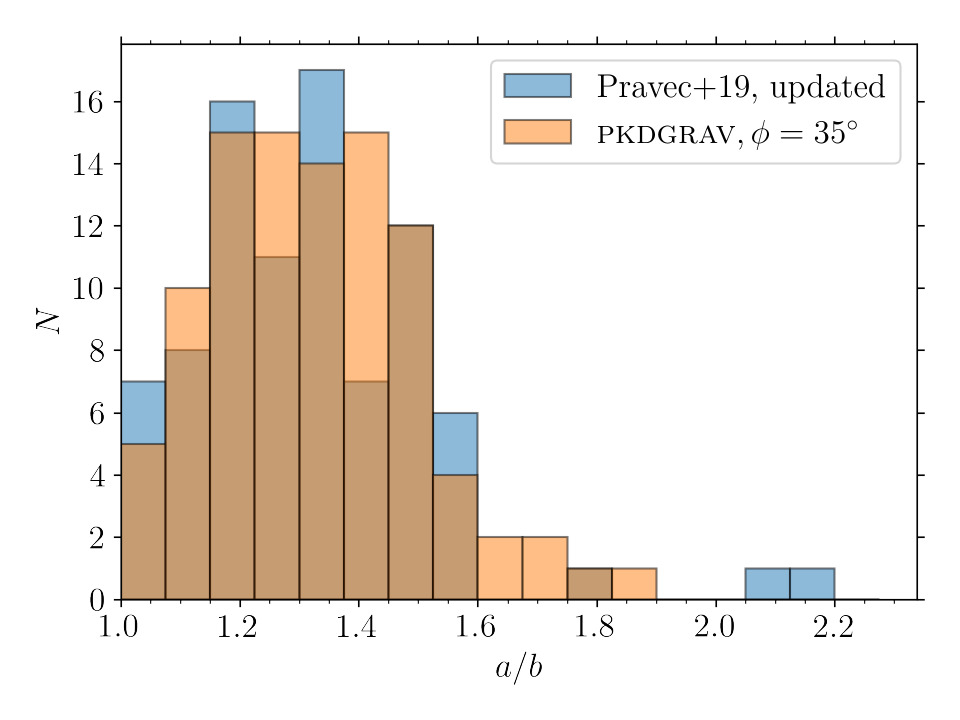}
        \caption{Lightcurve-derived shapes are only sensitive to $a/b$, so we construction histograms of $a/b$ from the simulations (defined by their physical extents) to compare with $a/b$ of satellites derived from lightcurves \citep{Pravec2016,Pravec2019}. The y axis is not normalized since we are comparing a similar number of real systems (87) and simulations (96).}\label{fig:finalAxisRatios_vs_lighcurve}
\end{figure}

We also compare our results to the lightcurve-derived shapes from \cite{Pravec2016,Pravec2019}. Fig.\ \ref{fig:finalAxisRatios_vs_lighcurve} shows a histogram of the simulated $a/b$ ratio compared to the dataset of \cite{Pravec2019} along with some new (not yet published) secondaries (Pravec, personal communication). Here, we only include lightcurve secondaries in which the primary is less than $20$ km in diameter and the secondary-to-primary size ratio is less than 0.6, to ensure that we are only comparing with binaries likely formed by YORP.

We find a remarkable agreement between the shapes of lightcurve secondaries and those simulated in this work. Both data sets show that most satellites have $a/b<1.6$, aside from a few outliers. The sharp drop-off in satellites high $a/b$ has been attributed to chaotic dynamics of satellites with $a/b\gtrapprox\sqrt{2}$ that ultimately destroys (or alters) them \citep{Cuk2010,Pravec2016}. While this is a real dynamical effect, we suggest that it may also be that secondaries simply don't form with extremely high elongations to begin with.

Interestingly, there are several secondaries with low elongations ($a/b\lesssim1.1$) measured by Pravec. Naively, it appears as if the simulations do a good job at matching this low elongation population. However, it is important to note that lightcurves are heavily biased \textit{against} detecting satellites with a $a/b\lesssim1.05$ (Pravec, personal communication). This fact, combined with Dimorphos apparently having $a/b<1.1$ suggests the existence of a significant population of secondaries with low $a/b$. If the simulations presented here reasonably capture the formation of these secondaries, we would expect the simulations to produce an \textit{overabundance} of low elongation secondaries compared to the observed population. This suggests that either (1) there are other longer-term processes at play that could reshape some satellites over time or (2), there is an effect not captured by our model resulting from our assumptions or initial conditions.

\subsection{Effect of friction angle}
We conducted an additional set of simulations where the disk mass was held constant at $M_\text{disk}=0.03M_\text{A}$ and the friction angle was varied between $29^\circ$ and $40^\circ$. For each friction angle ($29^\circ$, $32^\circ$, and $40^\circ$), we conducted 8 simulations with randomized initial conditions. With this smaller sample size, we are dealing with small number statistics but can still draw a few basic conclusions. 

\begin{figure}[H]
    \centering
    \subfloat[Final orbits\label{fig:subfig:finalOrbits_friction}]{\includegraphics[width=0.5\linewidth]{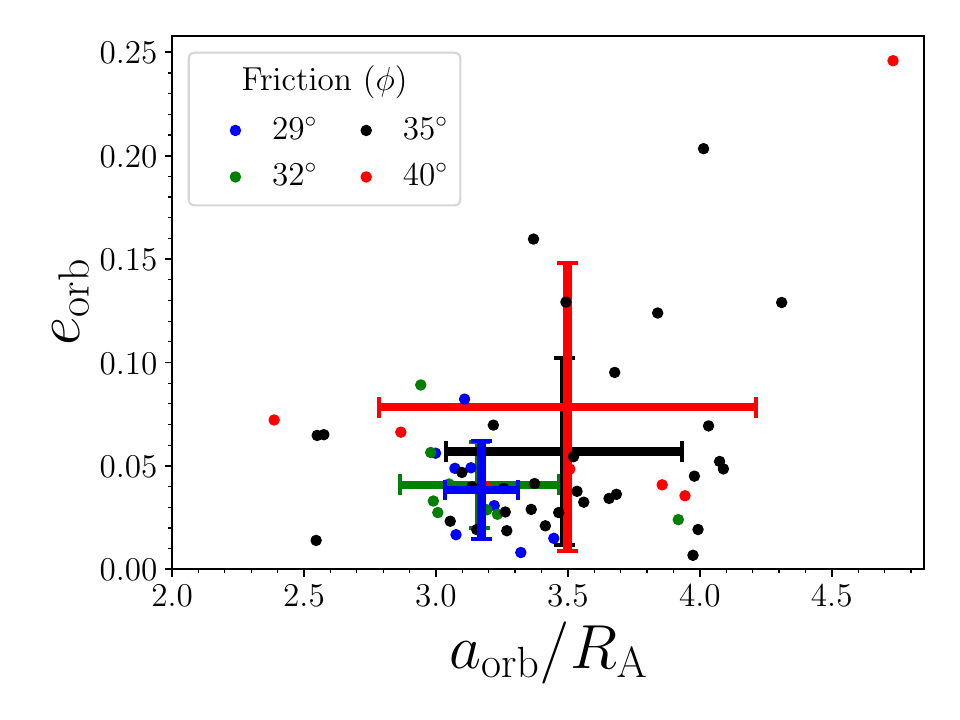}}
    \subfloat[Final shapes (extents)\label{fig:subfig:finalShape_friction}]{\includegraphics[width=0.5\linewidth]{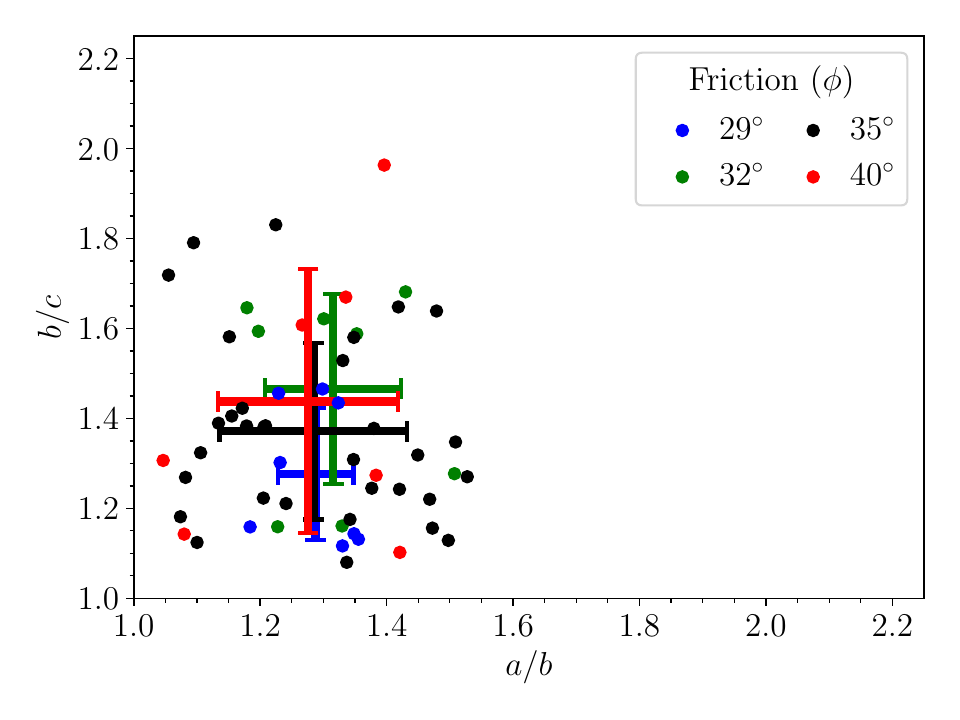}}\\
    \subfloat[Bulk Density\label{fig:subfig:rho_bullk_friction}]{\includegraphics[width=0.5\linewidth]{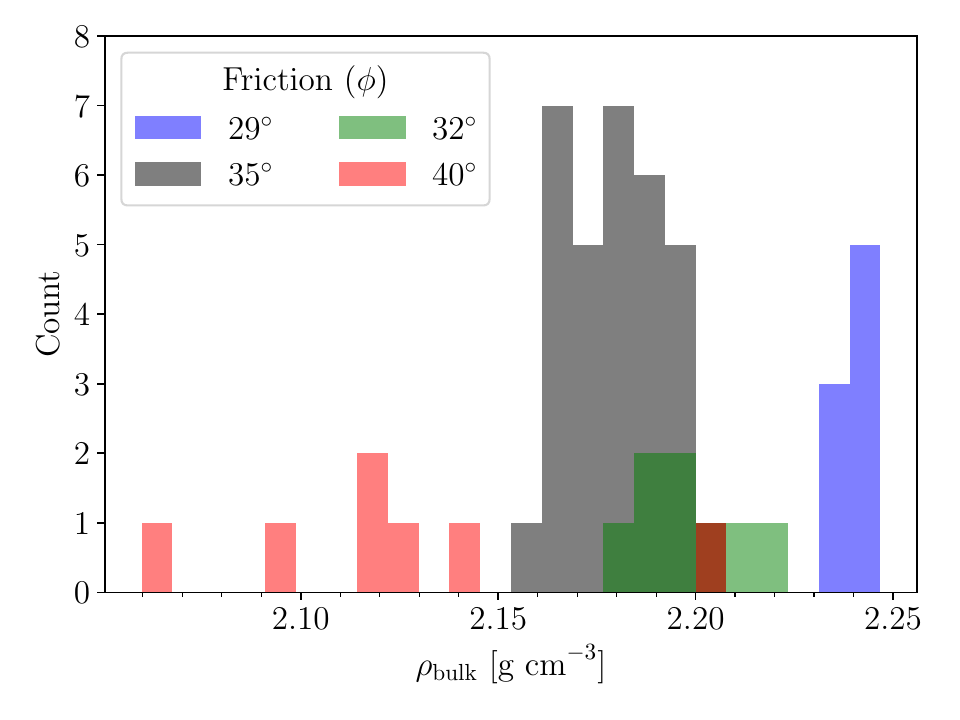}}
    \caption{Plots showing the effect of friction on the resulting satellite. (a) The final semimajor axis and eccentricity from each simulation. (b) The axis ratios of the satellite, based on its physical extents, as a function of friction angle. (c) The bulk density of each satellite as a function of its friction angle $\phi$. Generally, bodies with a lower effective friction are able to achieve a more efficient packing and thus have a higher bulk density.}
    \label{fig:plots_friction}
\end{figure}

As the friction angle is decreased, there is significantly less variation between simulations in general, including the satellite's formation distance, eccentricity, shape, and rotation state. In Fig.\ref{fig:subfig:finalOrbits_friction} we show the satellite's final semimajor axis and eccentricity, along with an average and standard deviation for each value of $\phi$. Although this is small number statistics, there is a clear trend with friction; satellites with a lower friction angle tend to form on a closer, more circular orbit. This is because the violent processes such as collisions and close gravitational encounters that lead to highly excited orbits tend to disrupt lower friction moonlets. This means that the moonlets, which eventually merge to form the final satellite, are necessarily on more circular orbits. Therefore, the final satellite will tend to have a less excited orbit. %Although, \textit{on average} the lower friction satellites tend to form closer to the primary, the higher friction satellites have a smaller Roche limit and thus have the ability to form closer \citep{Holsapple2006}. 

The axis ratios (based on the physical extents) are given in Fig.\ \ref{fig:subfig:finalShape_friction}. In general, there is a smaller variation in the secondary's shape as the friction is reduced. In addition, a lower friction tends to result in a less flattened object (i.e., more prolate). As the friction angle is lowered, the body is forced to reshape itself until it is structurally stable, and bodies with higher frictions are able to maintain a wider range of shapes. In the limit that the friction angle were reduced to $0^\circ$, the satellite effectively behaves like a fluid and would take on the shape of a Roche ellipsoid, at its given orbital distance \citep{Holsapple2006}. This is not demonstrated here, simply because a simulation with $\phi=0^\circ$ would take an excessive amount of time to form a satellite.

A histogram of the secondary's bulk density as a function of friction angle is shown in Fig.\ \ref{fig:subfig:rho_bullk_friction}, where there is a clear trend. Bodies with a lower effective friction are able to achieve a more efficient packing arrangement, thus having less void space and a higher bulk density. At a \textit{grain} density of 3.5 g cm$^{-3}$, this figure should be thought of as a lower limit for the secondary's \textit{bulk} density. In reality, some void space could be filled with boulders that are below the resolution of these simulations, so the ``true'' bulk density might be be higher.

\subsection{Matching the Shape of Dimorphos}
\begin{figure}
    \centering
    \subfloat[Disk 069\label{fig:subfig:timeSeries_disk069}]{\includegraphics[width=0.45\linewidth]{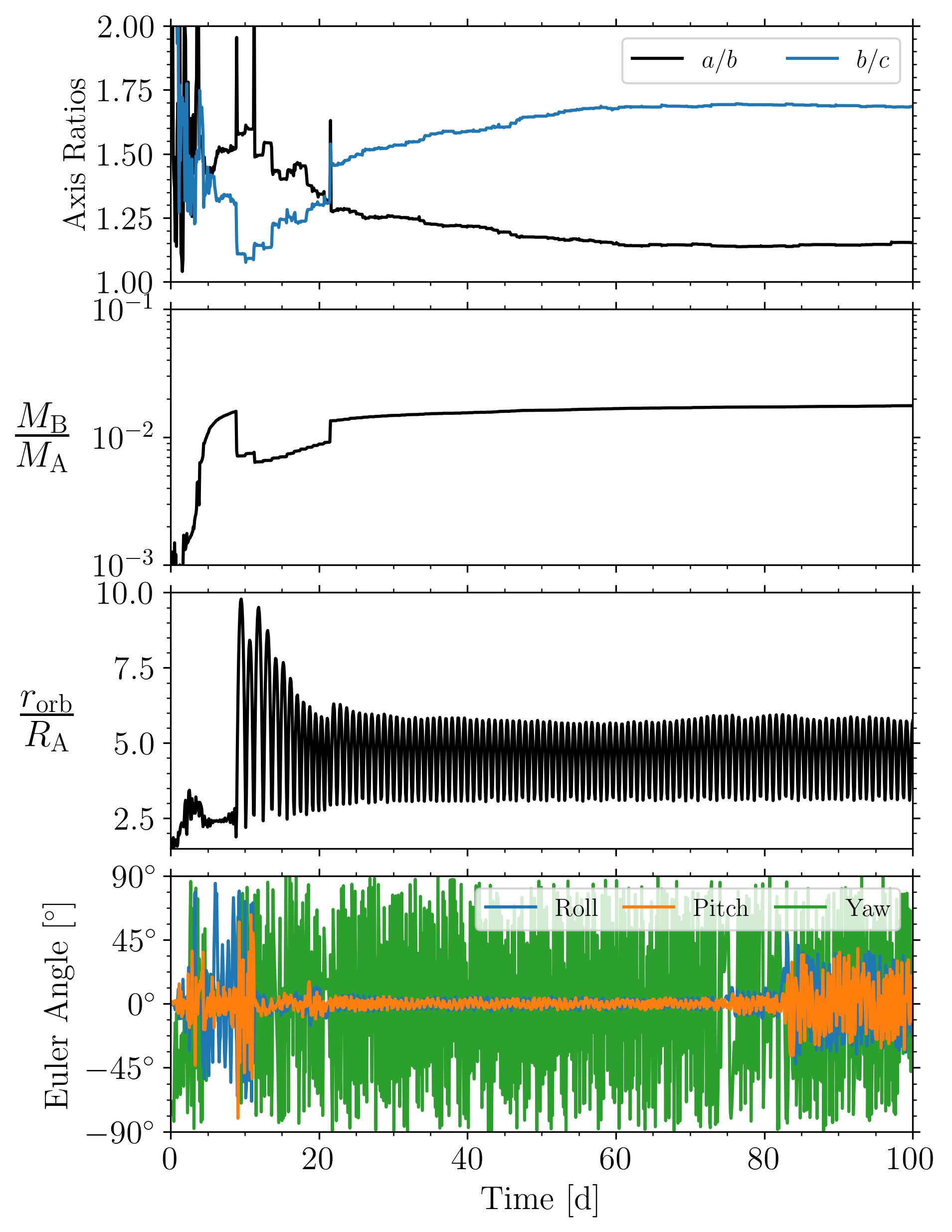}}
    \subfloat[Disk 081\label{fig:subfig:timeSeries_disk081}]{\includegraphics[width=0.45\linewidth]{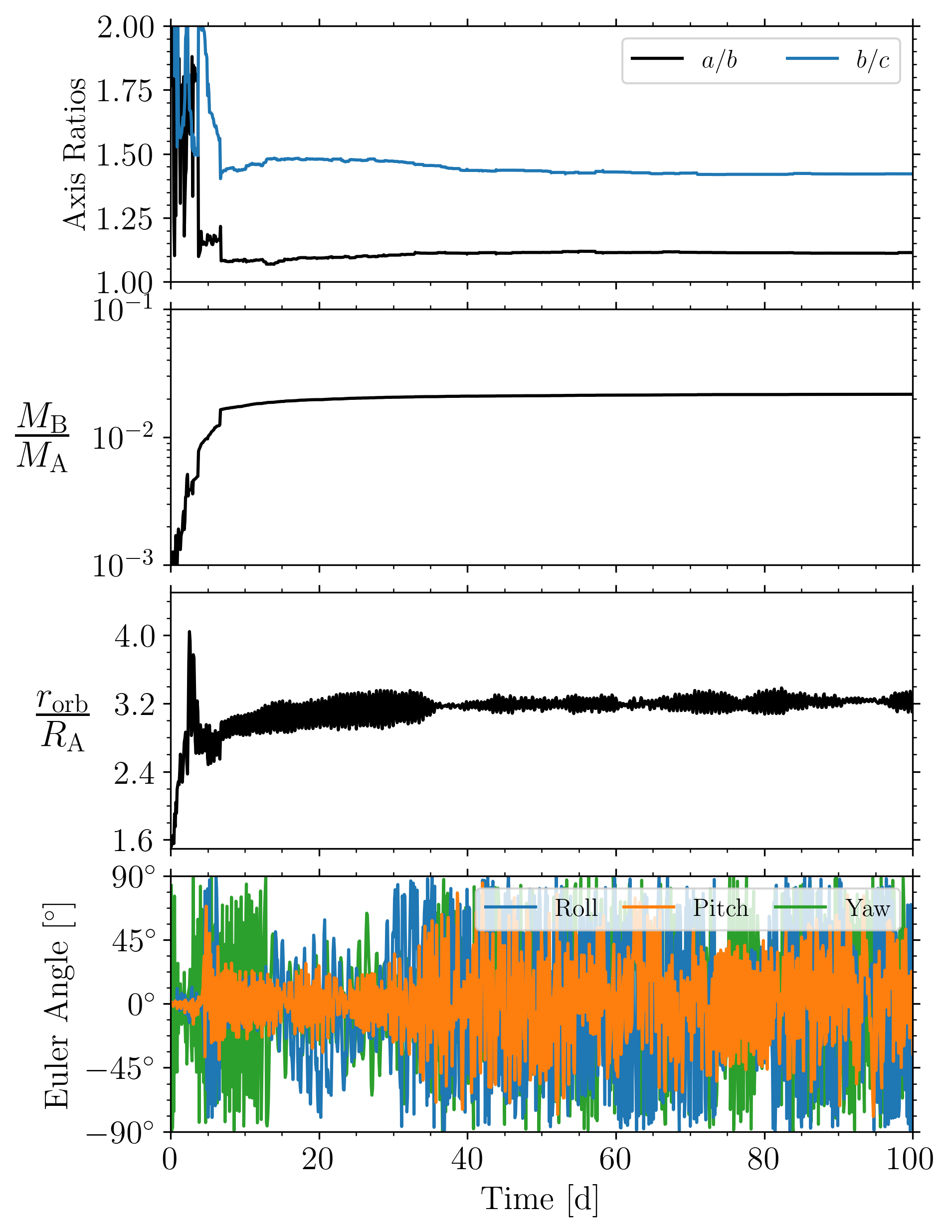}}
    \caption{Two examples that result in a Dimorphos-like shape (i.e., low $a/b$). (a) This satellite achieves an oblate shape due to a partial tidal disruption torquing the remaining mass onto a wide, eccentric orbit. Then, due to the body's asynchronous rotation, it accretes material in an azimuthally symmetric manner, as it sweeps through periapse with a different orientation every orbit. (b) This satellite achieves an oblate shape due to serendipitous mergers having a favorable geometry and is able to maintain its shape through the rest of the simulation. Movies of these simulations are provided}
    \label{fig:timeSeries_dimorphos_matches}
\end{figure}

Although these simulations demonstrate that immediately forming an oblate satellite is relatively rare, we show two cases with a $\phi=35^\circ$ that achieve shapes similar to Dimorphos. Although neither cases are a perfect match, they do suggest that an oblate spheroid-like shape can plausibly be formed from a single mass shedding event. In Fig.\ \ref{fig:timeSeries_dimorphos_matches}, we show the two examples, both of which have an initial disk mass of $M_\text{disk} = 0.04M_\text{A}$. In the first example (Fig. \ref{fig:subfig:timeSeries_disk069}), the satellite grows rapidly over the first several days until it undergoes a close encounter with another satellite (not shown), which causes the satellite to move inwards and undergo a partial tidal disruption event with the primary. This torques the remaining mass onto a much wider, eccentric orbit and also causes the satellite to rotate asynchronously yet maintain principal axis rotation (as demonstrated by the roll and pitch angles remaining small and the yaw angle circulating). As a result of the asynchronous rotation, the satellite begins accreting material in an azimuthally symmetric manner, as it has a different orientation at each periapse passage (where there is material available to accrete). The satellite also undergoes a merger with the last remaining moonlet at ${\sim}20$ d, which abruptly increases $b/c$. After the merger, the satellite continues accreting on all sides, leading to a gradual reduction in $a/b$ and increase in $b/c$, until there is almost no material left to accrete. This process suggests that an oblate shape like that of Dimorphos could plausibly be acquired if the satellite accretes material after some dynamical process triggers an asynchronous rotation state.

In the second example (Fig.\ \ref{fig:subfig:timeSeries_disk081}), the satellite achieves its oblate shape simply by undergoing two near-head-on mergers with other moonlets at early times (around ${\sim}3.5$ d and ${\sim}7$ d, respectively). These mergers serendipitously lead to a more oblate shape due to their favorable geometry. In addition, the satellite's relatively circular orbit, which is well outside the Roche limit, prevents its shape from undergoing significant further modification, despite its tumbling rotation state. Mergers among similar-sized moonlets have been proposed to explain the oblate shapes of some of Saturn's small moons \citep{Leleu2018}.

\begin{figure}[H]
\centering
\subfloat[Disk 069 \\ DEEVE: $a/b{\approx}1.15, b/c{\approx}1.69$ \\ Extents: $a/b{\approx}1.14, b/c{\approx}1.49$]{\includegraphics[width=0.33\linewidth]{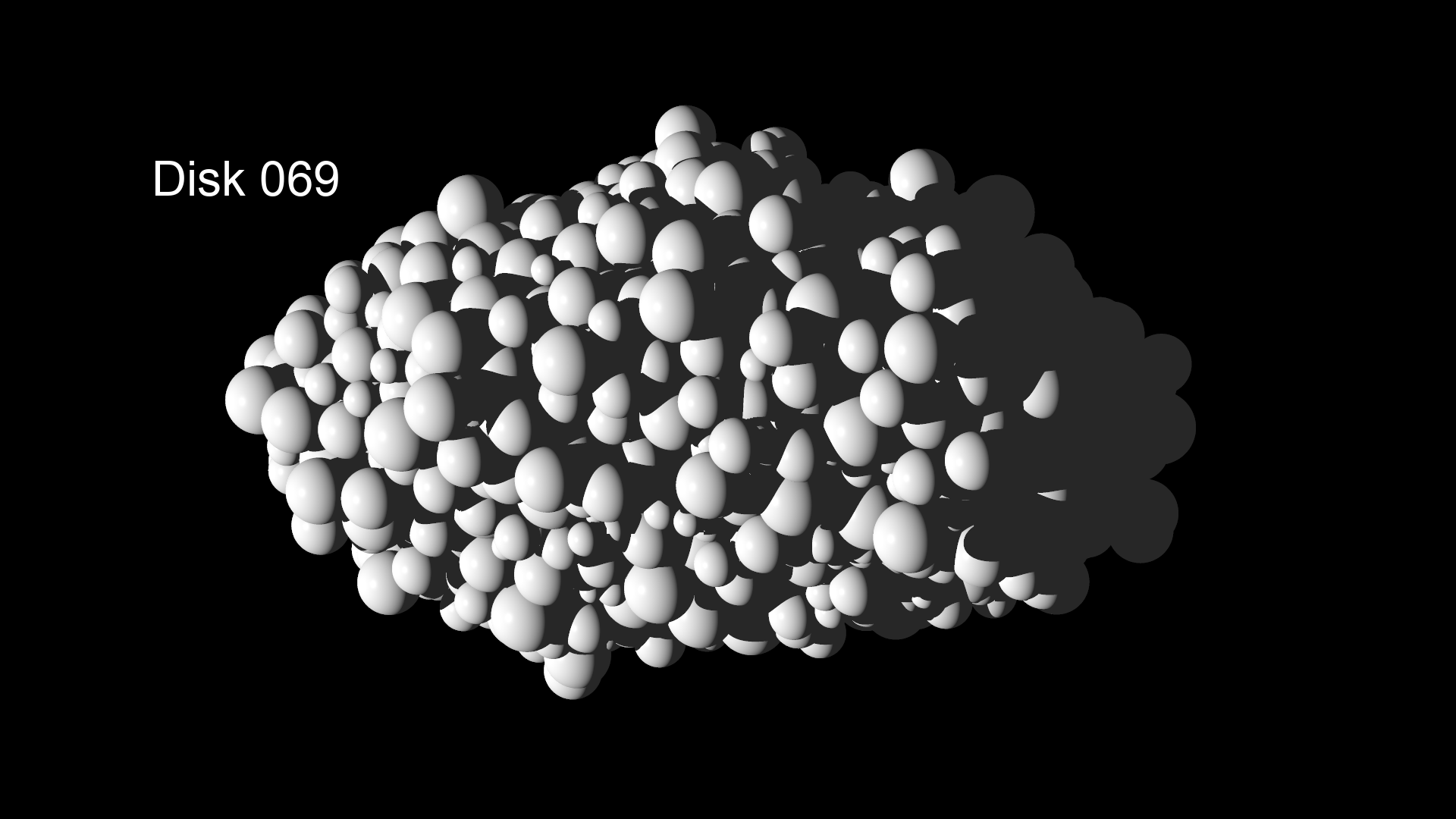}}
\subfloat[Disk 081 \\ DEEVE: $a/b{\approx}1.11$, $b/c{\approx}1.42$ \\ Extents: $a/b{\approx}1.08$, $b/c{\approx}1.37$]{\includegraphics[width=0.33\linewidth]{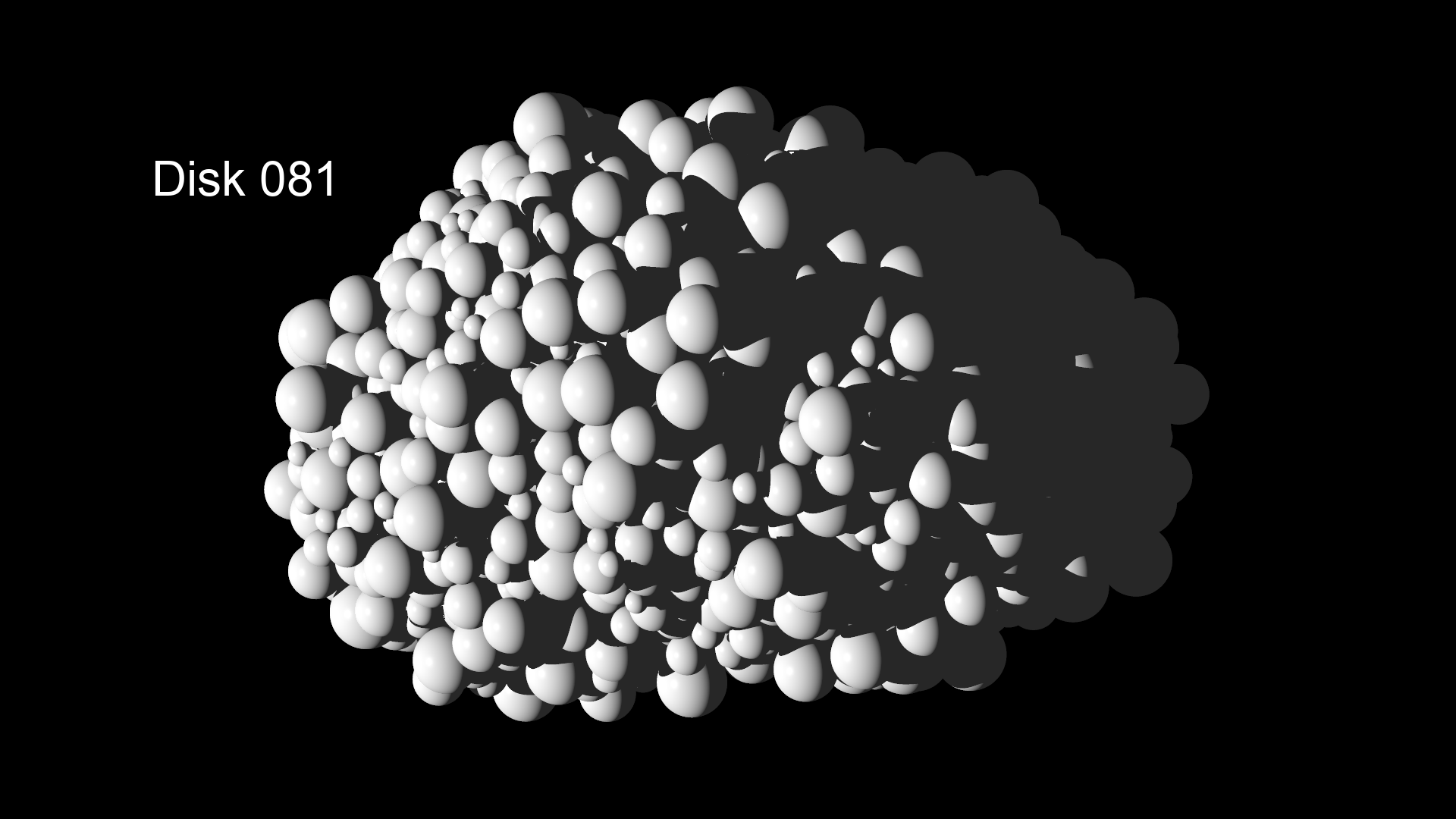}}
\subfloat[DRACO Image \\ DEEVE: $a/b{\approx}1.02$, $b/c{\approx}1.49$ \\ Extents: $a/b{\approx}1.06$, $b/c{\approx}1.47$]{\includegraphics[width=0.33\linewidth]{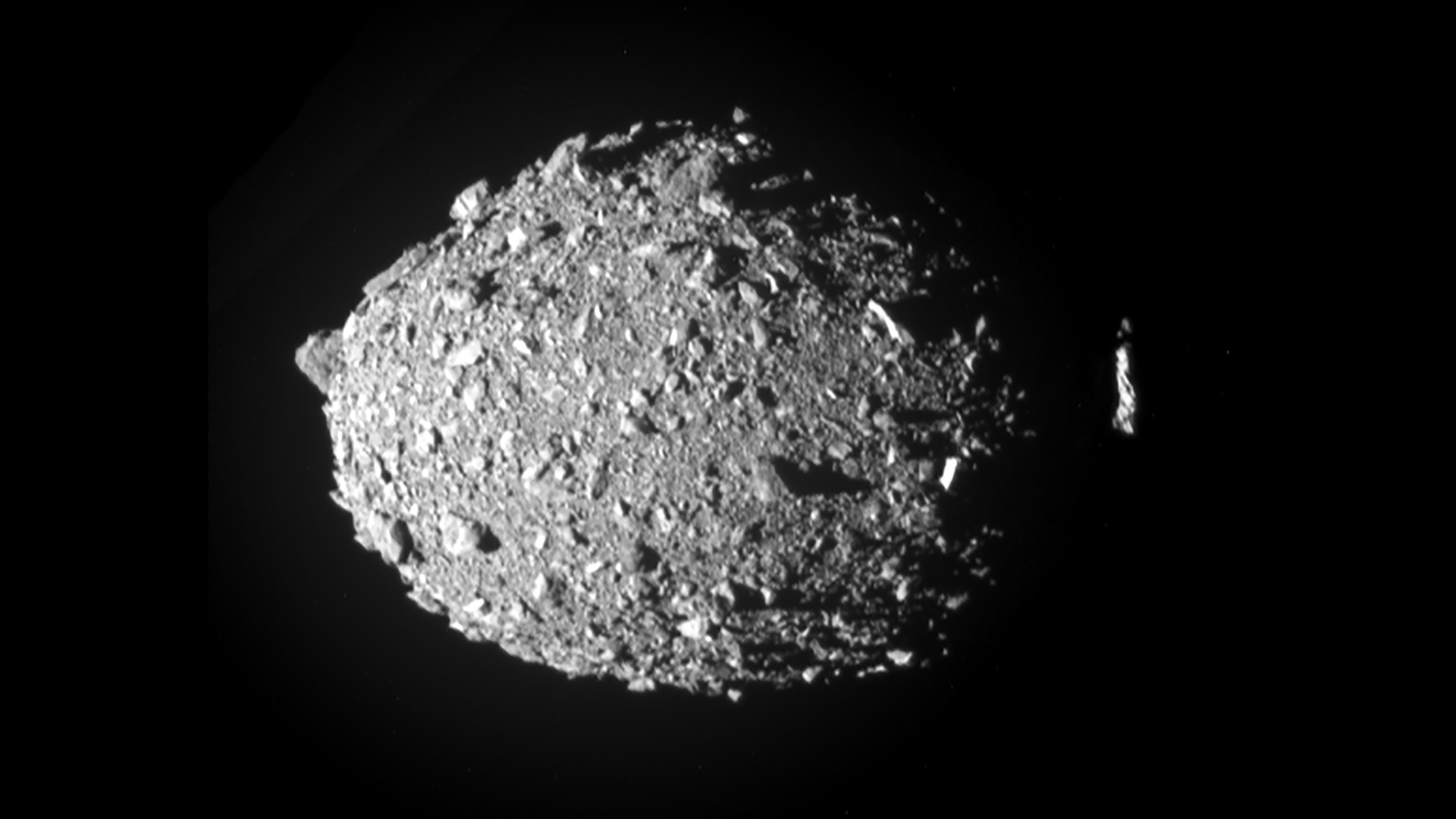}}
\caption{A rendering of two satellites that have similar shapes to Dimorphos, with the same lighting conditions and viewing geometry as DART's approach. }\label{fig:dimorphos_comparison}
\end{figure}

In Fig.\ \ref{fig:dimorphos_comparison}, we show renderings of these two examples in each body's respective body-fixed frame, using the same lighting conditions and viewing geometry as the DART spacecraft on its approach to Dimorphos. Although these two examples demonstrate that an oblate satellite can plausibly form by asynchronous rotation during accretion or by lucky mergers, { these events are relatively rare. Out of over 100 simulations, Dimorphos-like shapes are only formed a few percent of the time. Therefore, it is feasible that Dimorphos obtained its shape through some other unexplored mechanism.}

\subsection{Unique outcomes}
Due to the chaotic orbital and collisional evolution of the moonlets, some simulations resulted in outcomes that were not expected at the conception of this study. This includes the creation of asteroid pairs and triple systems. To be clear, we do not claim that all or even a majority of pairs or triples originate from a single mass shedding event, rather that this formation mechanism is plausible for some systems.

\subsubsection{Asteroid Pairs}\label{subsec:pairs}
An asteroid pair is where two unbound asteroids have similar enough orbital elements (and colors/spectra; see \cite{Pravec2019}, and references therein) such that they can be traced back to a common origin. An asteroid cluster is where multiple unbound bodies can be traced back to the same system. Many pairs and clusters are thought to form either through collisions, or through rotational fission \citep{Vokrouhlicky2008, Pravec2010, Pravec2018}. Because the dynamics of a newly formed binary system are chaotic and have a positive free energy for mass ratios below ${\sim}0.2$, the secondary can be ejected purely by the internal dynamics of the system \citep{Scheeres2007a, Jacobson2011a}. This phenomenon does an excellent job at reproducing the observed correlation between the primary's rotation period and the mass ratio of the asteroid pair \citep{Pravec2010, Pravec2019}. There is a distinct correlation between asteroid pair mass ratio and the primary's spin rate;  pairs with a large mass ratio tend to have slower primary rotation periods, presumably because more angular momentum was transferred from the rapidly rotating primary to the secondary in order eject it from the system.

Due to the azimuthal symmetry of the primary in these simulations, there is very weak coupling between the primary's rotation and the satellites orbit, preventing these simulations from exploring this correlation. However, our  simulations demonstrate that pairs/clusters can also form via three-body interactions between the moonlets during their formation. In all cases where we form a successful pair or cluster, there is always a secondary still bound to the primary. In other words, in any case where we form a pair, it is always a ``paired binary''. 

\begin{figure}[H]
    \centering
    \includegraphics[width=0.35\textwidth]{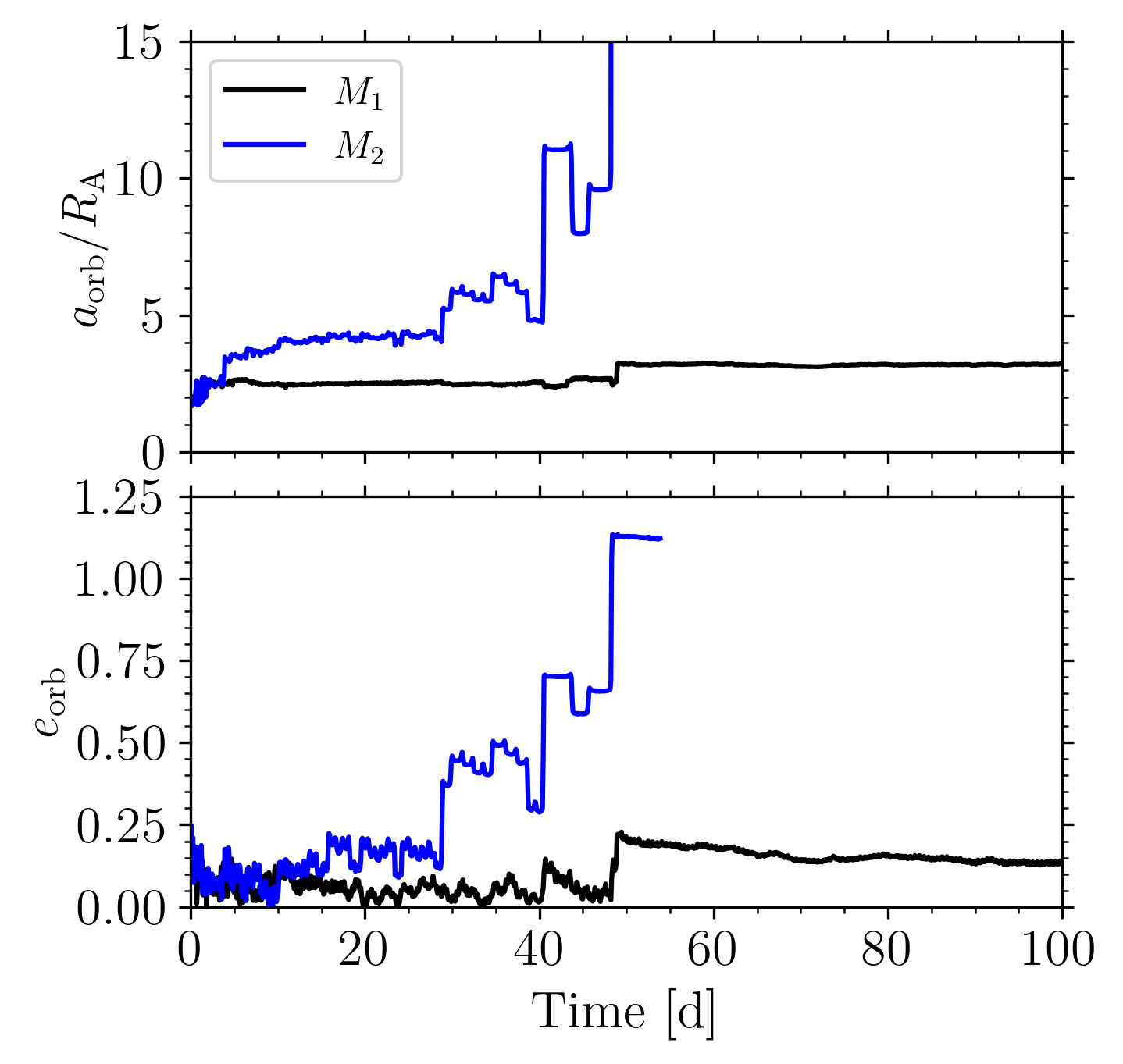}
    \caption{An example of a paired binary being formed as the larger satellite ejects the smaller satellite from the system. After an extremely close encounter, the smaller satellite is ejected from the system ($e_\text{orb}>1$) before it hits the simulation boundary at $100R_\text{A}^\text{eq}$ where it is then removed. This particular simulation is Disk 040 ($M_\text{disk}/M_\text{A}=0.03$, $\phi=35^\circ$) and a movie of this simulation is provided.}
    \label{fig:pair_example_timeseries}
\end{figure}

\begin{figure}[H]
\centering
\includegraphics[width=0.2\textwidth]{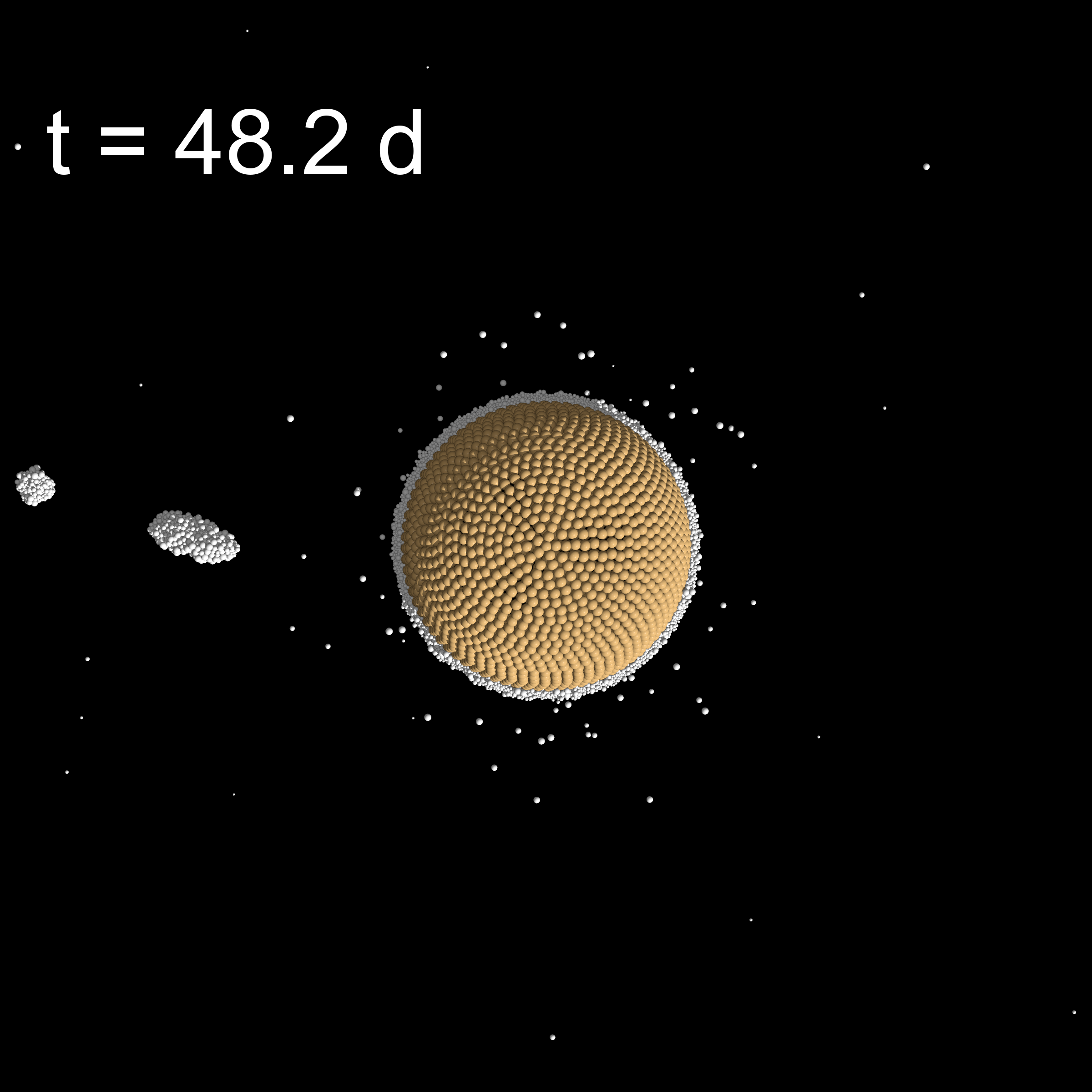}
\includegraphics[width=0.2\textwidth]{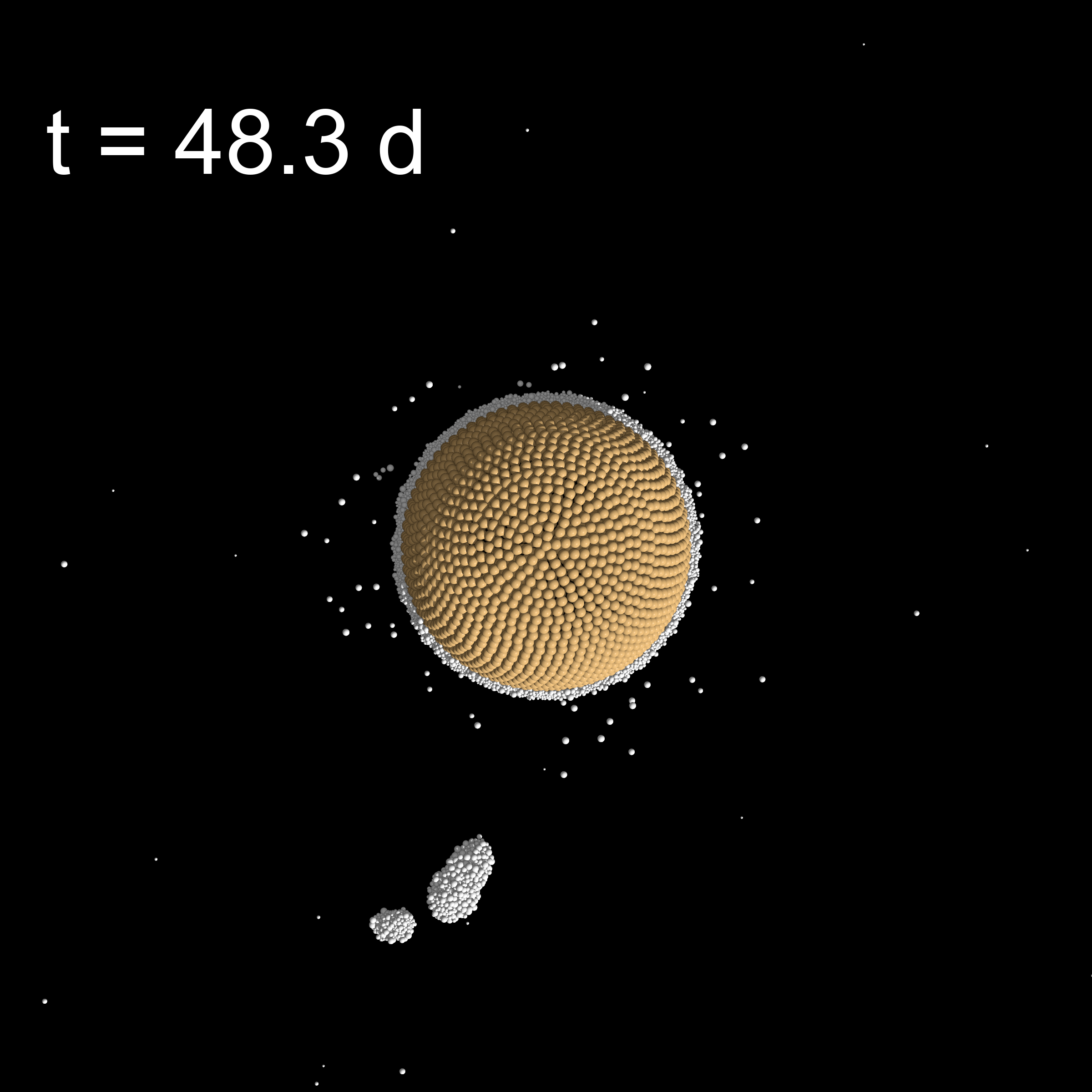}
\includegraphics[width=0.2\textwidth]{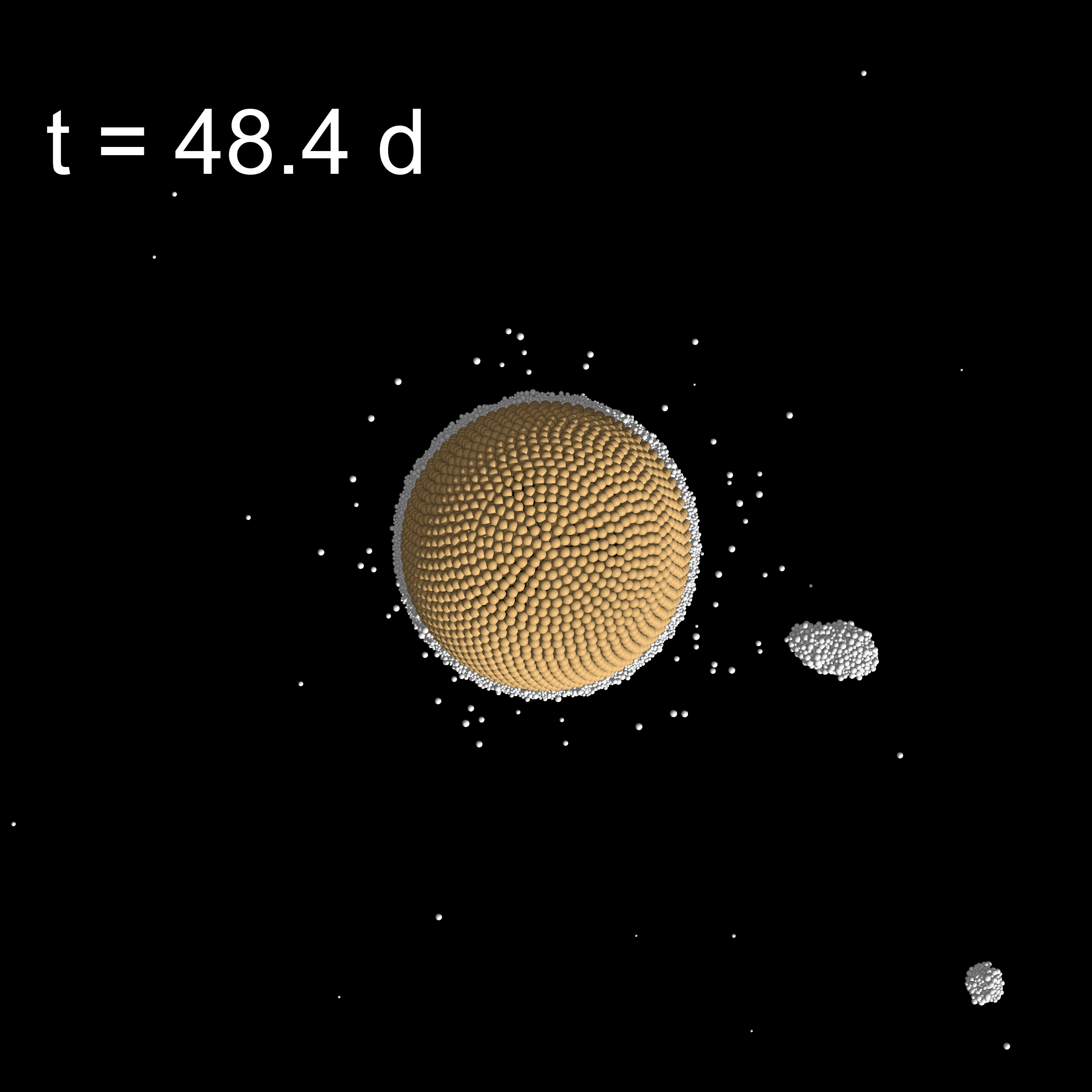}
\includegraphics[width=0.2\textwidth]{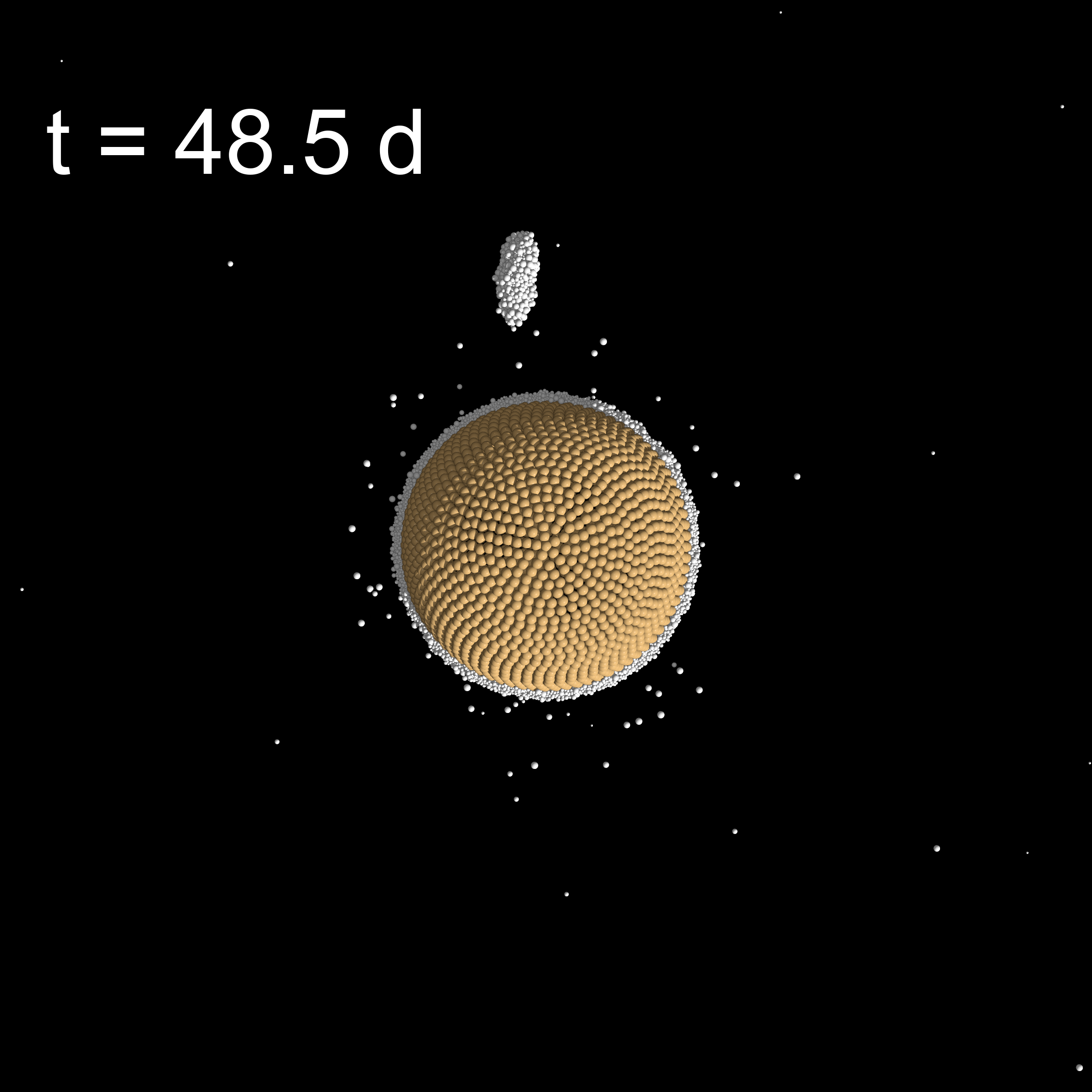}
\includegraphics[width=0.2\textwidth]{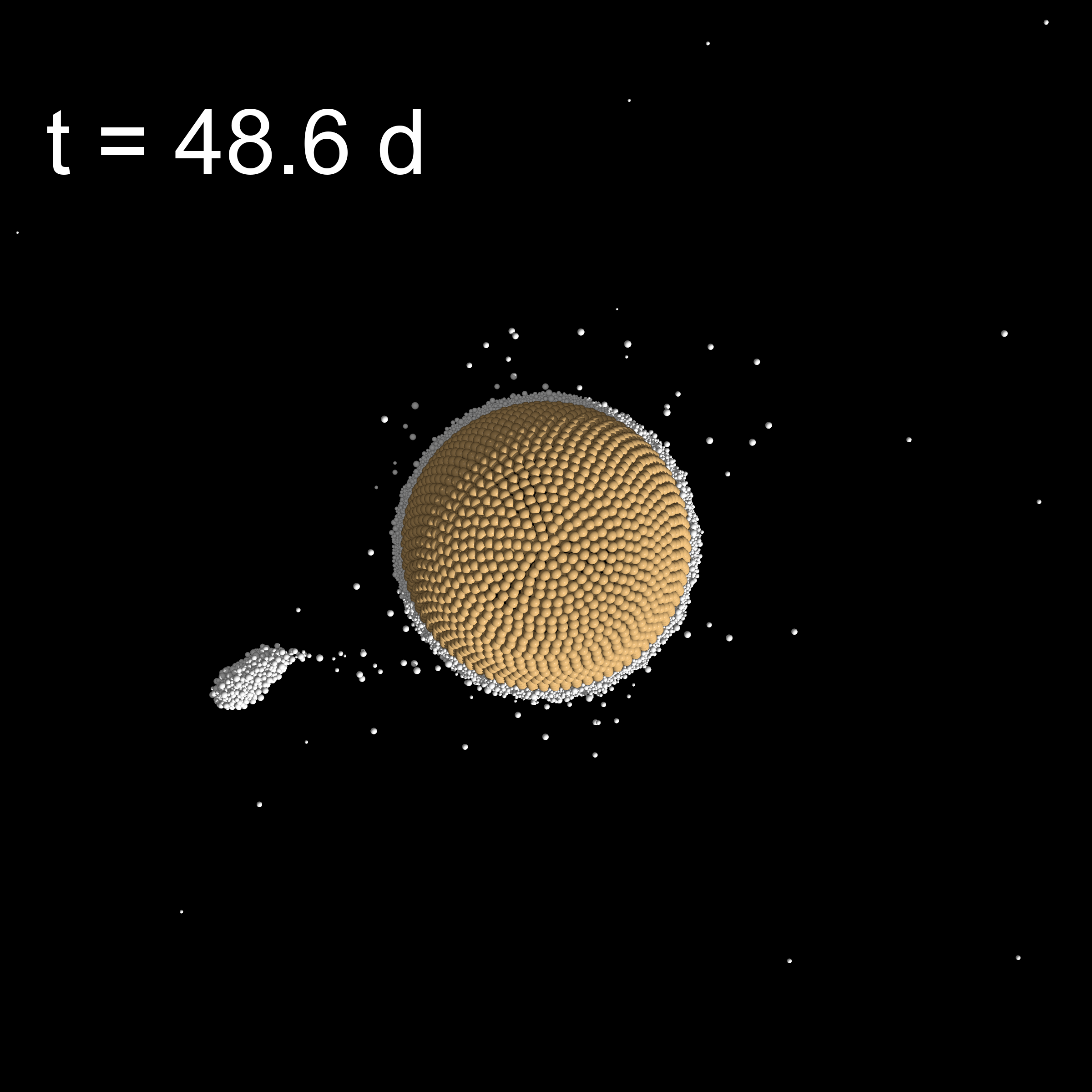}
\includegraphics[width=0.2\textwidth]{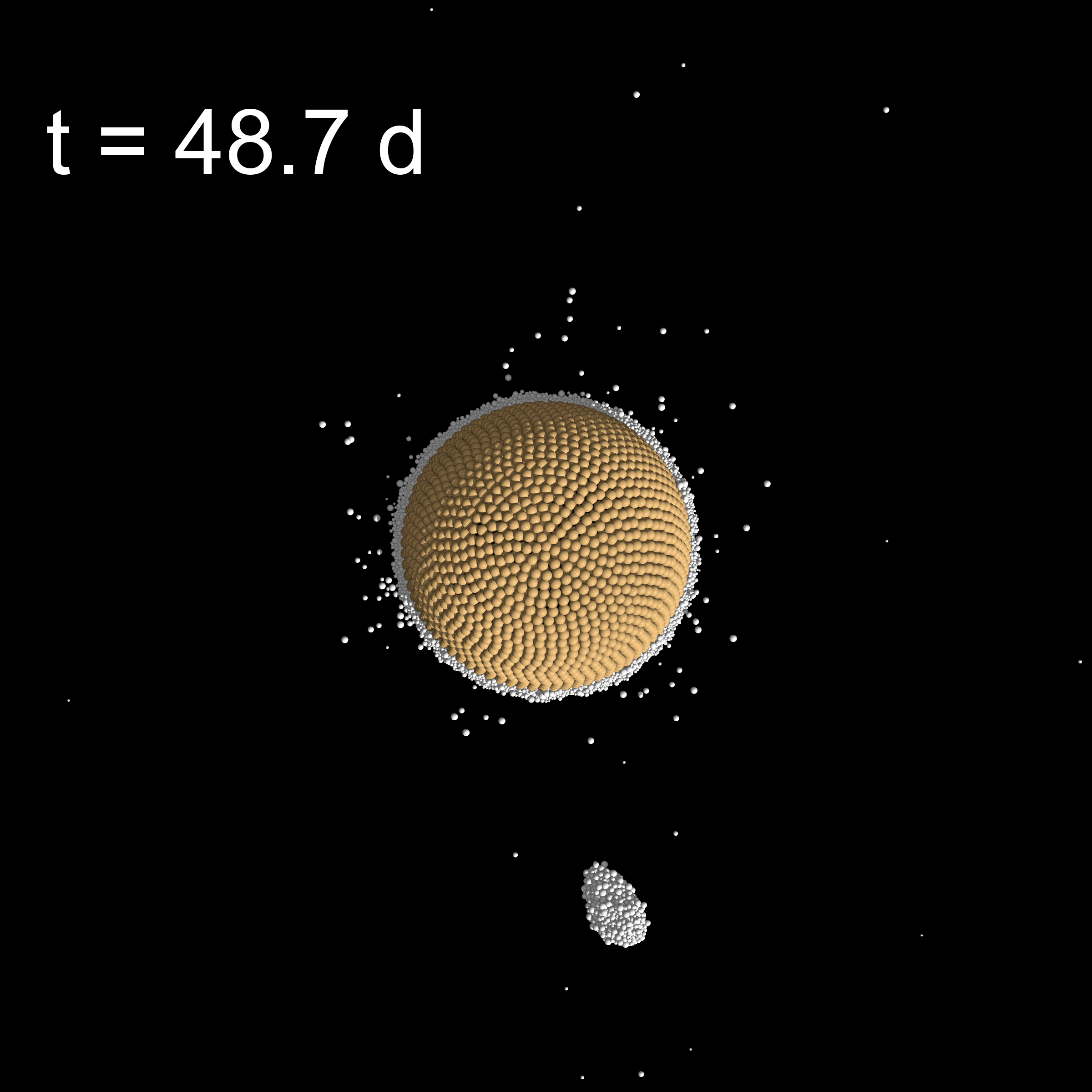}
\includegraphics[width=0.2\textwidth]{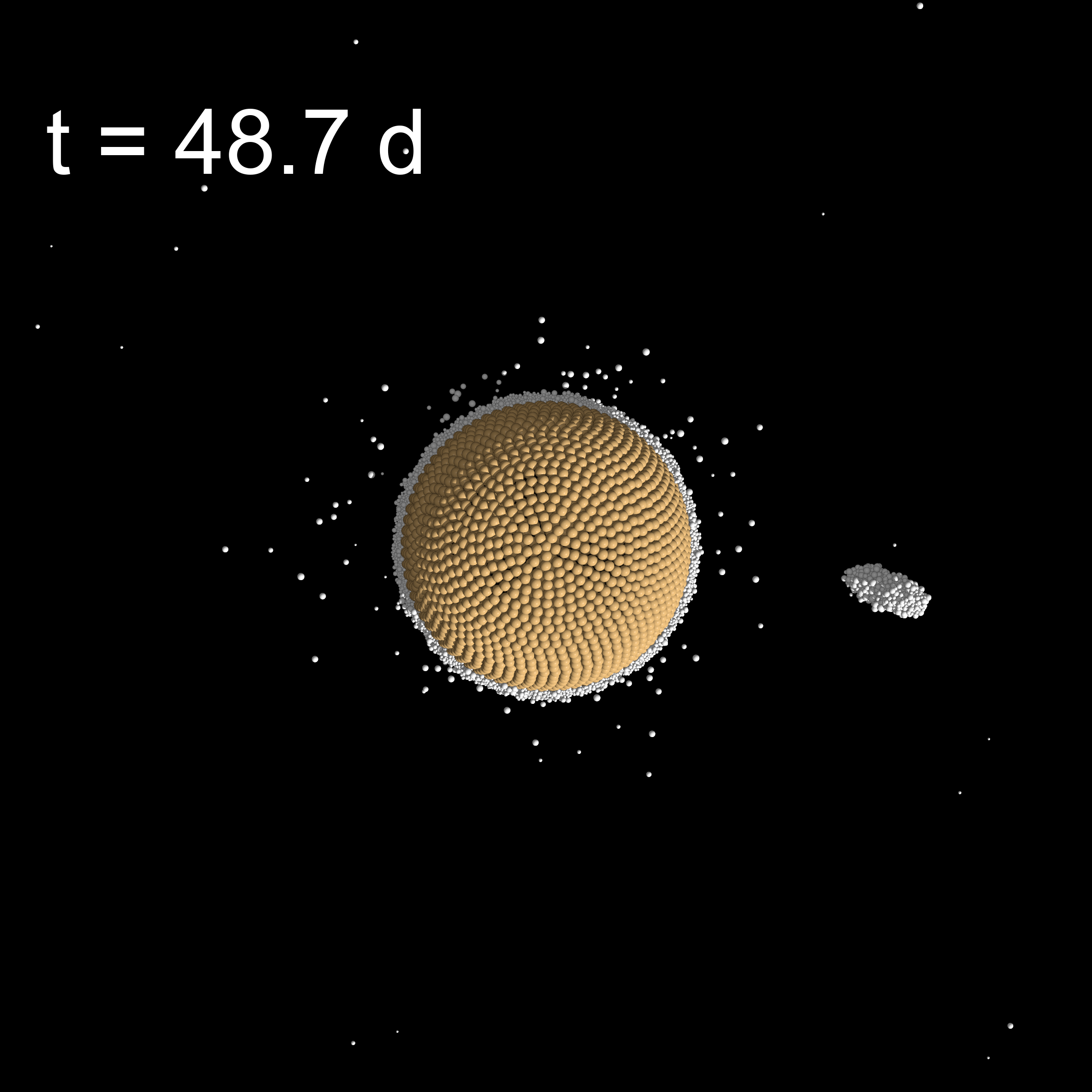}
\includegraphics[width=0.2\textwidth]{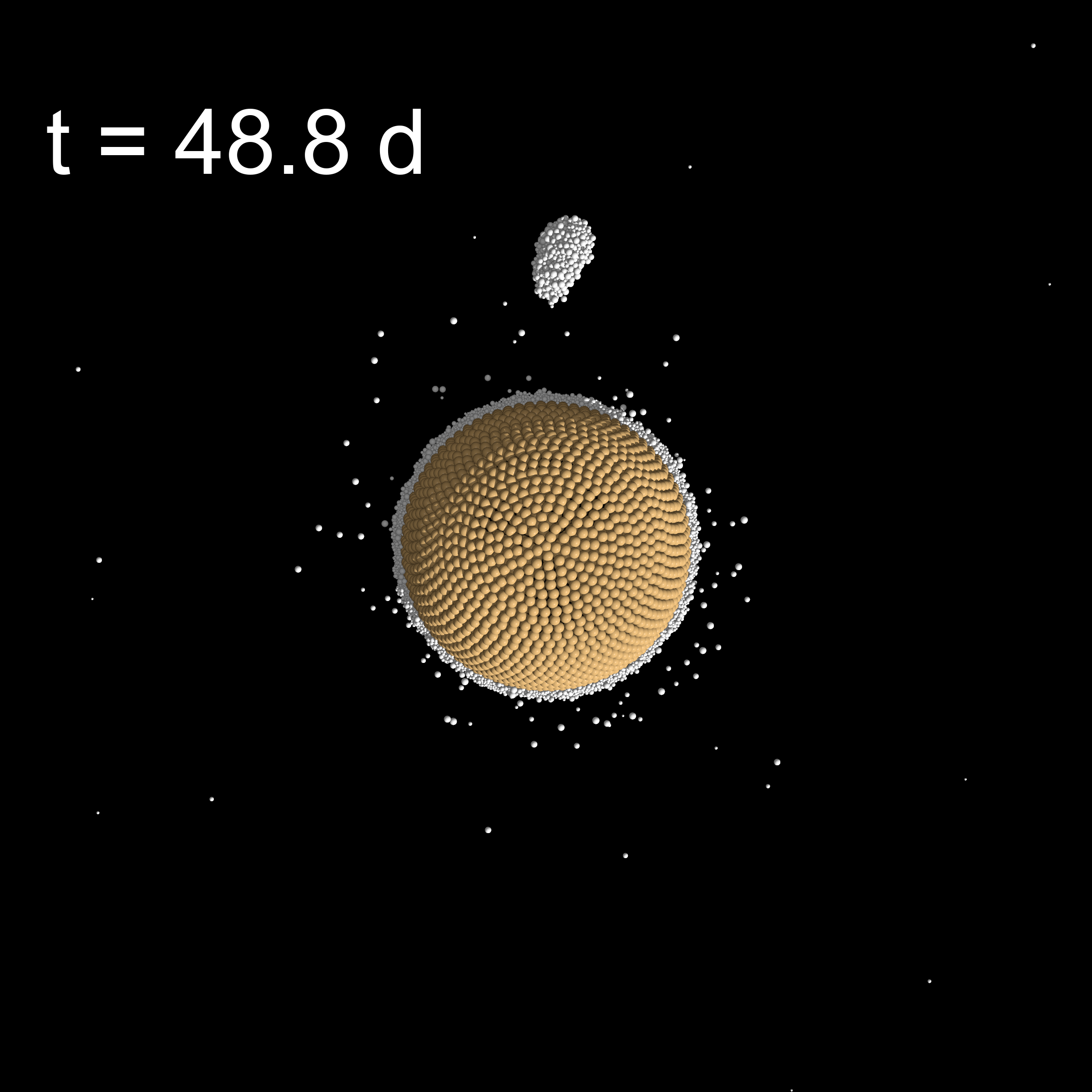}
\includegraphics[width=0.2\textwidth]{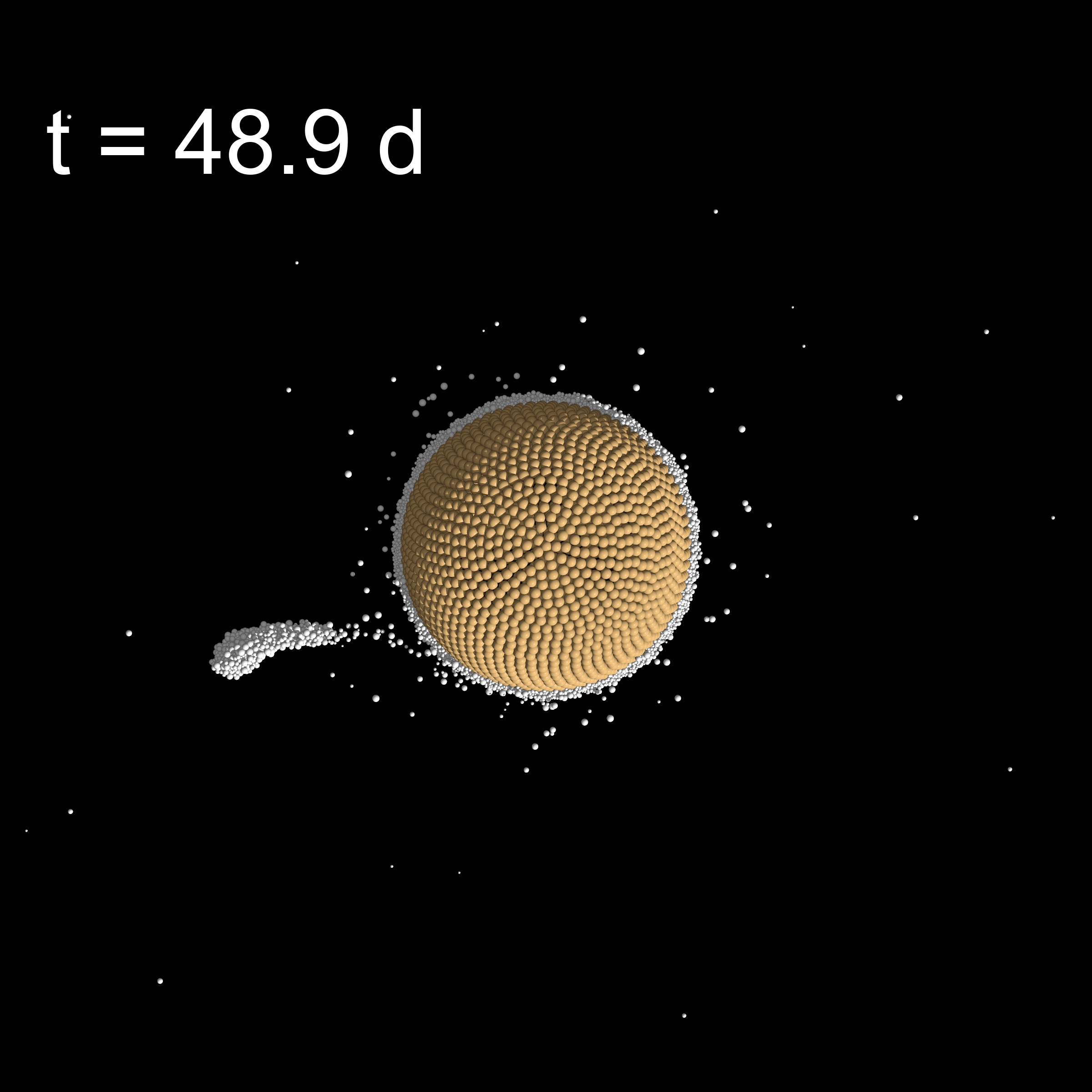}
\includegraphics[width=0.2\textwidth]{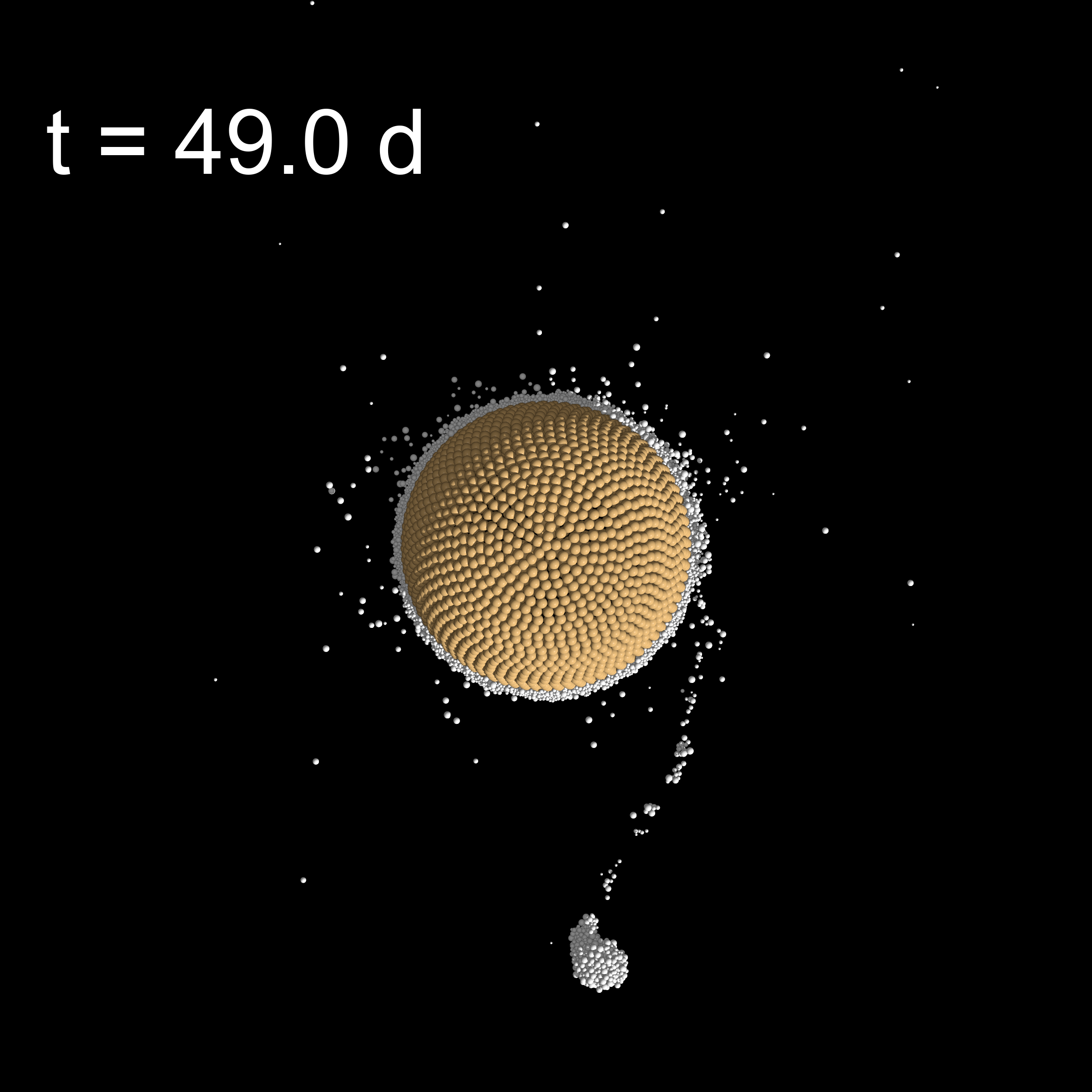}
\includegraphics[width=0.2\textwidth]{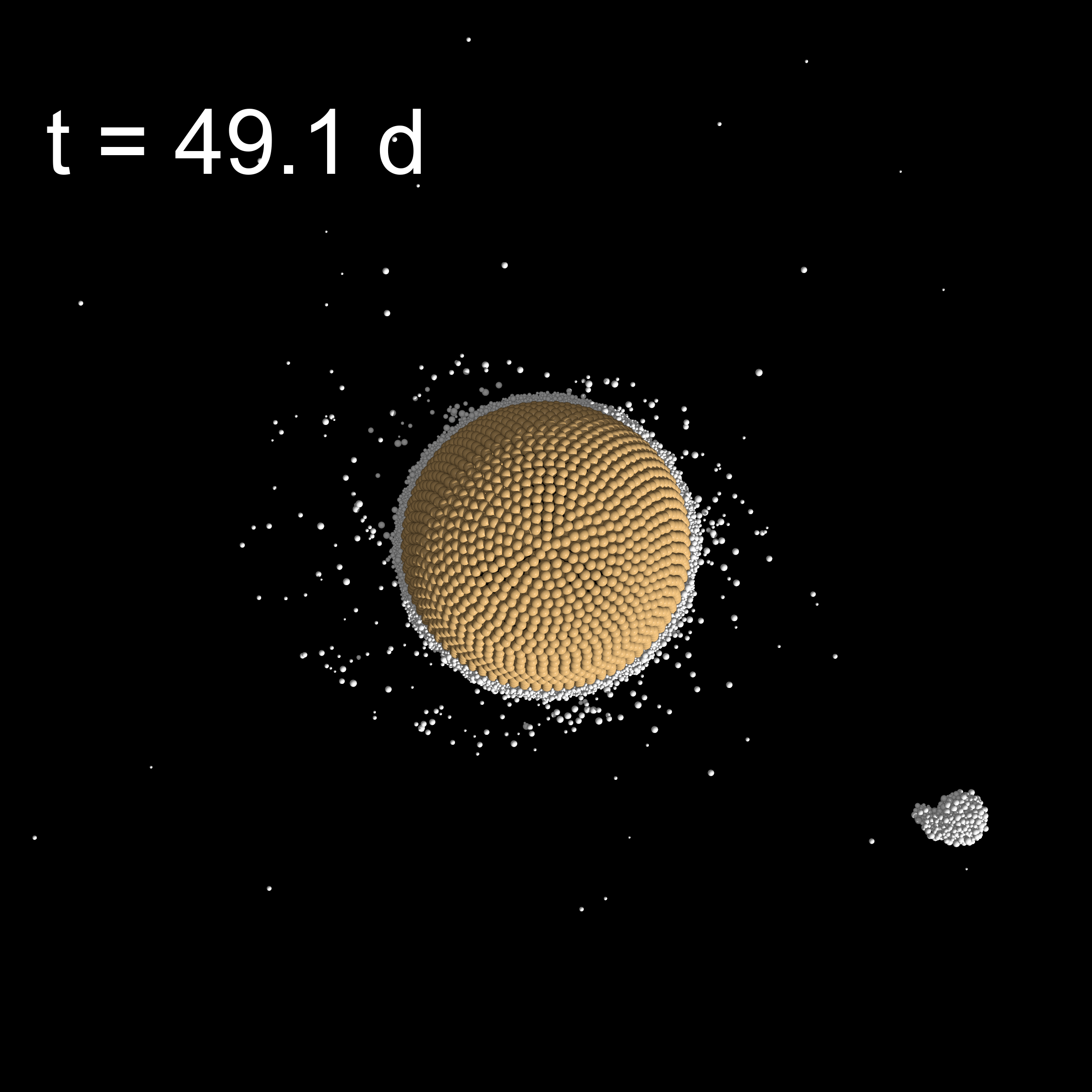}
\includegraphics[width=0.2\textwidth]{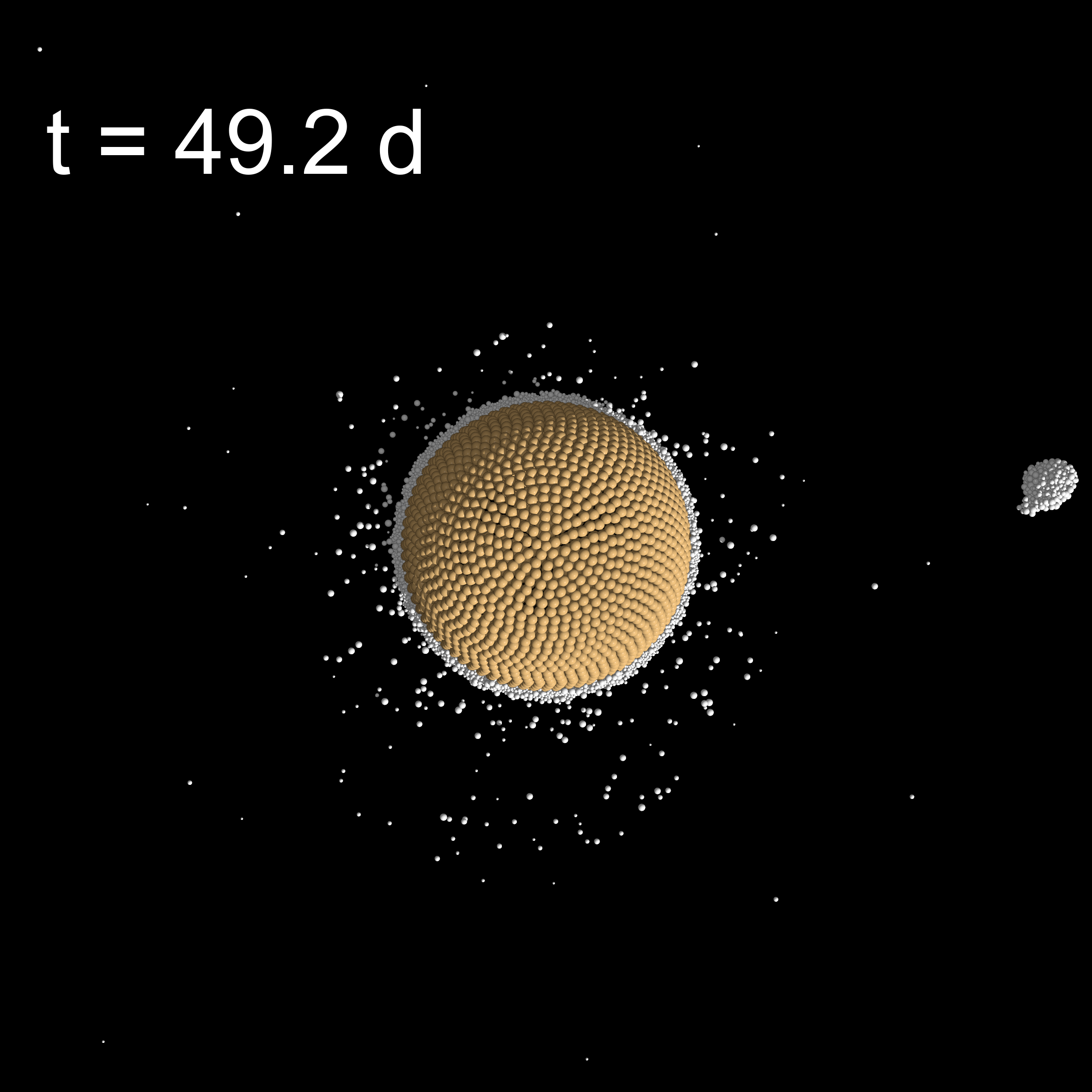}
\caption{Simulation snapshots corresponding to Fig.\ \ref{fig:pair_example_timeseries} of the final close encounter that ejects the smaller satellite from the system. When this occurs, larger satellite's periapse distance is lowered and its eccentricity is raised. This leads to a partial tidal disruption where the remaining satellite loses some mass and breaks from synchronous rotation. A movie of this simulation is provided.}\label{fig:pair_snapshopts}
\end{figure}

As an example, we show a time-series plot of the semimajor axis and eccentricity of the two largest satellites from Disk 040 ($M_\text{disk}=0.03M_{A}$, $\phi=35^\circ$) in Fig.\ \ref{fig:pair_example_timeseries}. In this case, the smaller satellite ($M_2$) has a series of close encounters with the larger satellite ($M_1$), which increase its semimajor axis and eccentricity until its eccentricity eventually exceeds 1, placing it on an escape trajectory. The particle then reaches the simulation boundary and is removed from the simulation. In Fig.\ \ref{fig:pair_snapshopts}, we show snapshots of this simulation at the moment when the $M_2$ is ejected. The two satellites have a close encounter at 48.3 d, where $M_1$ torques $M_2$ onto a hyperbolic orbit, ejecting it from the system. Meanwhile, as a consequence of angular momentum conservation, $M_1$'s eccentricity is raised (i.e., its periapse distance decreases) causing it to have a close approach with the primary at 48.9 d. $M_1$ then undergoes a partial tidal disruption, loosing some mass, reshaping, and torquing it back onto a higher orbit safe from further tidal disruption. This process also breaks $M_1$ from its previous synchronous rotation state.

\begin{figure}[H]
        \centering
        \subfloat[Fragment size\label{fig:subfig:pairs_size}]{\includegraphics[width=0.3\linewidth]{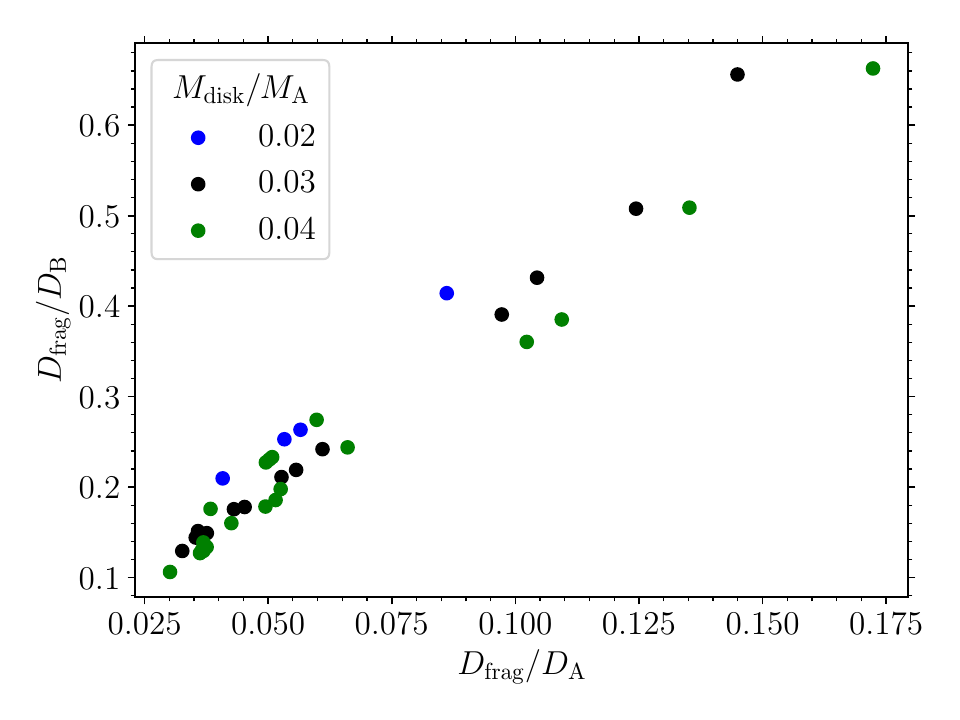}}
        \subfloat[Fragment spins\label{fig:subfig:pairs_spin}]{\includegraphics[width=0.3\linewidth]{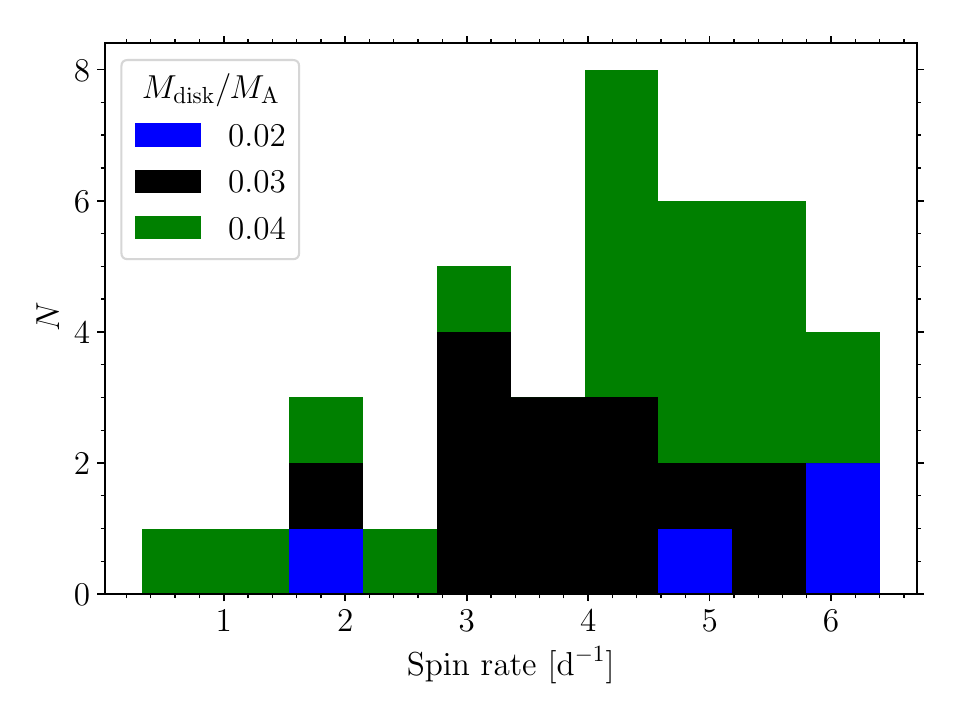}}\\
        \subfloat[Fragment shapes\label{fig:subfig:pairs_shape}]{\includegraphics[width=0.3\linewidth]{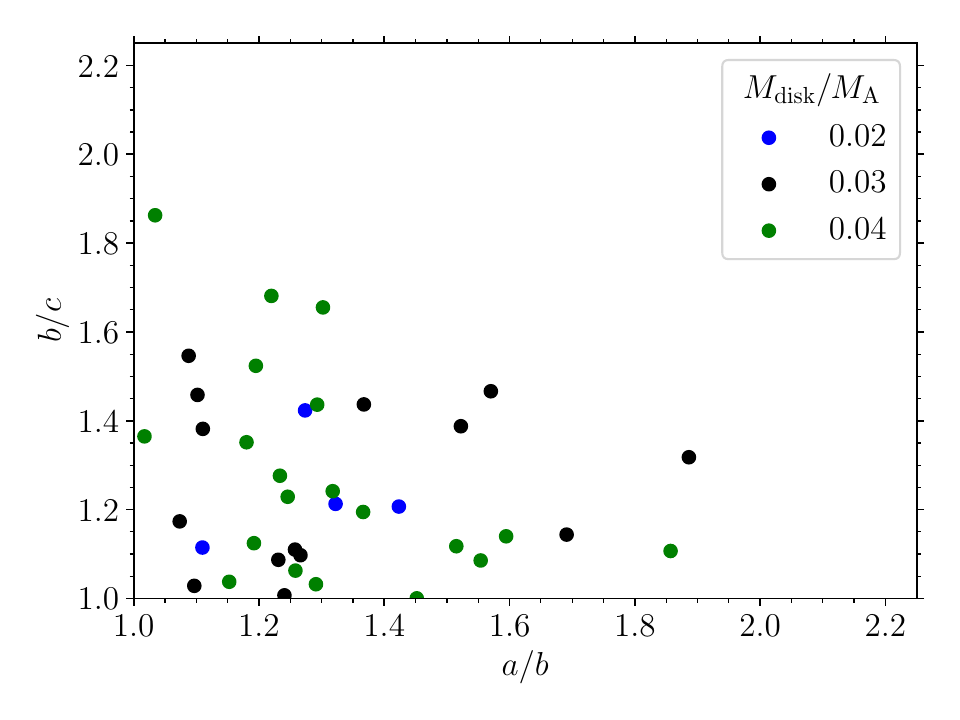}}
        \subfloat[Fragment eccentricity\label{fig:subfig:pairs_ecc}]{\includegraphics[width=0.3\linewidth]{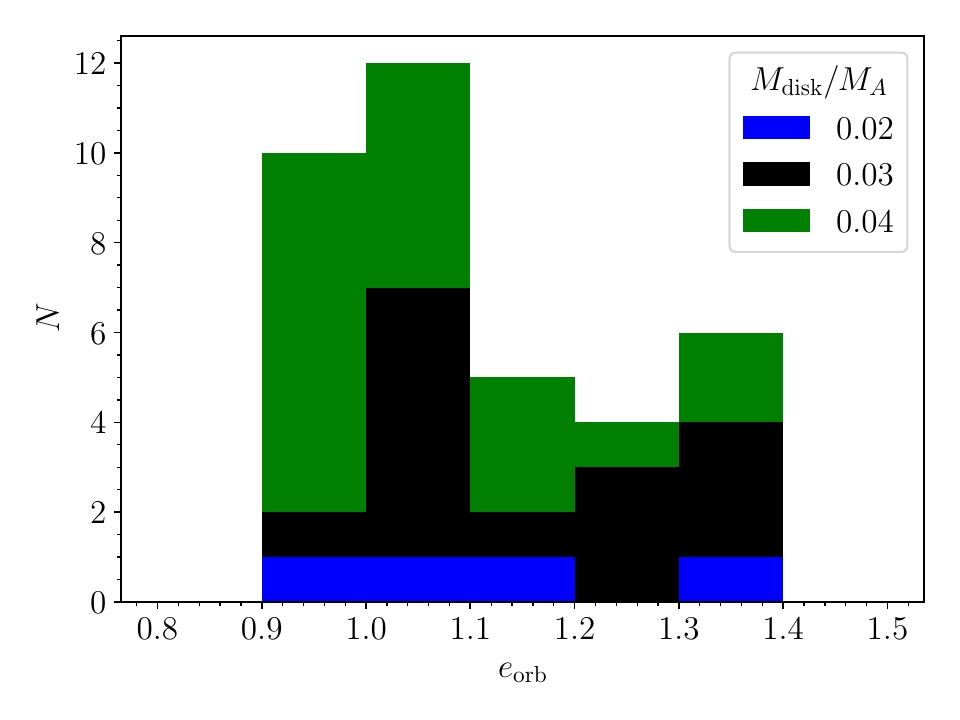}}
        \caption{Simulation results for all ejected fragments with $\phi=35^\circ$. These plots ignore all fragments containing less than 3 particles, of which there are many. The different colors show the initial mass of the disk. Only a minority of simulations eject fragments at all, but in some cases, multiple fragments are ejected in a single simulation. (a) The size of each ejected fragment as a function of both the fragment-to-primary and fragment-to-secondary diameter ratio. Unsurprisingly, a more massive initial disk ejects larger fragments. (b) A histogram of the spin rate of each ejected fragment. (c) The shapes of each ejected fragment, defined by their physical extents. (d) The eccentricity each fragment relative to the primary. Most fragments are on escape trajectories ($e_\text{orb}>1$) while the others have their apoapse outside of the systems Hill sphere if it were located at 1 au.}
        \label{fig:pairs}
\end{figure}

In Fig.\ \ref{fig:subfig:pairs_size}, we plot the size of each ejected fragment in relation to both the primary and the secondary, from all simulations where $\phi=35^\circ$.  It is important to note that this figure excludes fragments consisting of one or two particles, of which numerous are ejected from every simulation. Here, we only show ejected fragments that contain 3 or more particles. Also, this figure \textit{only} includes fragments that have been ejected within the ${\sim}100$ d \textsc{pkdgrav} integrations. However, as we will see in the next section, there are several systems which still contain two satellites after 100 d that will be ejected at a later time and would also count as a successful pair. In general, we find that a larger initial mass shedding event tends to eject larger fragments, as there is more material and angular momentum available. For context, the smallest size ratio observed for an asteroid pair is roughly ${\sim}0.1$, so this is a size range that would be potentially discoverable \citep{Pravec2019}. We also show the spin rate of each ejected fragment (Fig.\ \ref{fig:subfig:pairs_spin}), which demonstrates that most fragments are spinning relatively fast, due to the high angular momentum nature of ejecting the fragment. The observed population of asteroid pairs shows a much broader distribution of spin rates of ejected secondaries, ranging between ${\sim}1-9$ d$^{-1}$ \citep{Pravec2019}. In addition, many of the pairs formed in these simulations are in non-principal axis (NPA) rotation, whereas all observed secondaries of asteroid pairs for which there is sufficient lightcurve coverage appear to be in principal axis rotation. This suggests that these fragments undergo significant spin evolution after being ejected (through damping and/or YORP) or other mechanisms besides three body interactions are responsible for the production of most asteroid pairs. The axis ratios are also shown in Fig.\ \ref{fig:subfig:pairs_shape} and are consistent with the axis ratios of observed pairs inferred from their lightcurve amplitudes \citep{Pravec2019}. Finally, we also show the eccentricity of the fragment before it is deleted from the simulation. The vast majority are on hyperbolic (escape) trajectories, having $e_\text{orb}>1$. However, some of these fragments are still bound to the system. If these fragments were kept in the simulation, they would likely have been ejected or collided during a future close encounter. If these are near-Earth asteroids, these satellites would be lost anyways, having an apoapse outside of the system's Hill sphere. However, if these are main-belt asteroids with larger Hill spheres, there is a chance that these objects deemed ``pairs'' could remain bound with wide eccentric orbits. 

For example, asteroid (3749) Balam is a main-belt triple with a fast-rotating (2.8 h spin period), ${\sim}4$ km primary with a relatively close inner satellite, separated by ${\sim}13$ km on a near-circular orbit \citep{Pravec2019}. The outer satellite is on a wide orbit with an eccentricity between 0.35 and 0.77, \citep{Merline2002a, Vachier2012}. The inner and outer satellites have respective satellite-to-primary size ratios of ${\sim}0.46$ and ${\sim}0.24$ \citep{Pravec2019}, making them relatively large. In addition, Balam is a member of an asteroid pair \citep{Vokrouhlicky2009}, with backwards numerical integrations indicating that asteroid 2009 BR$_{60}$ separated from Balam less than one million years ago \citep{Vokrouhlicky2009, Pravec2019}, making the Balam system very young. Due to its high eccentricity, it has been suggested that Balam's outer satellite could have been formed as the result of a collision \citep{Durda2004} or by capture \citep{Marchis2008b}, while the inner satellite is consistent with YORP spin-up and rotational disruption due to the primary's fast rotation \citep{Polishook2011}. Since the simulations in this study considered much more mild mass shedding events and removed these eccentric satellites, there are no direct analogs to the Balam system in our study. However, our results suggest that a plausible explanation for the Balam system could be formation via a single, much more massive rotational disruption event. In such a scenario, an inner satellite, an outer eccentric satellite, and an unbound satellite could all be produced at the same time, although the details of such a scenario require further study.

\subsubsection{Triples}
In some cases, we find that 3-body interactions lead to the formation of systems with two satellites on non-crossing orbits. However, these simulations are run for less than 100 days, making it impossible to determine their stability with \textsc{pkdgrav} alone. In rare cases where two satellites are still remaining at the end of the \textsc{pkdgrav} simulation, we perform a brief test of the system's stability. We hand off the masses, positions, and velocities of the primary and its satellites to the REBOUND $N$-body code \citep{rebound,reboundias15,reboundsa}. These simulations are relatively simple; the two satellites are treated as spheres but we include the effect of the primary's $J_2$ using the REBOUNDx package \citep{reboundx}. As these simulations do not include higher-order effects such as spin-orbit coupling of the two satellites, tidal dissipation, BYORP, etc., we only integrate the system for $10^3$ years and then check if both satellites still remain. In reality, higher-order effects may strongly affect the system's evolution and stability. Additionally, the primary is largely azimuthally symmetric, and a more realistic primary with a non-negligible $C_{22}$ may make these close-in triple systems less stable. Here, we merely demonstrate that the formation of a triple system from a single rotational disruption event is plausible. In total, we find 11 triple systems after the 100 d \textsc{pkdgrav} simulations, of which 7 survive after further integration with REBOUND. 

We find that many of the stable triples achieved their stability due to capture into mean motion resonances (MMRs) during the \textsc{pkdgrav} stage. While moonlets are growing, their orbits are constantly perturbed by collisions, mergers, and close encounters, so capture into a MMR is effectively random. Out of the 7 stable triples, 5 are in MMRs, while two are not in resonance. 

\begin{figure}
        \centering
        \includegraphics[width=0.325\textwidth]{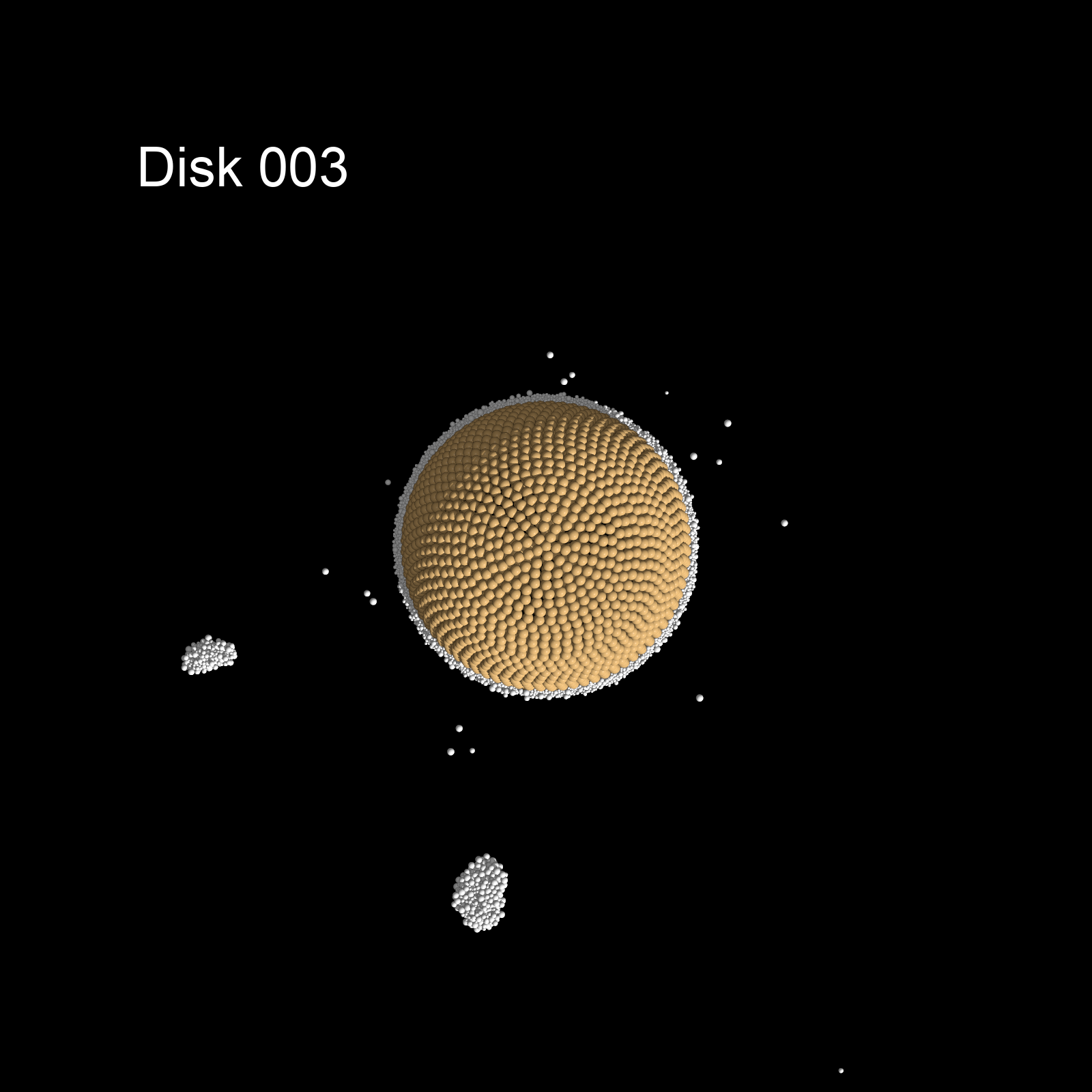}
        \includegraphics[width=0.325\textwidth]{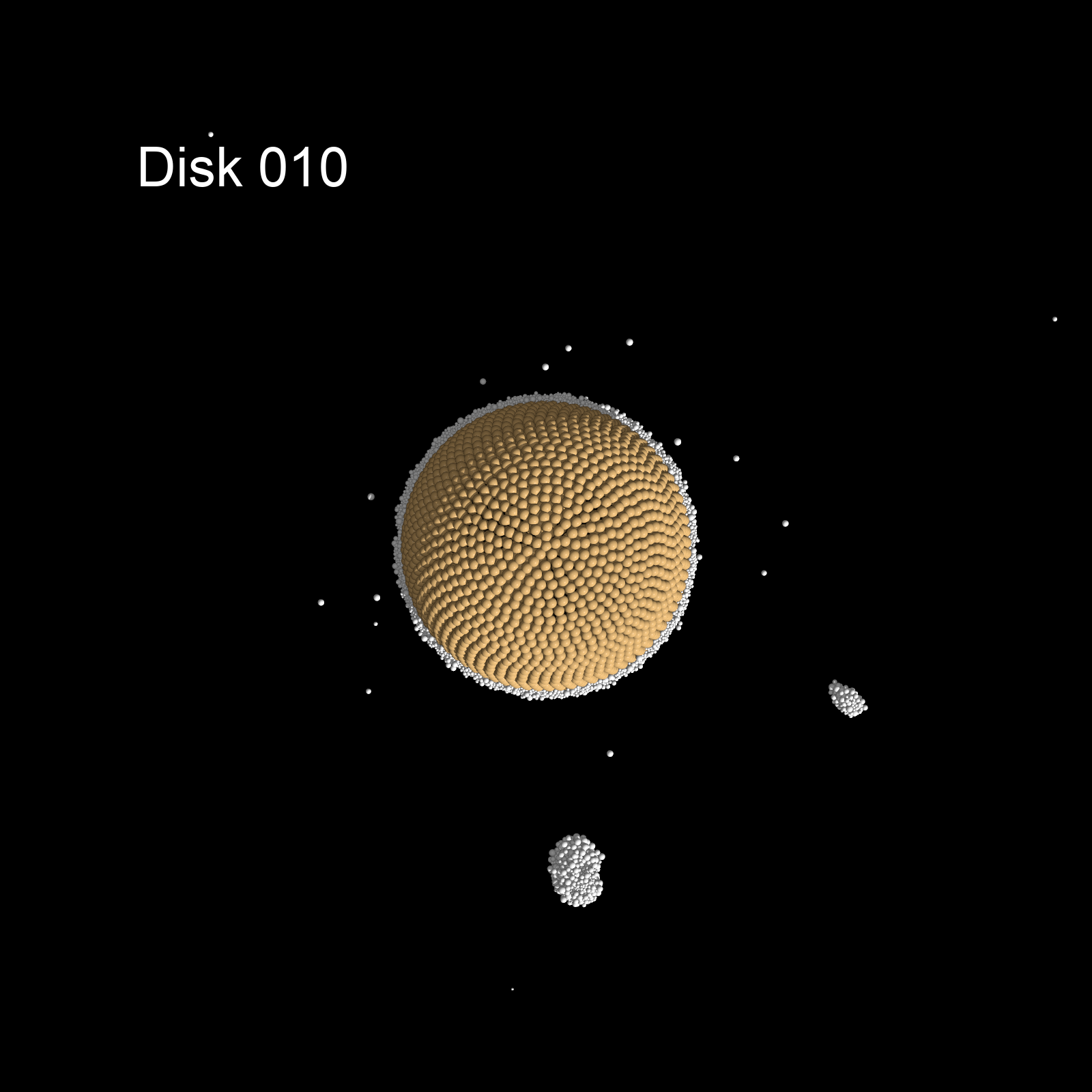}
        \includegraphics[width=0.325\textwidth]{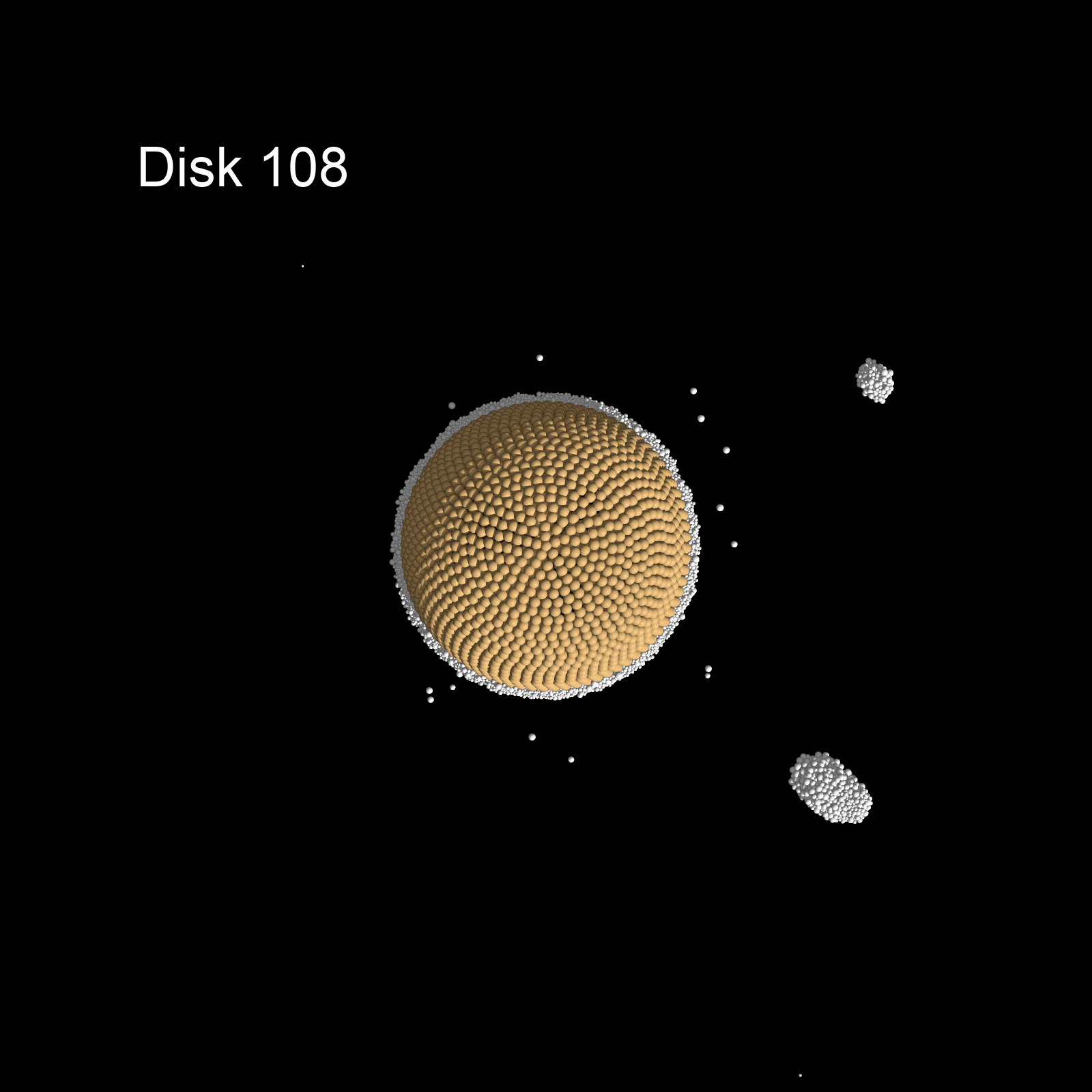}

        \includegraphics[width=0.325\textwidth]{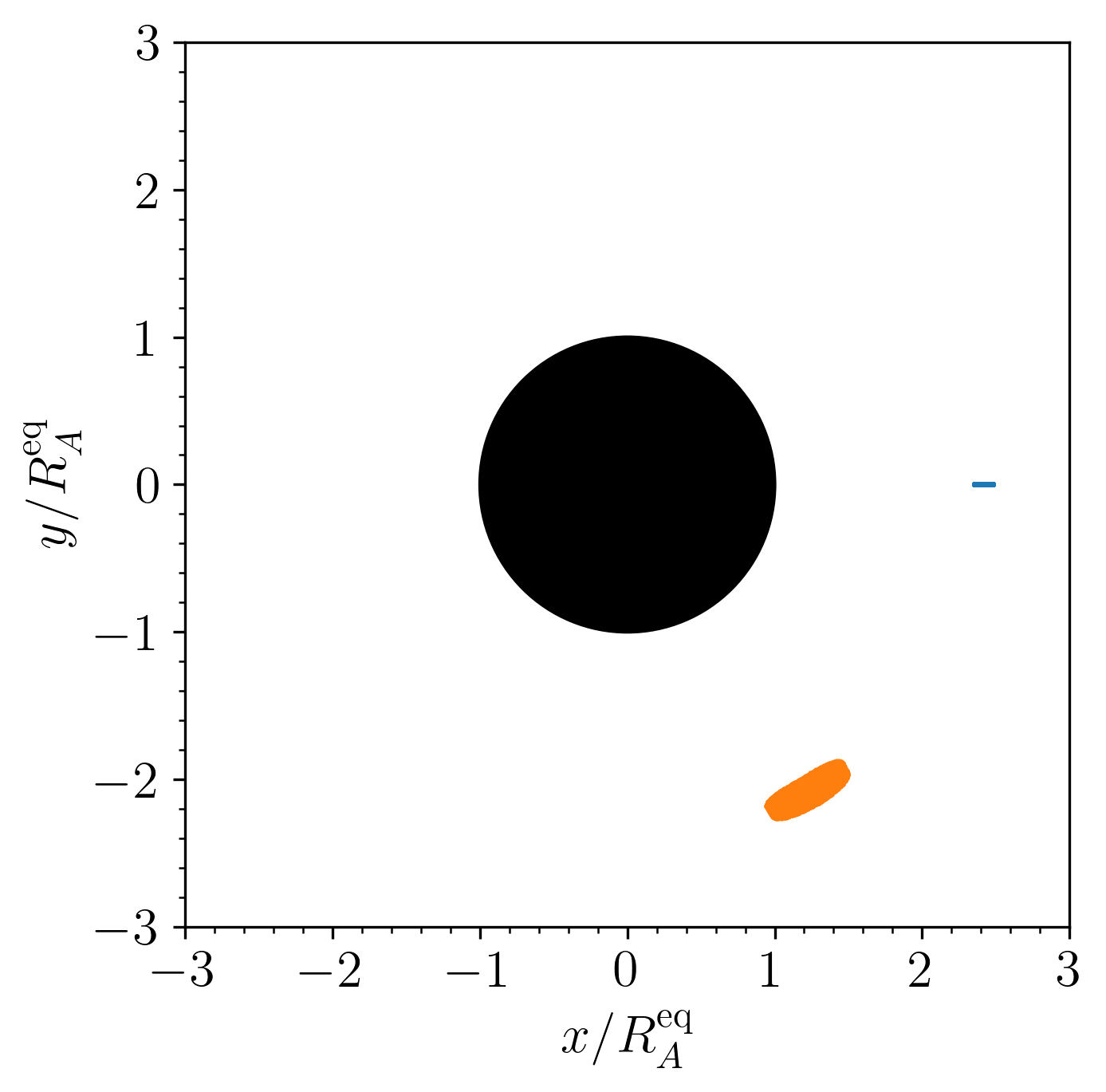}
        \includegraphics[width=0.325\textwidth]{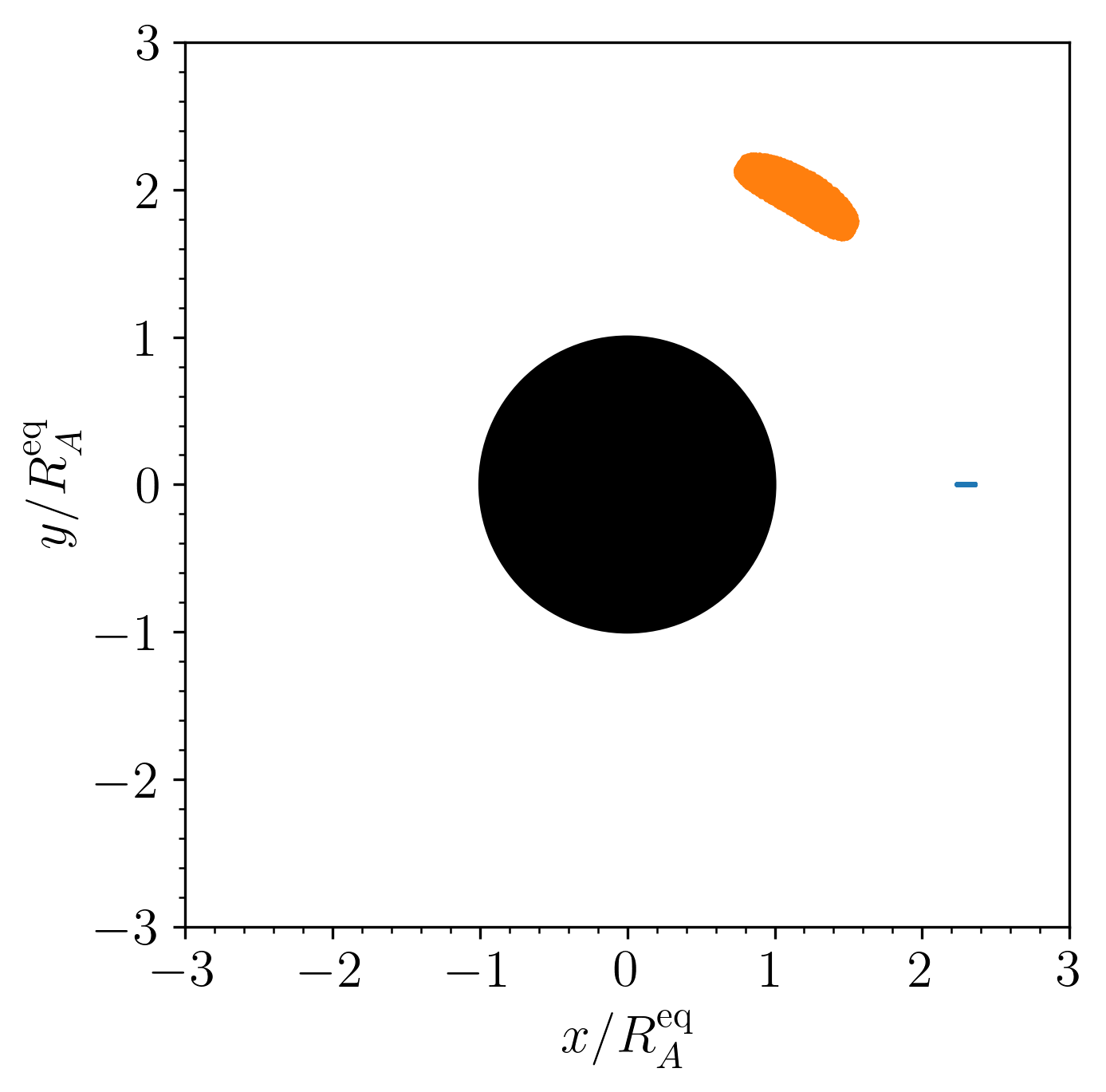}    
        \includegraphics[width=0.325\textwidth]{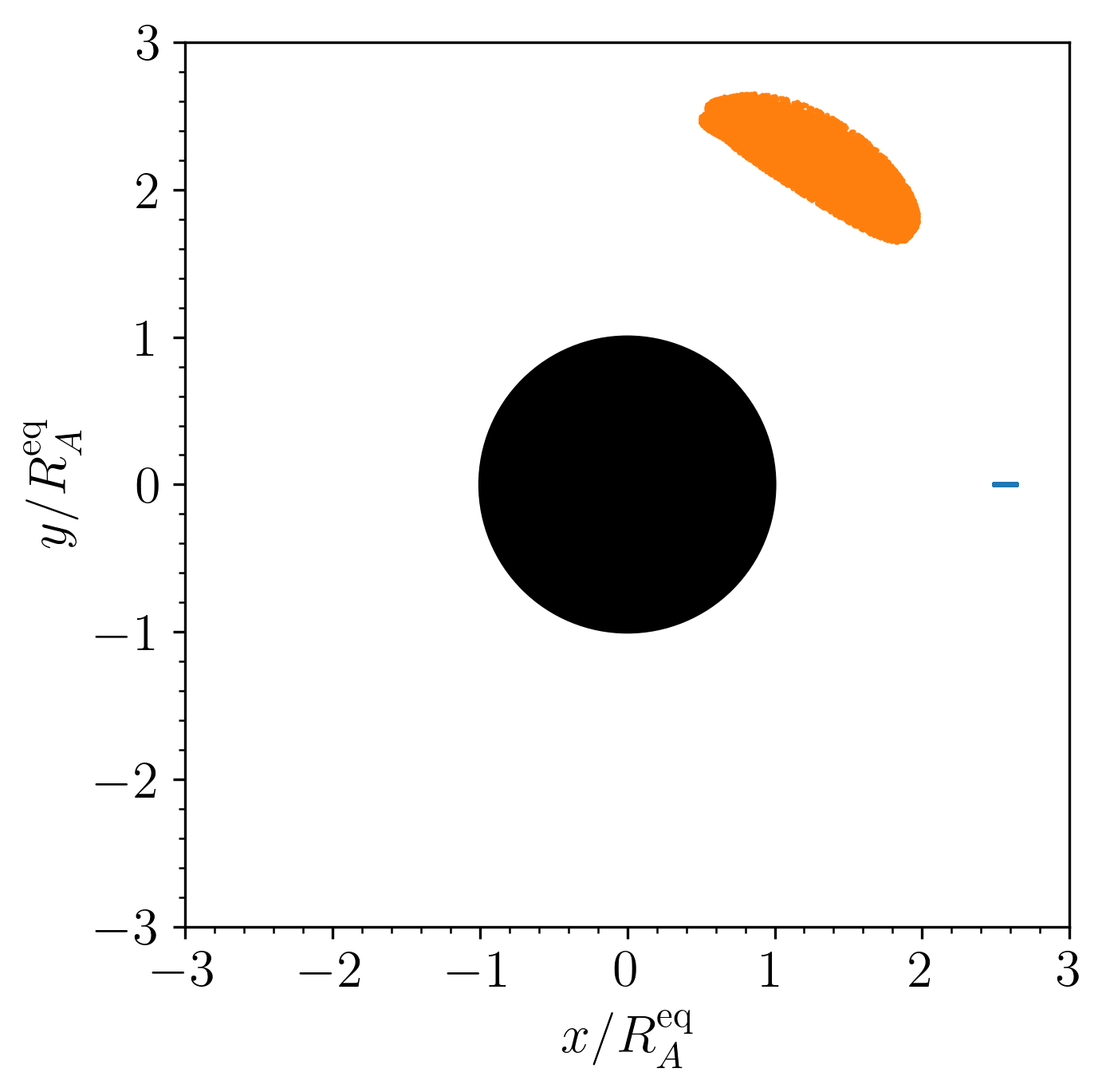}    
        \caption{Three cases where the co-orbital configuration of two satellites remains stable after a $10^3$ yr integration with REBOUND. The top panels show a view of the \textsc{pkdgrav} simulation after ${\sim}100$ d prior to handing off to REBOUND, looking down from the primary's spin pole. We encourage the reader to view the provided movies of these simulations. The corresponding bottom panels show the position of the two satellites in the rotating frame of the larger satellite at each output over the $10^3$ yr REBOUND integration, demonstrating that the smaller satellite resides in a Lagrange point (either $L_4$ or $L_5$). In the left case (Disk 003), the smaller satellite librates around the trailing Lagrange point ($L_5$) with an amplitude of ${\sim}7.2^\circ$, while in the middle and right cases (Disk 010 and Disk 108), the smaller satellite librates around the leading Lagrange point ($L_4$) with amplitudes of ${\sim}11.4^\circ$ and ${\sim}18.7^\circ$, respectively. Movies of these threes simulations are provided.}  \label{fig:coorbitals}
\end{figure}

In three of the resonant cases, the two satellites form in a co-orbital configuration (i.e., the 1:1 MMR) by co-accretion. When this occurs, the smaller satellite occupies the  $L_4$ or $L_5$ Lagrange point of the larger body. Snapshots of the three cases are shown in Fig.\ \ref{fig:coorbitals}. These systems satisfy the condition for stability, in which $\frac{M_1+M_2}{M+M_1+M_2}\lesssim 0.04$, where $M$ is the primary's mass and $M_1$ and $M_2$ are the two satellite masses \citep{Murray2000,Laughlin2002,Sicardy2010}. We also numerically test the system's stability using REBOUND and find that the smaller satellite remains in stable libration about the Lagrange point for at least $10^3$ yr. We restrict the REBOUND simulations to only $10^3$ yr, because we neglect tides and non-gravitational forces which can tighten or loosen the tadpole orbits, depending on the direction of migration \citep[e.g.,][]{Fleming2000}.  We find that co-orbital systems are only formed when the initial disk mass is small ($M_\text{disk}/M_\text{A}\lesssim0.03$), in agreement with the study of \cite{Hyodo2015} on the formation of multiple satellites with $N$-body simulations in circumplanetary disk.

We note that the idealized initial conditions of these simulations may increase the likelihood of forming co-orbital satellites and that the primary's azimuthal symmetry likely contributes to their long term stability. Here, we merely point out that forming a co-orbital system via a single mass shedding event is \textit{possible}, albeit \textit{unlikely}. A future detection of a triple asteroid with co-orbital satellites, however unlikely, could be explained by co-accretion following a single mass shedding event.

The remaining two resonant cases are capture into the 2:1 and 5:2 MMRs. This occurs after a satellite is first torqued onto a wider orbit (${\gtrsim}4R_A$), while there is still material near the Roche limit available to accrete. This allows a second body to grow closer in, which can capture into a nearby MMR with the outer satellite while it is accreting. Once the initial capture occurs, the resonance can further stabilize due to the smooth growth of the inner satellite.

\begin{figure}[H]
        \centering
        \subfloat[Final frame from pkdgrav simulation \label{fig:subfig:triple_topDown_disk014}]{\includegraphics[width=0.5\linewidth]{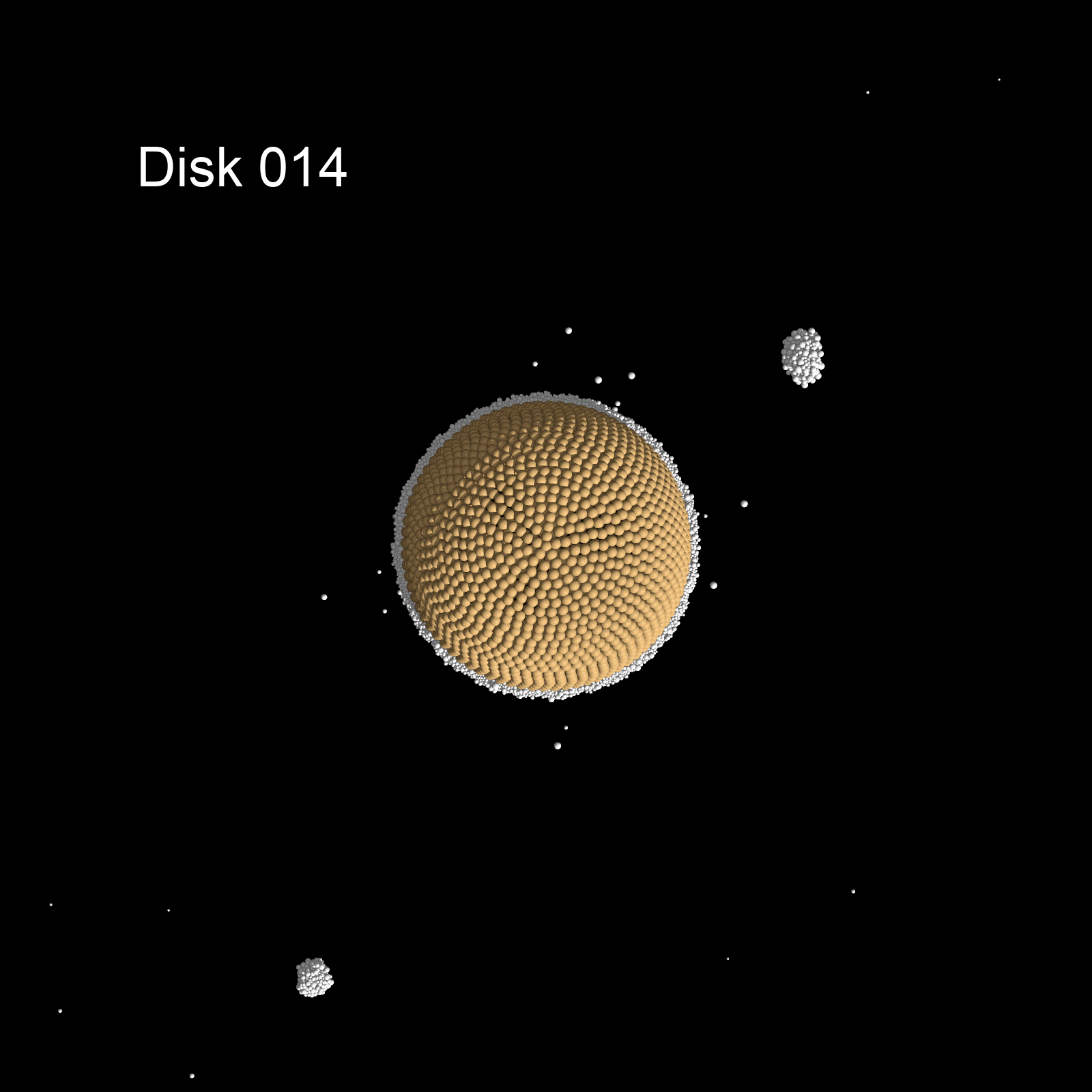}}
        \subfloat[Time series showing capture into 2:1 MMR.\label{fig:subfig:resonance_disk014}]{\includegraphics[width=0.5\linewidth]{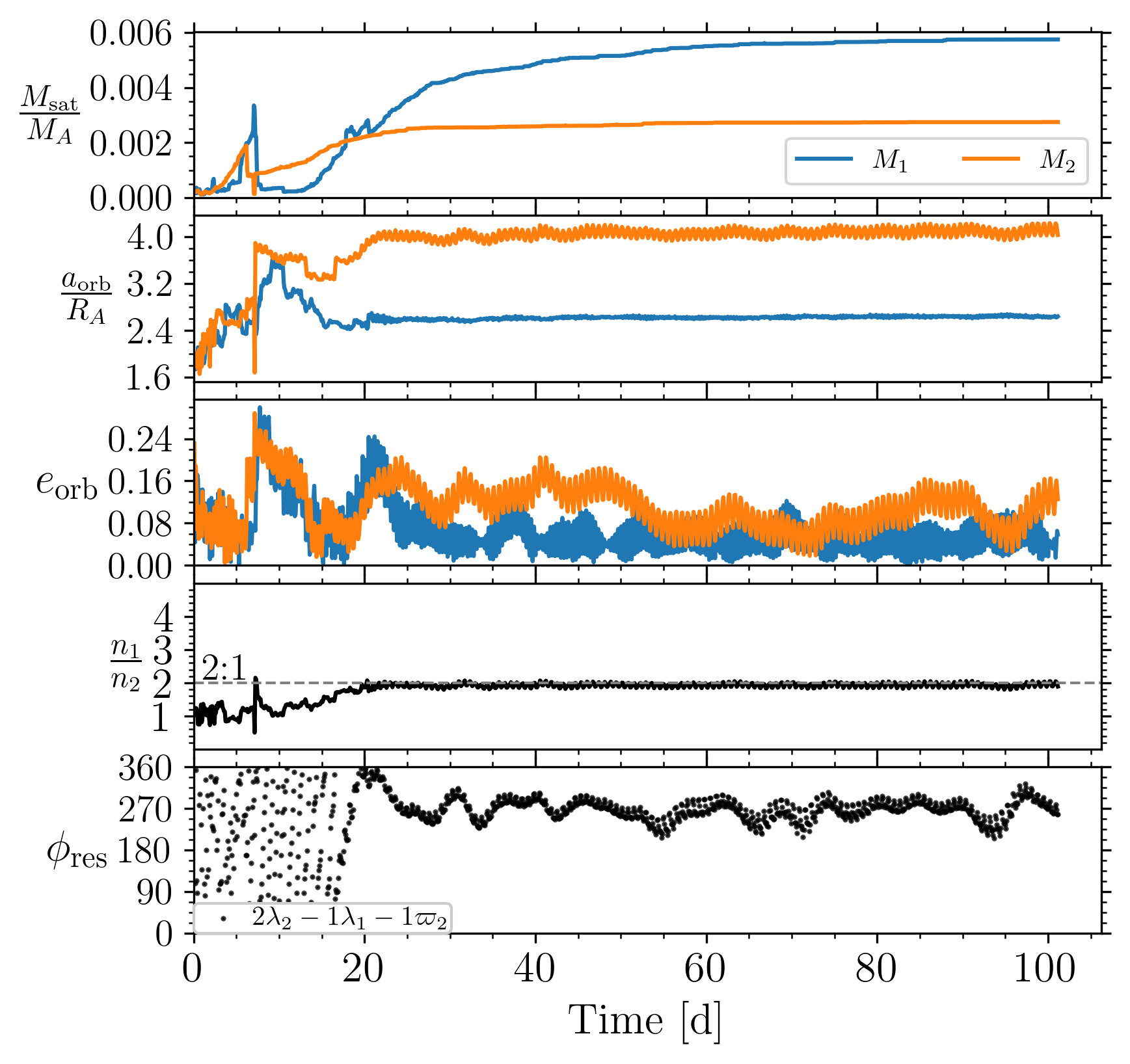}}
        \caption{An example of a stable triple system formed from a single mass shedding event. (a) A snapshot of the system at the end of the simulation. (b) A time-series plot showing the satellite mass, semimajor axis, eccentricity, and resonance argument, $\phi_\text{res}$. Capture into the 2:1 MMR enables the system to remain stable.}\label{fig:triple_disk014}
\end{figure}

In Fig.\ \ref{fig:triple_disk014} we show an example of capture into the 2:1 MMR, where the inner satellite is the most massive.  In Fig.\ \ref{fig:subfig:triple_topDown_disk014}, we show a top-down view of the triple system at the end of the 100 d \textsc{pkdgrav} integration along with a time-series plot in Fig.\ \ref{fig:subfig:resonance_disk014}. Early on, both $M_1$ and $M_2$ grow rapidly until they temporarily merge into a single body which is then quickly tidally disrupted. The tidal disruption torques one large fragment (now $M_2$) onto a wide orbit. Then, $M_2$ continues to grow and migrate outwards through a series of mergers and close encounters with other moonlets. Meanwhile a new inner satellite forms (now $M_1$) from one of the disrupted remnants and migrates inwards. This process places $M_1$ and $M_2$ close enough to the 2:1 MMR that they capture. The resonance is then further stabilized by the smooth growth of $M_1$, the inner satellite. We can confirm the existence of the resonance due to the libration of the resonance argument $\phi_\text{res} = 2\lambda_2-\lambda_1-\varpi_2$, where $\lambda_i$ are the respective body's mean longitudes and $\varpi_2$ is $M_2$'s longitude of pericenter. We attribute the drift in the resonance argument to the growth in the satellite masses and collisions, both of which make the process non-adiabatic. We confirmed that the resonance argument continues to librate when handed off to REBOUND. 

\begin{figure}[H]
    \centering
    \subfloat[Final frame from pkdgrav simulation]{\includegraphics[width=0.5\linewidth]{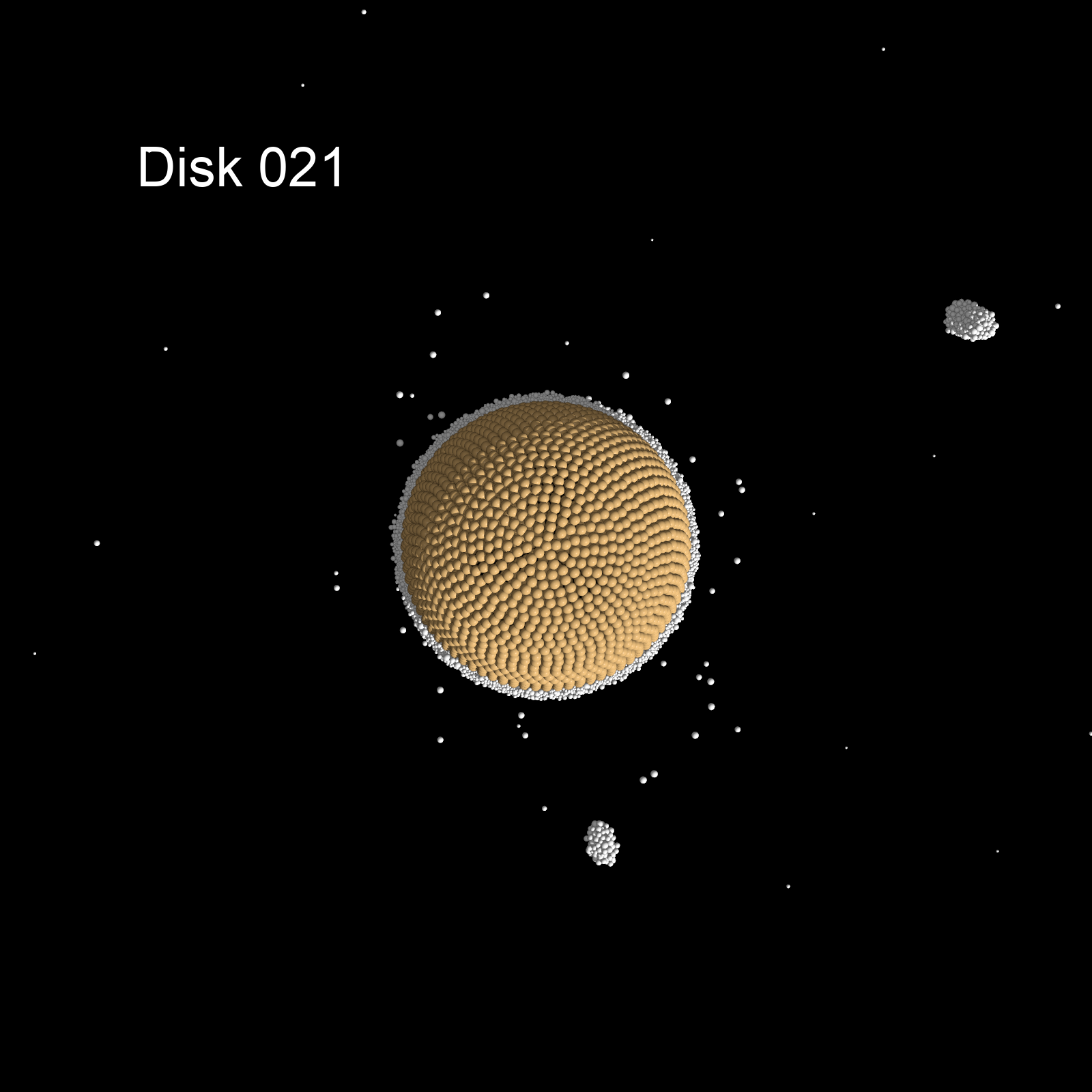}}
    \subfloat[Time series showing capture into 5:2 MMR]{\includegraphics[width=0.5\textwidth]{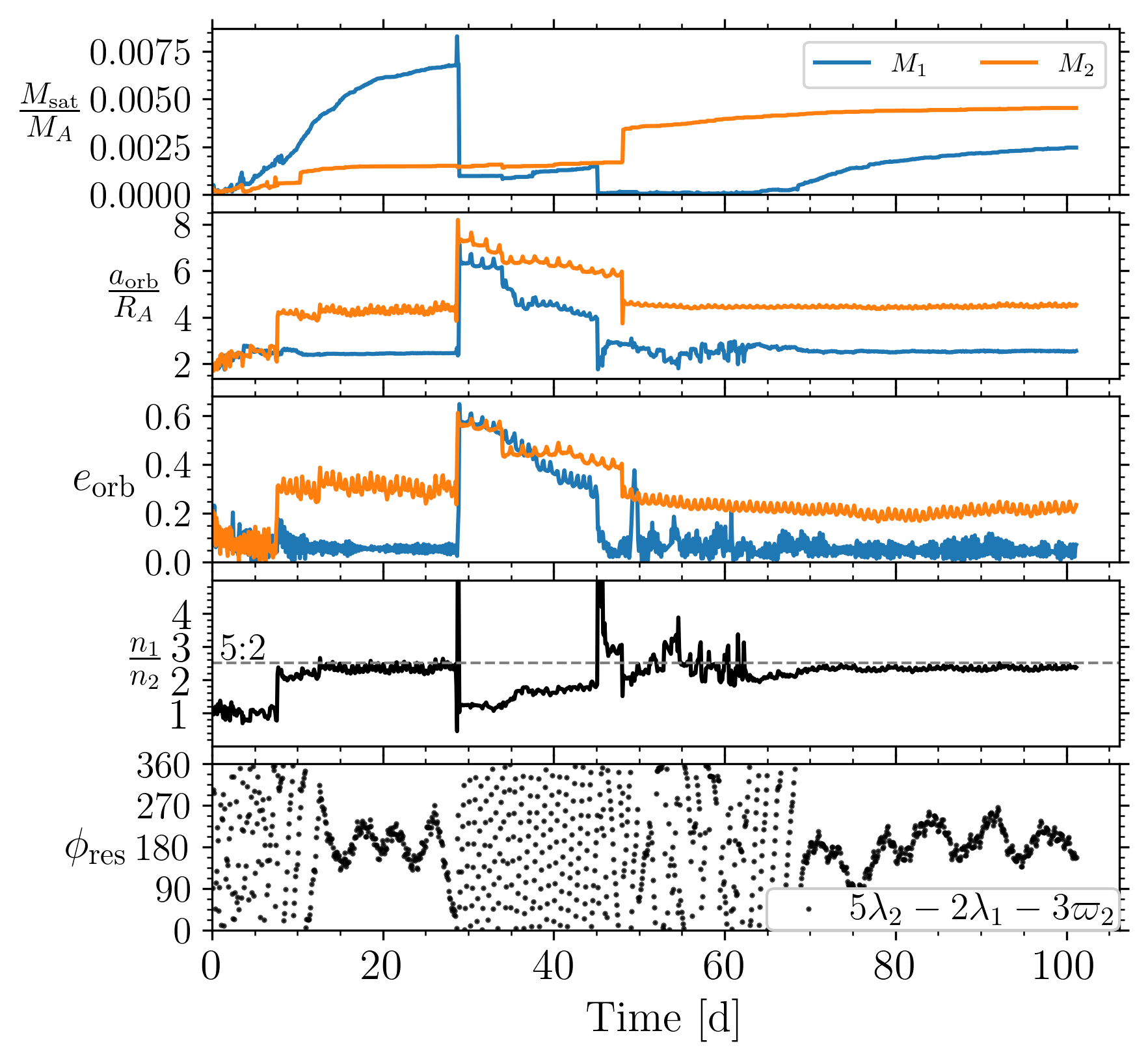}}
    \caption{The two satellites capture into the 5:2 MMR around 15 d. Due to $M_2$'s high eccentricity, the two satellites have a close encounter, sending $M_1$ inwards where it undergoes a tidal disruption around 30 d, which destroys both $M_1$ and the resonance. Around 60 d, a new inner satellite begins growing and quickly captures again into the 5:2 resonance again with the more massive outer satellite, this time remaining stable.}
    \label{fig:triple_disk021}
\end{figure}

We also find one example of capture into the 5:2 MMR, shown in Fig.\ \ref{fig:triple_disk021}. Capture into such a high-order resonance came as a surprise and may have been enabled by the high eccentricity of the outer satellite. If a triple asteroid were to be observed in such a resonance, then this process should be studied further. 

We find two other long-term stable triples (disk 075, 085) with a more massive outer satellite having period ratios close to 2:1 and 4:1 respectively, although we confirm that are \textit{not} in resonance. However, these systems have well-separated satellites and thus remain long-term stable in the REBOUND simulations.

Comparing our simulated triples to the known population is a problem of small number statistics. There are very few known triple asteroids, none of which are known to be in mean motion resonances. Out of the near-Earth asteroids, there are three confirmed triples: 1994 CC, 2001 SN$_{263}$, and 3122 Florence \citep{Brozovic2009, Nolan2008}. Both CC and SN$_{263}$ are well-characterized, not near MMRs, and also have some mutual inclination, which as been attributed a previous close planetary encounter \citep{Brozovic2009, Fang2012, Becker2015}. However, no detailed dynamical studies have been published for 3122 Florence. Preliminary estimates have put the orbit periods of the two satellites at 22 h and 7 h \citep{Brozovic2018b},  which is not far from the 3:1 MMR, although a more comprehensive analysis of the satellite orbits would be needed to confirm this. Despite our simulations occasionally resulting in a stable triple in a mean-motion resonance, a single mass shedding event is certainly not the only way to achieve such a configuration. For example, a second mass-shedding event occurring after the first satellite has migrated outwards, followed by convergent migration (due to tides and/or BYORP) is an equally plausible mechanism to form a MMR triple. 

\subsubsection{Simulation outcomes}
\begin{figure}[H]
        \centering
        \includegraphics[width=0.75\textwidth]{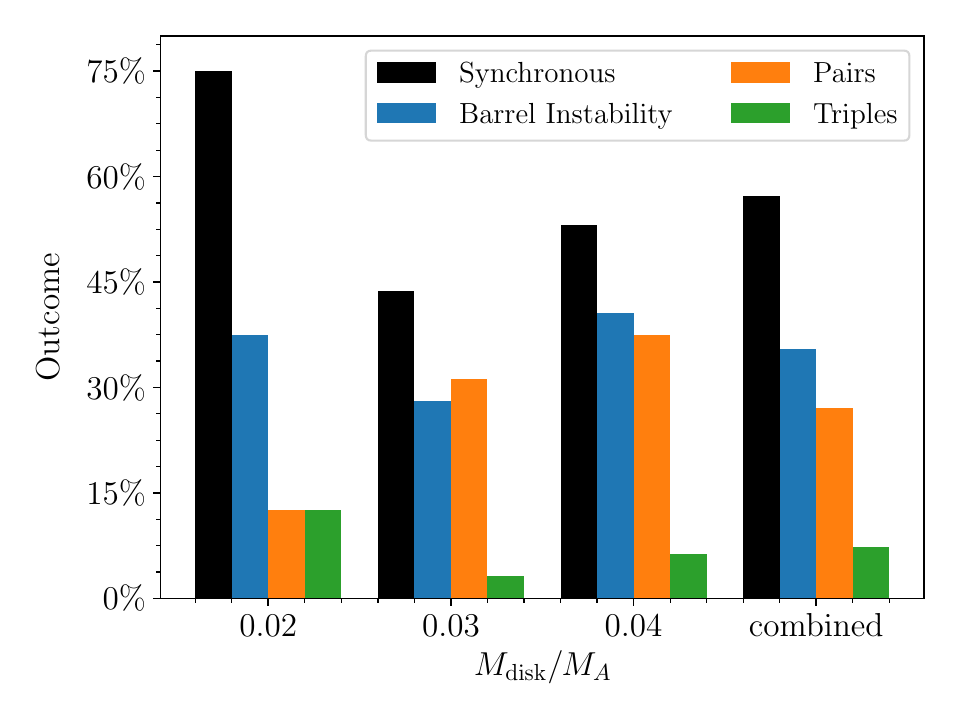}
        \caption{A summary of simulation outcomes showing the percentage of cases that have a given outcome, considering only the 96 simulations where $\phi=35^\circ$. A satellite is counted as a synchronous rotator if its yaw angle remains $<60^\circ$ over the final 10 d of simulation. Similarly, a satellite is considered to be in the barrel instability if its roll angle is $>60^\circ$ over the final 10 d. These thresholds are set arbitrarily and should only be considered notional.}\label{fig:simulation_outcomes}
\end{figure}

In Fig.\ \ref{fig:simulation_outcomes}, we show the simulation outcomes for all simulations with $\phi=35^\circ$, summarizing the resulting rotation state of the secondary and whether the system formed a pair or triple. We consider the satellite to be in synchronous rotation if its maximum yaw angle over the final 10 days of the simulation is $<60^\circ$. If a synchronous rotator has a roll angle $>60^\circ$ over the final 10 d, we also consider it to be in the ``barrel instability'' (see Fig.\ \ref{fig:max_roll_libration}). If a fragment consisting of 3 or more particles is ejected from the system, we count the simulation as successfully forming a pair. Triples are only counted if the system remains stable after further integration with REBOUND. Each category is not exclusive, in the sense that a triple could have also formed a pair, for example.

We find that a lower disk mass leads to a higher likelihood of the satellite forming with synchronous rotation, likely due to fewer violent collisions and mergers. For similar reasons, we find that more massive initial disks are more likely to form asteroid pairs as there is more mass (and angular momentum) that can launch  fragments onto escape trajectories. Given that the final outcomes are highly dependent on the initial angular momentum in the disk, we caution against over-interpreting Fig.\ \ref{fig:simulation_outcomes}. A higher fidelity study with more realistic initial conditions would provide better estimates for the fraction of triples, pairs, and synchronous rotators that could plausibly form from a single mass shedding event.

\section{Conclusions}\label{sec:conclusions}

We demonstrate that satellites can rapidly form by accretion following a single mass shedding event. The formation time is relatively fast, with the satellite reaching its terminal mass in a matter of days. Satellites tend to form on near-circular orbits within ${\sim}4$ primary radii, although 3-body interactions with other moonlets can sometimes put the final satellite onto wider and more eccentric orbits. Due to the satellite's accretion near the Roche limit, the satellite tends to be prolate in shape and is oftentimes in synchronous rotation. Many of these synchronous rotators, however, are in a rolling state about their long axis while remaining tidally locked, which would terminate or significantly weaken any BYORP evolution until this mode is dissipated \citep{Cuk2021, Quillen2022a}.

The simulated secondary shapes are broadly consistent with the radar-derived shapes of other asteroid satellites, given their uncertainties \citep{Ostro2006, Naidu2015b, Becker2015}. In addition, the distribution of secondary elongations ($a/b$) agrees well with the lightcurve-derived data set from \cite{Pravec2019}. Our simulations rarely produce satellites with $a/b\gtrsim1.5$ in strong agreement with the observed population of secondaries. Despite the strong preference to form elongated satellites, oblate shapes like Dimorphos are occasionally formed in our simulations. However, the fact that lightcurves are heavily biased against measuring satellite shapes with low $a/b$, combined with Dimorphos's low $a/b$ suggests the existence of a significant population of oblate secondaries (unless Dimorphos is an outlier). This implies that there may be a longer term, yet-unidentified pathway that can reshape satellites to become oblate. 

We tested four different friction angles ($29^\circ, 32^\circ, 35^\circ, 40^\circ$), finding only subtle differences. Generally, higher friction satellites have more irregular shapes. In addition, they are more resistant to disruption during close gravitational encounters and collisions, allowing them to form on wider, more excited orbits, whereas a lower friction satellite would disrupt and reaccrete during this process. The most noticeable difference is that higher friction tends to form satellites with lower bulk densities, due to their ability to maintain larger interior void spaces.

The accretion process is highly chaotic, resulting in a wide range of outcomes. In some cases, moonlets are ejected from the system via three-body interactions, effectively forming asteroid pairs. We note that this process does not explain the population of asteroid pairs in general, as there are many other processes that can form pairs. Here, we merely demonstrate that 3-body interactions may play some role in the production of asteroid pairs.

In a few percent of cases, we form stable triple systems, including satellites in mean motion resonances and co-orbital satellites. Although unlikely, a future discover of a system with co-orbital satellites would be strong evidence for formation from a single mass shedding event, as co-accretion would naturally explain their configuration.

There are some caveats and limitations of our model. These simulations are only run for 100 days due to their computational cost and therefore cannot capture longer term effects. In addition, the primary is azimuthally symmetric which may affect the likelihood of forming pairs and triples. Finally, this study was originally focused solely on forming a Dimorphos-mass satellite, therefore we only explored a narrow range of disk masses. Although the initial conditions of these simulations were intentionally simplified, we would expect the general results and trends to hold. 

There are a wide range of parameters available to explore in future work, and we list a few avenues to explore. With stronger spin-orbit coupling, an asymmetric primary may change the fraction of systems that form asteroid pairs, and would also allow the observed correlation between pair mass ratio and primary spin rate to be explored. In addition, this may affect the likelihood of forming a triple, as well as their long-term stability. In addition, a larger disk mass should be explored to probe the formation of systems with larger mass ratios such as Balam, and would also strongly affect the production of unbound fragments. Rather than a single mass shedding event, its plausible that these systems form from multiple events. If this is the case, simulated additional mass shedding events after the satellite is initially formed may change the distribution of secondary shapes. 

The Hera mission will investigate the Didymos system for 6 months in early 2027 and will provide crucial information for binary formation models \citep{Michel2022}. In particular, it will measure the mass of Dimorphos and its internal properties, allowing us to check some of the predictions from our modeling. Through the landing and possible bouncing of its CubeSats on Dimorphos' surface, we will  be able to access some of the mechanical parameters used in these simulations. However, since the shape of Dimorphos may have changed as a result of the DART impact \citep{Raducan2023_submitted}, we may not be able to put constraints on the formation models leading to either oblate or prolate secondaries, although all efforts will be done to relate the measured shape by Hera to the original one. 

\section*{Acknowledgments}
% \begin{acknowledgments}
This work was supported in part by the DART mission, NASA Contract \#80MSFC20D0004 to JHU/APL. H.F.A. was supported by the French government, through the UCA J.E.D.I. Investments in the Future project managed by the National Research Agency (ANR) with the reference number ANR-15-IDEX-01. P.P. was supported by the Grant Agency of the Czech Republic, grant 20-04431S. P. M. acknowledges funding support from the French space agency CNES and the University of Tokyo. M.P., A.L., and F.T. were supported by the Italian Space Agency (ASI) within the LICIACube project (ASI-INAF agreement n. 2019-31-HH.0) and Hera project (ASI-INAF agreement n. 2022-8-HH.0). S.D.R. acknowledges support from the Swiss National Science Foundation (project number 200021\_207359). We thank Adriano Campo Bagatin, Kate Minker, and John Wimarsson for helpful discussions and feedback. We are grateful to David Minton and the anonymous referee, whose feedback substantially improved the quality of this manuscript.

\software{\textsc{pkdgrav} \citep{Richardson2000, Schwartz2012, Zhang2017}, REBOUND \citep{rebound}, REBOUNDx \citep{reboundx}, Persistence of Vision Raytracer (\href{https://www.povray.org/}{https://www.povray.org/}), Alpha Shape Toolbox (\href{https://alphashape.readthedocs.io/}{https://alphashape.readthedocs.io/}), NumPy \citep{Harris2020numpy}, SciPy \citep{Virtanen2020scipy}, Matplotlib \citep{Caswell2020matplotlib}.}

\facility{Simulations and analysis were performed on the ASTRA cluster administered by the Center for Theory and Computation, part of the Department of Astronomy at the University of Maryland and on Mésocentre SIGAMM hosted at the Observatoire de la Côte d'Azur.}
% \end{acknowledgments}

\appendix

\section{Notation}

For reference, Table \ref{tab:notation} lists the notation used in this paper. 

\begin{deluxetable*}{c|c}[h!]
\tablenum{2}
\tablecaption{Notation table }
\tablehead{\colhead{Variable} & \colhead{Definition} }
\startdata
$M_\text{A}$ & Primary's mass  \\
$M_\text{B}$ & Secondary's mass\textsuperscript{a} \\
$M_\text{disk}$ & Initial mass of disk placed in orbit around primary  \\
$R_\text{A}$, $D_\text{A}$ & Primary's radius and diameter\textsuperscript{b} \\
$R_\text{A}^\text{eq}$ & Primary's equatorial radius \\
$R_\text{B}$, $D_\text{B}$ & Secondary's radius and diameter\textsuperscript{b}  \\
$a$ & Secondary's semimajor (longest) axis length\textsuperscript{c}   \\
$b$ & Secondary's semi-intermediate axis\textsuperscript{c}  \\
$c$ & Secondary's semi-minor axis (i.e., principal spin axis)\textsuperscript{c}    \\
$r_\text{orb}$ & Instantaneous binary orbital separation  \\
$a_\text{orb}$ & Semimajor axis\textsuperscript{d}   \\
$e_\text{orb}$ & Orbital eccentricity\textsuperscript{d}  \\
$P_\text{orb}$ & Orbital period\textsuperscript{d}   \\
$\theta$ & Libration angle\textsuperscript{e} \\
$\theta_1, \theta_2, \theta_3$ & The roll, pitch, and yaw angles\textsuperscript{f} \\
$\phi$ & Friction angle \\
\enddata
\tablecomments{\textsuperscript{a}{$M_1$ and $M_2$ are used when there is more than one satellite.}
               \textsuperscript{b}{Based on the body's volume-equivalent sphere.}
               \textsuperscript{c}{This length is measured either based on the body's physical extents, or by the axis lengths of the corresponding dynamically equivalent equal-volume ellipsoid.}
               \textsuperscript{d}{Keplerian element, based on the instantaneous body positions and velocities.}
               \textsuperscript{e}{The angle between the secondary's long axis and the line connecting the body centers}
               \textsuperscript{f}{The 1-2-3 Euler angle set coordinated in the secondary's rotating orbital frame (see \cite{Agrusa2021}).}
               }\label{tab:notation}
\end{deluxetable*}

% \newpage
\section{Simulation Results}

Table \ref{tab:sim_results} shows the results of each simulation in this study.

\startlongtable
% \begin{longrotatetable}
\begin{deluxetable*}{ccc|cccccccccc}
\tablenum{3}
\tablecaption{Tabulated simulation outcomes for each simulation.\label{tab:sim_results}}
\tablewidth{0pt}
\tablehead{
\colhead{ID} &
\colhead{$\frac{M_\text{disk}}{M_\text{A}}$} & 
\colhead{$\phi$} & 
\colhead{$M_\text{B}/M_\text{A}$} & 
\colhead{$D_\text{B}/D_\text{A}$} &
\colhead{$a/b$, $b/c$} & 
\colhead{$a/b$, $b/c$} & 
% \colhead{$b/c$} & 
% \colhead{$b/c$} & 
\colhead{$\frac{a_\text{orb}}{R_\text{A}}$} & 
\colhead{$e_\text{orb}$} & 
\colhead{$P_\text{orb}$} &
\colhead{$\theta_\text{max}$} &
\colhead{$N_\text{frag}$} &
\colhead{Stable?} \\
\colhead{} &
\colhead{} & 
\colhead{[$^\circ$]} & 
\colhead{} & 
\colhead{(DEEVE)} &
\colhead{(DEEVE)} & 
\colhead{(Extents)} & 
% \colhead{(DEEVE)} & 
% \colhead{(Extent)} & 
\colhead{} & 
\colhead{} & 
\colhead{[H]} &
\colhead{[$^\circ$]} &
\colhead{} &
\colhead{}
}
\startdata
001 & 0.02 & $35^{\circ}$ & 0.0081 & 0.206 & 1.299, 1.384 & 1.29, 1.279 & 2.774 & 0.0407 & 9.8 & 26.1 & 0 &   \\
002 & 0.02 & $35^{\circ}$ & 0.0069 & 0.197 & 1.241, 1.242 & 1.237, 1.248 & 3.177 & 0.0387 & 12.02 & 51.8 & 0 &   \\
003 & 0.02 & $35^{\circ}$ & 0.0069 & 0.198 & 1.501, 1.656 & 1.474, 1.464 & 2.872 & 0.0403 & 10.33 & 12.7 & 0 &   \\
  &   &   & 0.0028 & 0.146 & 1.673, 1.298 & 1.612, 1.213 & 2.889 & 0.0371 & 10.44 & 12.7 &   & Yes \\
004 & 0.02 & $35^{\circ}$ & 0.0074 & 0.203 & 1.849, 1.177 & 1.793, 1.131 & 2.706 & 0.0603 & 9.45 & 18 & 0 &   \\
  &   &   & 0.0011 & 0.106 & 1.115, 1.178 & 1.09, 1.197 & 39.539 & 0.9324 & 529.15 & 88.8 &   & No \\
005 & 0.02 & $35^{\circ}$ & 0.0062 & 0.19 & 1.452, 1.156 & 1.447, 1.165 & 3.86 & 0.0126 & 16.1 & 89.3 & 0 &   \\
006 & 0.02 & $35^{\circ}$ & 0.0099 & 0.226 & 1.308, 1.853 & 1.268, 1.606 & 2.873 & 0.0232 & 10.32 & 48 & 0 &   \\
007 & 0.02 & $35^{\circ}$ & 0.0082 & 0.208 & 1.326, 1.503 & 1.305, 1.409 & 3.346 & 0.0296 & 12.98 & 16.2 & 0 &   \\
008 & 0.02 & $35^{\circ}$ & 0.0085 & 0.21 & 1.318, 1.202 & 1.267, 1.177 & 3.155 & 0.0345 & 11.88 & 16.6 & 0 &   \\
009 & 0.02 & $35^{\circ}$ & 0.0087 & 0.215 & 1.623, 1.422 & 1.51, 1.334 & 3.081 & 0.0134 & 11.47 & 23.5 & 0 &   \\
010 & 0.02 & $35^{\circ}$ & 0.0076 & 0.202 & 1.411, 1.504 & 1.254, 1.506 & 2.743 & 0.0689 & 9.64 & 22.5 & 0 &   \\
  &   &   & 0.0014 & 0.116 & 1.723, 1.06 & 1.511, 1.127 & 2.756 & 0.0426 & 9.74 & 20.8 &   & Yes \\
011 & 0.02 & $35^{\circ}$ & 0.0089 & 0.214 & 1.557, 1.133 & 1.294, 1.146 & 2.87 & 0.0175 & 10.31 & 17.6 & 0 &   \\
012 & 0.02 & $35^{\circ}$ & 0.0078 & 0.204 & 1.535, 1.241 & 1.414, 1.251 & 3.32 & 0.0361 & 12.83 & 18.6 & 0 &   \\
013 & 0.02 & $35^{\circ}$ & 0.0091 & 0.215 & 1.576, 1.183 & 1.419, 1.24 & 3.033 & 0.0775 & 11.2 & 30.3 & 0 &   \\
014 & 0.02 & $35^{\circ}$ & 0.0058 & 0.184 & 1.188, 1.204 & 1.164, 1.168 & 2.639 & 0.0584 & 9.1 & 89.1 & 0 &   \\
  &   &   & 0.0027 & 0.143 & 1.207, 1.042 & 1.28, 1.005 & 4.029 & 0.1254 & 17.2 & 90 &   & Yes \\
015 & 0.02 & $35^{\circ}$ & 0.0081 & 0.206 & 1.217, 1.148 & 1.154, 1.11 & 3.474 & 0.0782 & 13.74 & 89.5 & 0 &   \\
016 & 0.02 & $35^{\circ}$ & 0.0092 & 0.217 & 1.473, 1.743 & 1.342, 1.701 & 2.807 & 0.0429 & 9.97 & 12.5 & 0 &   \\
017 & 0.02 & $35^{\circ}$ & 0.0069 & 0.196 & 1.39, 1.191 & 1.322, 1.161 & 3.654 & 0.0315 & 14.82 & 87.9 & 0 &   \\
018 & 0.02 & $35^{\circ}$ & 0.0069 & 0.197 & 1.316, 1.149 & 1.228, 1.163 & 3.459 & 0.0833 & 13.65 & 90 & 0 &   \\
019 & 0.02 & $35^{\circ}$ & 0.0072 & 0.2 & 1.78, 1.231 & 1.585, 1.161 & 3.4 & 0.0233 & 13.31 & 49.1 & 0 &   \\
020 & 0.02 & $35^{\circ}$ & 0.0089 & 0.215 & 1.356, 1.676 & 1.336, 1.586 & 2.77 & 0.0556 & 9.78 & 21.3 & 1 &   \\
021 & 0.02 & $35^{\circ}$ & 0.0045 & 0.172 & 1.445, 1.261 & 1.411, 1.141 & 4.54 & 0.2352 & 20.56 & 89.9 & 0 &   \\
  &   &   & 0.0025 & 0.137 & 1.509, 1.049 & 1.446, 0.976 & 2.551 & 0.0716 & 8.67 & 62.5 &   & Yes \\
022 & 0.02 & $35^{\circ}$ & 0.0068 & 0.194 & 1.293, 1.201 & 1.228, 1.23 & 3.741 & 0.1173 & 15.36 & 89.9 & 0 &   \\
023 & 0.02 & $35^{\circ}$ & 0.0072 & 0.2 & 1.523, 1.207 & 1.382, 1.156 & 3.545 & 0.0188 & 14.16 & 25.7 & 0 &   \\
024 & 0.02 & $35^{\circ}$ & 0.0076 & 0.203 & 1.669, 1.121 & 1.559, 1.128 & 3.405 & 0.0221 & 13.33 & 12.7 & 0 &   \\
025 & 0.02 & $35^{\circ}$ & 0.0094 & 0.218 & 1.411, 1.723 & 1.385, 1.658 & 2.875 & 0.0467 & 10.33 & 7.5 & 0 &   \\
026 & 0.02 & $35^{\circ}$ & 0.0087 & 0.218 & 1.609, 1.329 & 1.411, 1.34 & 3.061 & 0.036 & 11.35 & 87.5 & 0 &   \\
027 & 0.02 & $35^{\circ}$ & 0.0082 & 0.208 & 1.583, 1.251 & 1.474, 1.171 & 2.708 & 0.0516 & 9.45 & 11.7 & 1 &   \\
028 & 0.02 & $35^{\circ}$ & 0.0092 & 0.216 & 1.436, 1.638 & 1.414, 1.441 & 2.653 & 0.0196 & 9.16 & 36.8 & 0 &   \\
029 & 0.02 & $35^{\circ}$ & 0.0068 & 0.195 & 1.188, 1.192 & 1.127, 1.24 & 3.432 & 0.0261 & 13.5 & 35.1 & 1 &   \\
030 & 0.02 & $35^{\circ}$ & 0.0081 & 0.206 & 1.522, 1.484 & 1.447, 1.472 & 2.623 & 0.0692 & 9.01 & 47.6 & 0 &   \\
031 & 0.02 & $35^{\circ}$ & 0.0084 & 0.211 & 1.718, 1.331 & 1.615, 1.304 & 3.074 & 0.0624 & 11.43 & 27.1 & 1 &   \\
032 & 0.02 & $35^{\circ}$ & 0.0071 & 0.198 & 1.445, 1.357 & 1.377, 1.335 & 3.531 & 0.0395 & 14.08 & 25.4 & 0 &   \\
033 & 0.03 & $35^{\circ}$ & 0.013 & 0.242 & 1.411, 1.318 & 1.348, 1.309 & 3.263 & 0.0277 & 12.47 & 35.9 & 1 &   \\
034 & 0.03 & $35^{\circ}$ & 0.0137 & 0.246 & 1.726, 1.381 & 1.509, 1.348 & 3.465 & 0.0274 & 13.65 & 88.9 & 0 &   \\
035 & 0.03 & $35^{\circ}$ & 0.0125 & 0.238 & 1.228, 1.445 & 1.134, 1.389 & 3.974 & 0.0068 & 16.77 & 89.9 & 0 &   \\
036 & 0.03 & $35^{\circ}$ & 0.0139 & 0.249 & 1.568, 1.335 & 1.468, 1.22 & 3.136 & 0.0399 & 11.75 & 35 & 1 &   \\
037 & 0.03 & $35^{\circ}$ & 0.0123 & 0.238 & 1.317, 1.208 & 1.337, 1.08 & 4.089 & 0.0486 & 17.51 & 89.6 & 0 &   \\
038 & 0.03 & $35^{\circ}$ & 0.0151 & 0.254 & 1.515, 1.172 & 1.473, 1.156 & 3.256 & 0.0391 & 12.42 & 44.7 & 1 &   \\
039 & 0.03 & $35^{\circ}$ & 0.0135 & 0.245 & 1.16, 1.539 & 1.155, 1.405 & 4.014 & 0.2034 & 17.02 & 89.6 & 1 &   \\
040 & 0.03 & $35^{\circ}$ & 0.0099 & 0.221 & 1.249, 1.252 & 1.205, 1.223 & 3.37 & 0.1597 & 13.11 & 89.2 & 1 &   \\
041 & 0.03 & $35^{\circ}$ & 0.0159 & 0.261 & 1.15, 1.968 & 1.095, 1.79 & 3.054 & 0.0233 & 11.28 & 56.3 & 0 &   \\
042 & 0.03 & $35^{\circ}$ & 0.0136 & 0.246 & 1.198, 1.427 & 1.179, 1.383 & 3.677 & 0.0952 & 14.92 & 88.1 & 0 &   \\
043 & 0.03 & $35^{\circ}$ & 0.0136 & 0.245 & 1.191, 1.511 & 1.206, 1.381 & 2.55 & 0.0647 & 8.61 & 89.6 & 2 &   \\
044 & 0.03 & $35^{\circ}$ & 0.0146 & 0.253 & 1.449, 1.355 & 1.377, 1.245 & 3.535 & 0.0377 & 14.05 & 77.6 & 0 &   \\
045 & 0.03 & $35^{\circ}$ & 0.0122 & 0.237 & 1.499, 1.455 & 1.38, 1.378 & 2.575 & 0.0651 & 8.75 & 22 & 0 &   \\
  &   &   & 0.0026 & 0.142 & 1.522, 1.188 & 1.297, 1.238 & 17.675 & 0.7959 & 158.06 & 89.1 &   & No \\
046 & 0.03 & $35^{\circ}$ & 0.0134 & 0.245 & 1.573, 1.135 & 1.498, 1.129 & 3.684 & 0.0363 & 14.96 & 46.2 & 0 &   \\
047 & 0.03 & $35^{\circ}$ & 0.0128 & 0.241 & 1.206, 1.561 & 1.172, 1.423 & 3.84 & 0.1239 & 15.93 & 89 & 0 &   \\
048 & 0.03 & $35^{\circ}$ & 0.0119 & 0.235 & 1.255, 1.232 & 1.241, 1.211 & 4.033 & 0.0693 & 17.15 & 89.5 & 0 &   \\
049 & 0.03 & $35^{\circ}$ & 0.0156 & 0.259 & 1.435, 1.751 & 1.419, 1.648 & 3.154 & 0.0192 & 11.84 & 18.9 & 0 &   \\
050 & 0.03 & $35^{\circ}$ & 0.014 & 0.248 & 1.148, 1.184 & 1.082, 1.269 & 3.56 & 0.0325 & 14.21 & 89.7 & 0 &   \\
051 & 0.03 & $35^{\circ}$ & 0.0121 & 0.237 & 1.225, 1.54 & 1.208, 1.384 & 3.993 & 0.0192 & 16.89 & 77.1 & 1 &   \\
052 & 0.03 & $35^{\circ}$ & 0.0133 & 0.245 & 1.467, 1.308 & 1.42, 1.243 & 3.656 & 0.0343 & 14.79 & 89.7 & 0 &   \\
053 & 0.03 & $35^{\circ}$ & 0.015 & 0.255 & 1.462, 1.737 & 1.348, 1.58 & 3.218 & 0.0697 & 12.2 & 28.6 & 0 &   \\
054 & 0.03 & $35^{\circ}$ & 0.0147 & 0.252 & 1.269, 1.963 & 1.225, 1.83 & 2.546 & 0.014 & 8.59 & 32.7 & 3 &   \\
055 & 0.03 & $35^{\circ}$ & 0.0142 & 0.25 & 1.066, 1.912 & 1.055, 1.719 & 3.521 & 0.0544 & 13.98 & 88.4 & 2 &   \\
056 & 0.03 & $35^{\circ}$ & 0.0147 & 0.252 & 1.544, 1.443 & 1.449, 1.319 & 3.415 & 0.021 & 13.34 & 45.6 & 0 &   \\
057 & 0.03 & $35^{\circ}$ & 0.0121 & 0.238 & 1.359, 1.324 & 1.342, 1.175 & 4.075 & 0.0521 & 17.41 & 88 & 0 &   \\
  &   &   & 0.0019 & 0.129 & 1.386, 1.091 & 1.288, 1.1 & 2.397 & 0.0143 & 7.89 & 46.2 &   & Yes \\
058 & 0.03 & $35^{\circ}$ & 0.0119 & 0.235 & 1.112, 1.259 & 1.074, 1.181 & 3.98 & 0.045 & 16.81 & 89.7 & 0 &   \\
059 & 0.03 & $35^{\circ}$ & 0.0112 & 0.23 & 1.133, 1.13 & 1.1, 1.124 & 4.31 & 0.129 & 18.95 & 89.9 & 0 &   \\
060 & 0.03 & $35^{\circ}$ & 0.0145 & 0.252 & 1.347, 1.779 & 1.331, 1.529 & 3.374 & 0.0415 & 13.1 & 53.1 & 0 &   \\
061 & 0.03 & $35^{\circ}$ & 0.0148 & 0.254 & 1.182, 1.693 & 1.151, 1.581 & 3.361 & 0.029 & 13.03 & 53 & 1 &   \\
062 & 0.03 & $35^{\circ}$ & 0.0153 & 0.257 & 1.149, 1.389 & 1.106, 1.324 & 3.269 & 0.0187 & 12.49 & 30 & 0 &   \\
063 & 0.03 & $35^{\circ}$ & 0.0158 & 0.26 & 1.625, 1.71 & 1.479, 1.639 & 3.098 & 0.0468 & 11.52 & 12.9 & 0 &   \\
064 & 0.03 & $35^{\circ}$ & 0.0144 & 0.252 & 1.552, 1.306 & 1.528, 1.27 & 3.493 & 0.1292 & 13.8 & 89.6 & 0 &   \\
065 & 0.04 & $35^{\circ}$ & 0.0216 & 0.288 & 1.309, 2.051 & 1.26, 1.881 & 3.282 & 0.0721 & 12.53 & 57 & 0 &   \\
066 & 0.04 & $35^{\circ}$ & 0.0192 & 0.277 & 1.328, 1.215 & 1.236, 1.178 & 3.863 & 0.0762 & 16.02 & 75.5 & 1 &   \\
067 & 0.04 & $35^{\circ}$ & 0.0225 & 0.292 & 1.372, 1.972 & 1.337, 1.839 & 3.056 & 0.0118 & 11.26 & 22.7 & 0 &   \\
068 & 0.04 & $35^{\circ}$ & 0.0174 & 0.267 & 1.288, 1.366 & 1.26, 1.326 & 4.359 & 0.1232 & 19.22 & 89.9 & 0 &   \\
069 & 0.04 & $35^{\circ}$ & 0.0176 & 0.27 & 1.153, 1.686 & 1.137, 1.492 & 4.549 & 0.3122 & 20.49 & 89.8 & 0 &   \\
070 & 0.04 & $35^{\circ}$ & 0.0209 & 0.285 & 1.109, 1.408 & 1.156, 1.325 & 3.627 & 0.1367 & 14.56 & 89.8 & 2 &   \\
071 & 0.04 & $35^{\circ}$ & 0.021 & 0.284 & 1.289, 1.64 & 1.275, 1.496 & 3.119 & 0.0458 & 11.61 & 51.4 & 2 &   \\
072 & 0.04 & $35^{\circ}$ & 0.0215 & 0.29 & 1.142, 1.768 & 1.133, 1.634 & 3.336 & 0.0675 & 12.84 & 27.2 & 0 &   \\
073 & 0.04 & $35^{\circ}$ & 0.0199 & 0.279 & 1.413, 1.143 & 1.302, 1.159 & 3.619 & 0.0251 & 14.52 & 89.3 & 0 &   \\
074 & 0.04 & $35^{\circ}$ & 0.0172 & 0.266 & 1.395, 1.657 & 1.342, 1.541 & 4.115 & 0.0788 & 17.64 & 89.8 & 1 &   \\
075 & 0.04 & $35^{\circ}$ & 0.0172 & 0.266 & 1.355, 1.094 & 1.314, 1.09 & 3.456 & 0.0287 & 13.57 & 38.2 & 3 &   \\
076 & 0.04 & $35^{\circ}$ & 0.0219 & 0.292 & 1.356, 1.901 & 1.208, 1.762 & 3.043 & 0.0631 & 11.18 & 41.2 & 0 &   \\
077 & 0.04 & $35^{\circ}$ & 0.0221 & 0.292 & 1.816, 1.46 & 1.709, 1.289 & 3.037 & 0.0187 & 11.15 & 26.1 & 0 &   \\
078 & 0.04 & $35^{\circ}$ & 0.0206 & 0.284 & 1.708, 1.191 & 1.612, 1.139 & 3.3 & 0.0697 & 12.64 & 63.9 & 1 &   \\
079 & 0.04 & $35^{\circ}$ & 0.0196 & 0.28 & 1.209, 1.992 & 1.187, 1.781 & 3.944 & 0.0686 & 16.52 & 26.1 & 1 &   \\
080 & 0.04 & $35^{\circ}$ & 0.0218 & 0.29 & 1.566, 1.827 & 1.516, 1.69 & 3.193 & 0.0948 & 12.02 & 25.2 & 0 &   \\
081 & 0.04 & $35^{\circ}$ & 0.0216 & 0.289 & 1.114, 1.422 & 1.075, 1.367 & 3.311 & 0.0385 & 12.69 & 90 & 0 &   \\
082 & 0.04 & $35^{\circ}$ & 0.0187 & 0.275 & 1.704, 1.195 & 1.484, 1.236 & 3.995 & 0.0806 & 16.85 & 87.2 & 0 &   \\
083 & 0.04 & $35^{\circ}$ & 0.0164 & 0.26 & 1.164, 1.306 & 1.149, 1.235 & 2.718 & 0.0535 & 9.47 & 84.9 & 1 &   \\
  % &   &   & 0.0011 & 0.109 & 1.028, 1.137 & 0.984, 1.149 & 61.277 & 0.8622 & 1021.27 & 89.7 &   & Yes \\
084 & 0.04 & $35^{\circ}$ & 0.0205 & 0.283 & 1.671, 1.424 & 1.501, 1.36 & 3.704 & 0.0179 & 15.03 & 51.7 & 0 &   \\
085 & 0.04 & $35^{\circ}$ & 0.0095 & 0.218 & 1.089, 1.38 & 1.067, 1.339 & 6.324 & 0.1577 & 33.73 & 89.3 & 5 &   \\
  &   &   & 0.005 & 0.176 & 1.666, 1.364 & 1.736, 1.235 & 2.488 & 0.0612 & 8.34 & 12.2 &   & Yes \\
086 & 0.04 & $35^{\circ}$ & 0.0208 & 0.286 & 1.116, 1.994 & 1.12, 1.864 & 3.705 & 0.0962 & 15.04 & 89.4 & 0 &   \\
087 & 0.04 & $35^{\circ}$ & 0.0193 & 0.276 & 1.365, 1.445 & 1.243, 1.462 & 3.83 & 0.07 & 15.82 & 59.8 & 0 &   \\
088 & 0.04 & $35^{\circ}$ & 0.0183 & 0.273 & 1.894, 1.232 & 1.744, 1.131 & 3.983 & 0.038 & 16.78 & 42.1 & 0 &   \\
089 & 0.04 & $35^{\circ}$ & 0.0192 & 0.279 & 2.228, 1.143 & 1.877, 1.215 & 3.513 & 0.0379 & 13.89 & 26.9 & 0 &   \\
090 & 0.04 & $35^{\circ}$ & 0.02 & 0.285 & 1.101, 1.443 & 1.063, 1.411 & 3.665 & 0.0486 & 14.8 & 89.9 & 0 &   \\
091 & 0.04 & $35^{\circ}$ & 0.0188 & 0.274 & 1.589, 1.147 & 1.49, 1.12 & 3.941 & 0.0211 & 16.52 & 39.4 & 0 &   \\
092 & 0.04 & $35^{\circ}$ & 0.0183 & 0.271 & 1.405, 1.363 & 1.349, 1.302 & 4.023 & 0.0136 & 17.03 & 40.4 & 1 &   \\
093 & 0.04 & $35^{\circ}$ & 0.0196 & 0.277 & 1.189, 1.655 & 1.157, 1.542 & 3.862 & 0.1091 & 16.01 & 89.6 & 0 &   \\
094 & 0.04 & $35^{\circ}$ & 0.0202 & 0.281 & 1.359, 1.386 & 1.292, 1.33 & 3.607 & 0.0639 & 14.45 & 54.6 & 1 &   \\
095 & 0.04 & $35^{\circ}$ & 0.0194 & 0.277 & 1.285, 1.443 & 1.202, 1.414 & 3.943 & 0.1736 & 16.52 & 89.6 & 1 &   \\
096 & 0.04 & $35^{\circ}$ & 0.0189 & 0.279 & 1.683, 1.485 & 1.535, 1.317 & 4.18 & 0.2117 & 18.04 & 89.6 & 0 &   \\
097 & 0.03 & $29^{\circ}$ & 0.0131 & 0.24 & 1.396, 1.174 & 1.355, 1.131 & 3.322 & 0.0081 & 12.81 & 53.3 & 0 &   \\
098 & 0.03 & $29^{\circ}$ & 0.0136 & 0.243 & 1.357, 1.111 & 1.33, 1.117 & 3.221 & 0.0308 & 12.23 & 48.9 & 0 &   \\
099 & 0.03 & $29^{\circ}$ & 0.0145 & 0.249 & 1.416, 1.507 & 1.324, 1.434 & 3.076 & 0.0167 & 11.41 & 17.1 & 0 &   \\
100 & 0.03 & $29^{\circ}$ & 0.0139 & 0.244 & 1.268, 1.564 & 1.229, 1.456 & 3.108 & 0.0823 & 11.59 & 20.6 & 0 &   \\
101 & 0.03 & $29^{\circ}$ & 0.0143 & 0.247 & 1.365, 1.266 & 1.232, 1.302 & 2.998 & 0.0562 & 10.98 & 31 & 0 &   \\
102 & 0.03 & $29^{\circ}$ & 0.0148 & 0.249 & 1.312, 1.164 & 1.184, 1.159 & 3.133 & 0.0491 & 11.72 & 38.2 & 0 &   \\
103 & 0.03 & $29^{\circ}$ & 0.014 & 0.245 & 1.428, 1.46 & 1.299, 1.465 & 3.072 & 0.0488 & 11.39 & 21.1 & 0 &   \\
104 & 0.03 & $29^{\circ}$ & 0.0129 & 0.238 & 1.355, 1.126 & 1.348, 1.144 & 3.447 & 0.015 & 13.54 & 55 & 0 &   \\
105 & 0.03 & $32^{\circ}$ & 0.0152 & 0.255 & 1.501, 1.743 & 1.43, 1.681 & 2.943 & 0.0891 & 10.67 & 54.4 & 2 &   \\
106 & 0.03 & $32^{\circ}$ & 0.0145 & 0.252 & 1.444, 1.174 & 1.33, 1.161 & 3.233 & 0.0266 & 12.29 & 89.7 & 1 &   \\
107 & 0.03 & $32^{\circ}$ & 0.0113 & 0.231 & 1.265, 1.22 & 1.228, 1.159 & 3.919 & 0.024 & 16.43 & 81.5 & 0 &   \\
108 & 0.03 & $32^{\circ}$ & 0.0136 & 0.245 & 1.614, 1.323 & 1.507, 1.277 & 3.049 & 0.0412 & 11.26 & 27.1 & 0 &   \\
  &   &   & 0.0023 & 0.137 & 1.176, 1.23 & 1.155, 1.153 & 3.003 & 0.0491 & 11.07 & 89.6 &   & Yes \\
109 & 0.03 & $32^{\circ}$ & 0.0158 & 0.258 & 1.386, 1.74 & 1.301, 1.621 & 2.99 & 0.033 & 10.93 & 37 & 0 &   \\
110 & 0.03 & $32^{\circ}$ & 0.015 & 0.253 & 1.386, 1.721 & 1.353, 1.588 & 3.007 & 0.0274 & 11.02 & 20.7 & 1 &   \\
111 & 0.03 & $32^{\circ}$ & 0.015 & 0.253 & 1.241, 1.735 & 1.179, 1.646 & 3.193 & 0.0288 & 12.06 & 13 & 0 &   \\
112 & 0.03 & $32^{\circ}$ & 0.0154 & 0.257 & 1.277, 1.607 & 1.197, 1.594 & 2.98 & 0.0565 & 10.87 & 52.3 & 0 &   \\
113 & 0.03 & $40^{\circ}$ & 0.015 & 0.257 & 1.439, 1.685 & 1.376, 1.462 & 3.662 & 0.0502 & 14.81 & 54.3 & 0 &   \\
114 & 0.03 & $40^{\circ}$ & 0.0116 & 0.244 & 1.639, 1.219 & 1.383, 1.274 & 4.732 & 0.2459 & 21.8 & 89.5 & 1 &   \\
115 & 0.03 & $40^{\circ}$ & 0.0155 & 0.26 & 1.417, 1.924 & 1.336, 1.67 & 2.867 & 0.0663 & 10.26 & 36.8 & 0 &   \\
  &   &   & 0.0013 & 0.115 & 1.11, 1.089 & 1.007, 1.062 & 40.748 & 0.9502 & 553.62 & 89.9 &   & No \\
116 & 0.03 & $40^{\circ}$ & 0.0156 & 0.261 & 1.307, 1.741 & 1.267, 1.608 & 3.507 & 0.0484 & 13.88 & 89.7 & 1 &   \\
117 & 0.03 & $40^{\circ}$ & 0.0134 & 0.249 & 1.509, 1.162 & 1.421, 1.102 & 3.944 & 0.0356 & 16.57 & 89.3 & 0 &   \\
118 & 0.03 & $40^{\circ}$ & 0.0074 & 0.2 & 1.051, 1.412 & 1.047, 1.307 & 2.387 & 0.0722 & 7.83 & 89.5 & 3 &   \\
  &   &   & 0.007 & 0.199 & 1.08, 1.492 & 1.04, 1.444 & 7.467 & 0.5582 & 43.31 & 89.7 &   & No \\
119 & 0.03 & $40^{\circ}$ & 0.0149 & 0.257 & 1.113, 1.231 & 1.08, 1.143 & 3.857 & 0.0408 & 16.02 & 89.9 & 0 &   \\
120 & 0.03 & $40^{\circ}$ & 0.0174 & 0.274 & 1.494, 2.091 & 1.396, 1.963 & 3.197 & 0.0396 & 12.07 & 20.5 & 0 &   \\
\enddata
\tablecomments{Column $M_\text{disk}/M_\text{A}$ is the initial disk-to-primary mass ratio, $\phi$ is the material's friction angle. The rest of the columns are the simulation outcome: the resulting secondary-to-primary mass ratio ($M_\text{B}/M_\text{A}$), the diameter ratio $D_\text{B}/D\text{A}$, the secondary's axis ratios $a/b$ and $b/c$ (both DEEVE and extent-derived), the secondary's semimajor axis normalized to the primary's volume-equivalent radius ($a_\text{orb}/R_\text{A}$), the secondary's eccentricity ($e_\text{orb}$), orbit period ($P_\text{orb}$). $\theta_\text{max}$ is the secondary's maximum libration angle over the final 10 days of the simulation. If this number is close to 90$^\circ$, then the secondary in an asynchronous or tumbling rotation state. $N_\text{frag}$ is the number of fragments ejected from the system. In cases where there is a second satellite, we include a second row in the table containing the parameters of the satellite. The final column specifies whether that satellite was stable after the handoff to REBOUND. If the satellite was unstable, it was either ejected from the system, or collided with the primary or secondary.}
\end{deluxetable*}

\newpage
\section{Final renderings}\label{appendix:sec:satellites}
Here we show renderings of a subset of the simulations to give the reader a sense of the variance in satellite shape between similar simulations and as a function of the friction angle. All renderings are cases where $M_\text{disk}=0.03M_{A}$ and we show 8 cases for each value of friction ($\phi$). Each frame is rendered from a view looking down from the body's principal rotation axis. 

\centering

$\phi=29^\circ$:\\
\includegraphics[width=0.24\textwidth]{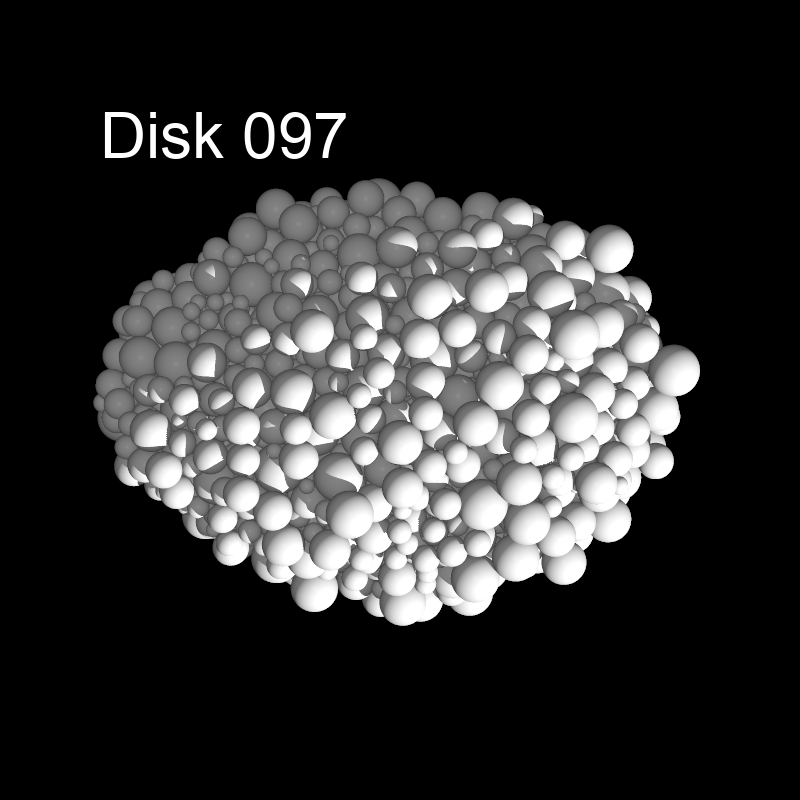}
\includegraphics[width=0.24\textwidth]{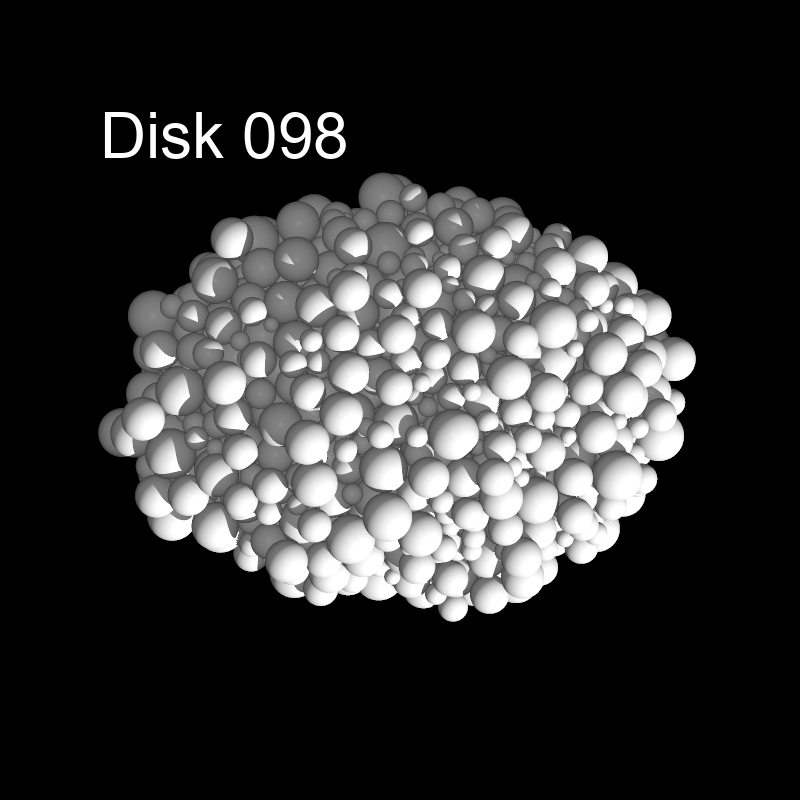}
\includegraphics[width=0.24\textwidth]{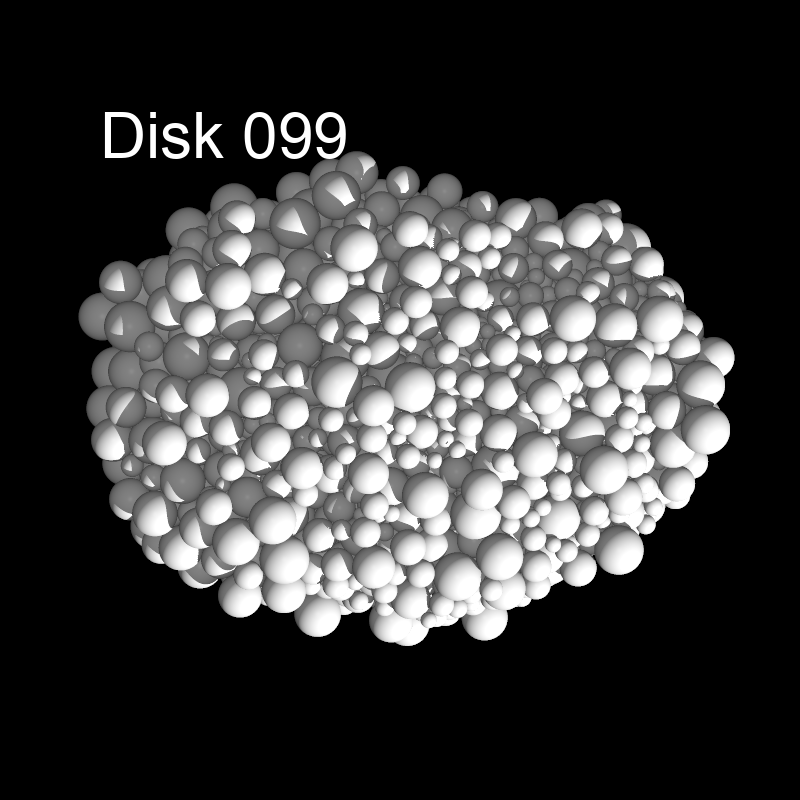}
\includegraphics[width=0.24\textwidth]{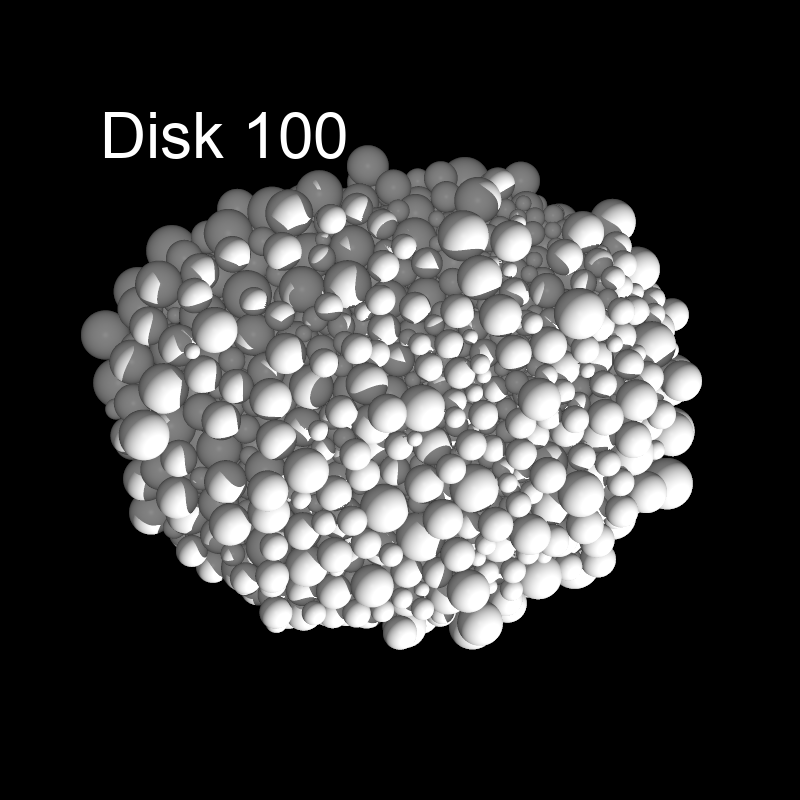}

\includegraphics[width=0.24\textwidth]{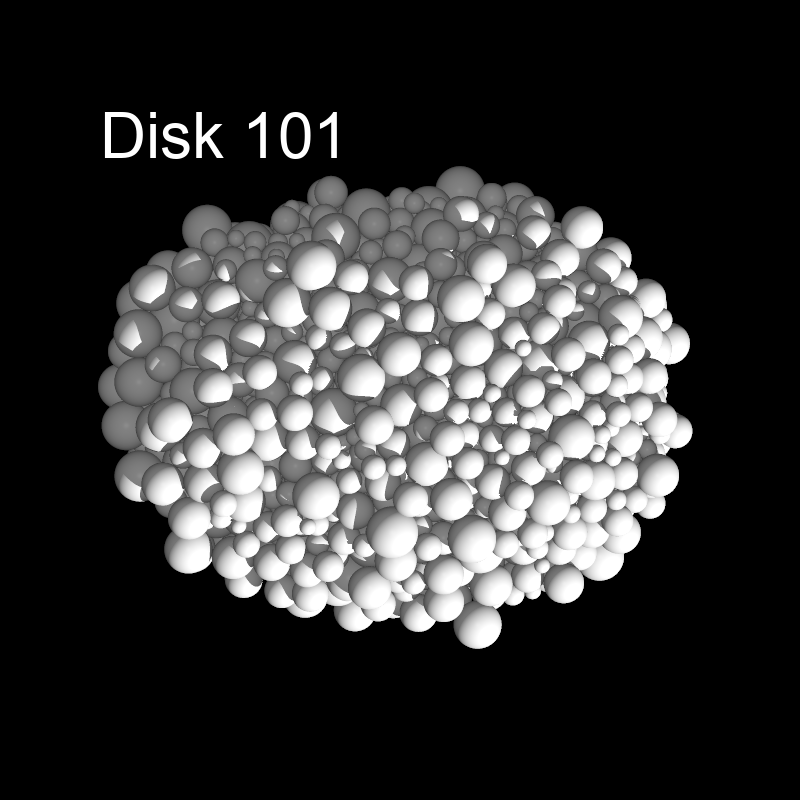}
\includegraphics[width=0.24\textwidth]{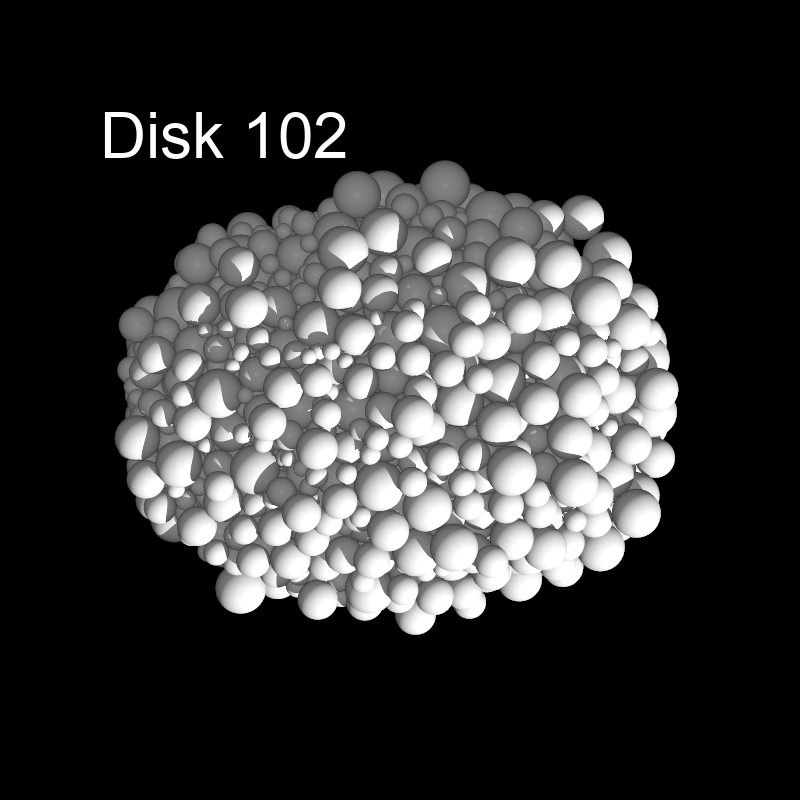}
\includegraphics[width=0.24\textwidth]{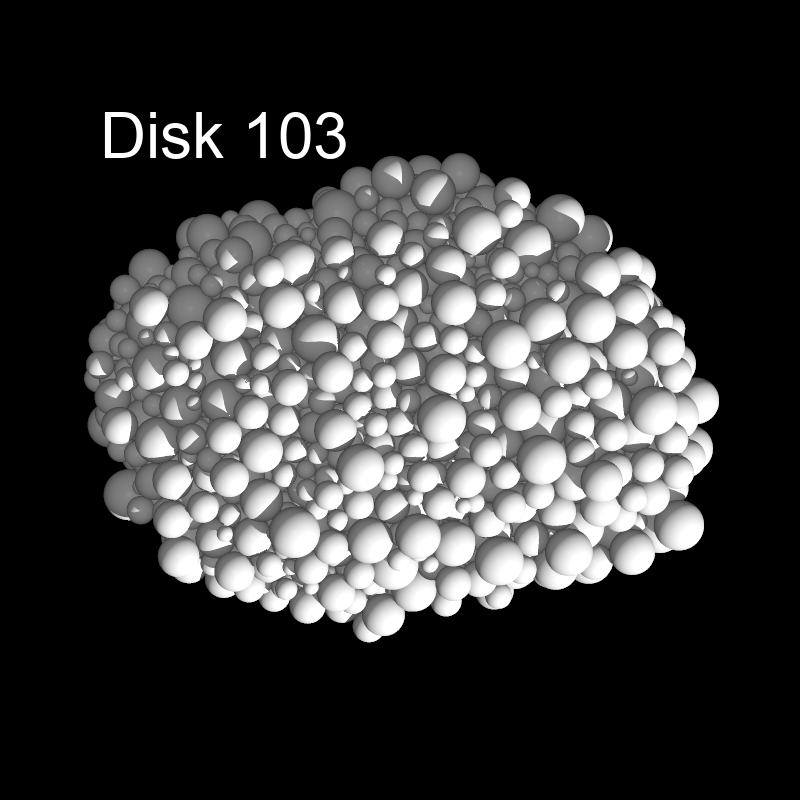}
\includegraphics[width=0.24\textwidth]{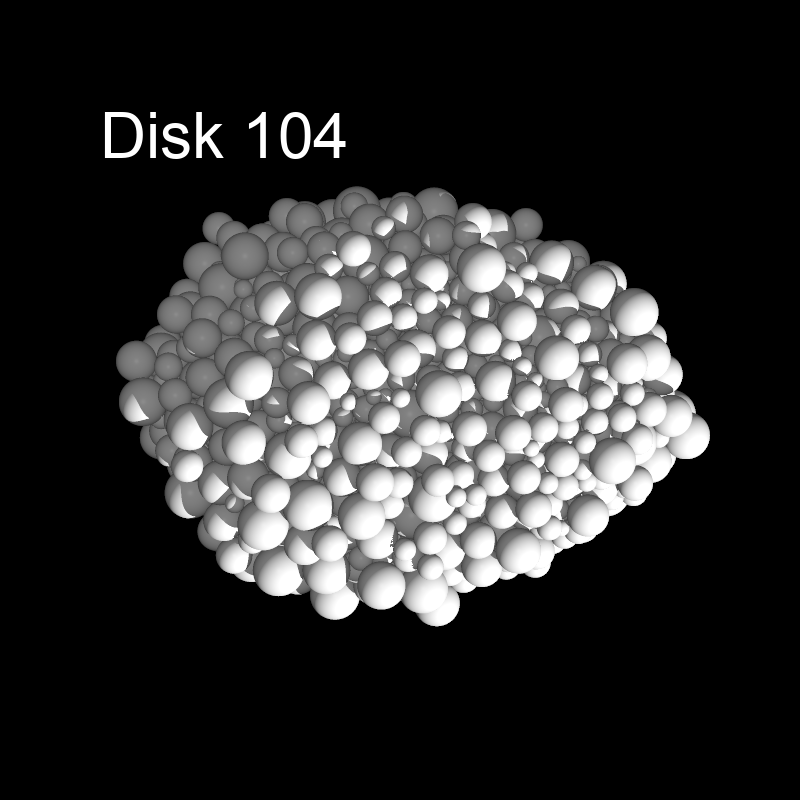}
$\phi=32^\circ$:\\
\includegraphics[width=0.24\textwidth]{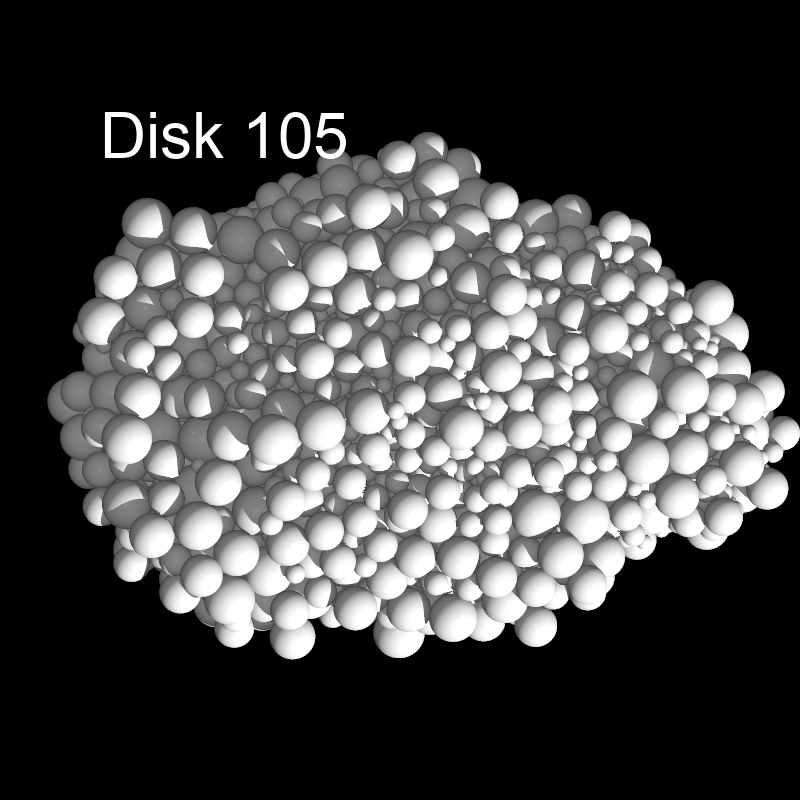}
\includegraphics[width=0.24\textwidth]{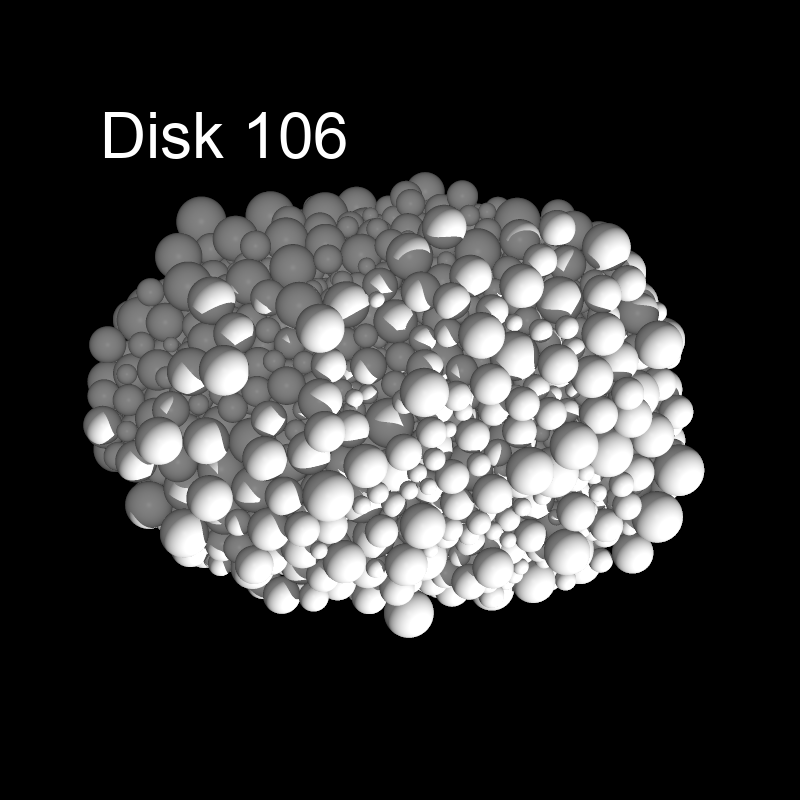}
\includegraphics[width=0.24\textwidth]{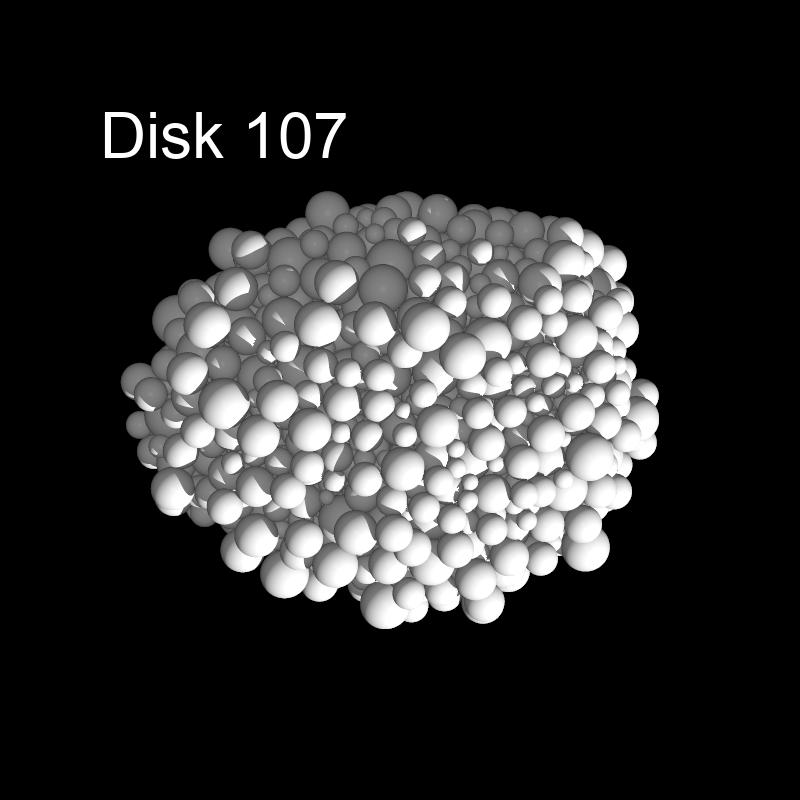}
\includegraphics[width=0.24\textwidth]{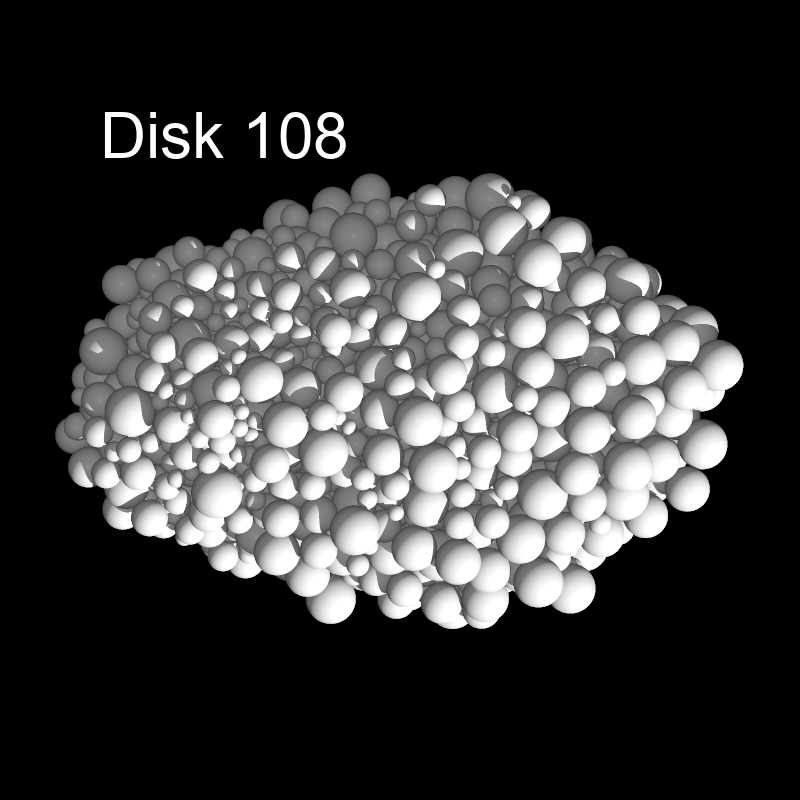}
\includegraphics[width=0.24\textwidth]{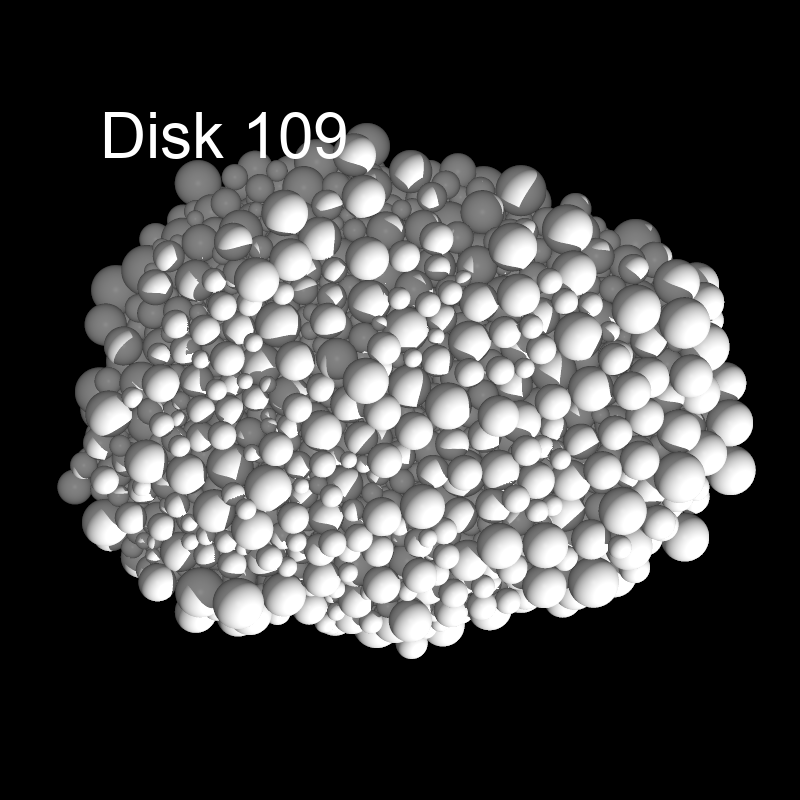}
\includegraphics[width=0.24\textwidth]{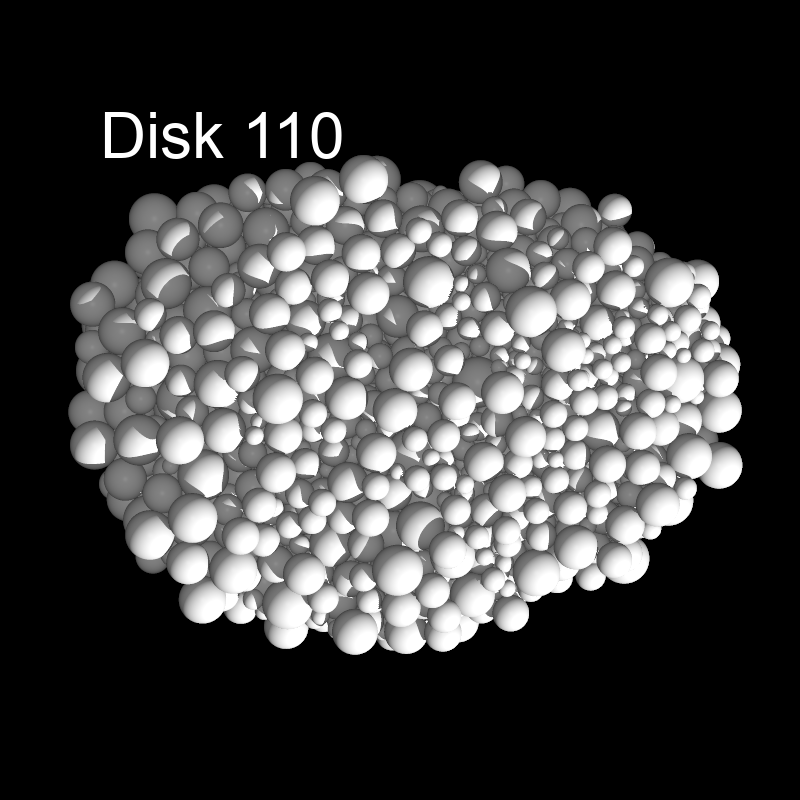}
\includegraphics[width=0.24\textwidth]{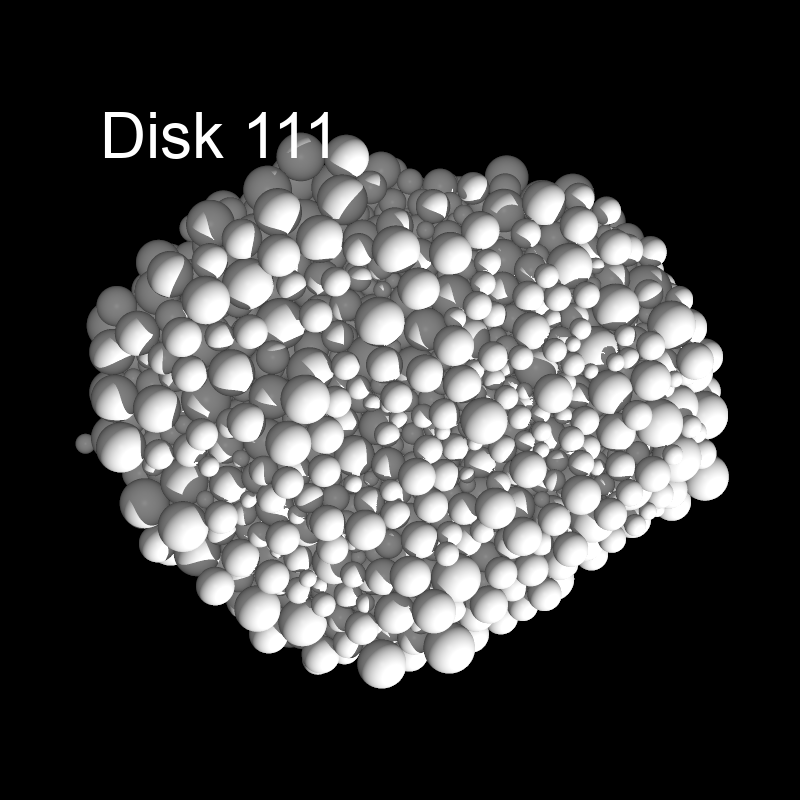}
\includegraphics[width=0.24\textwidth]{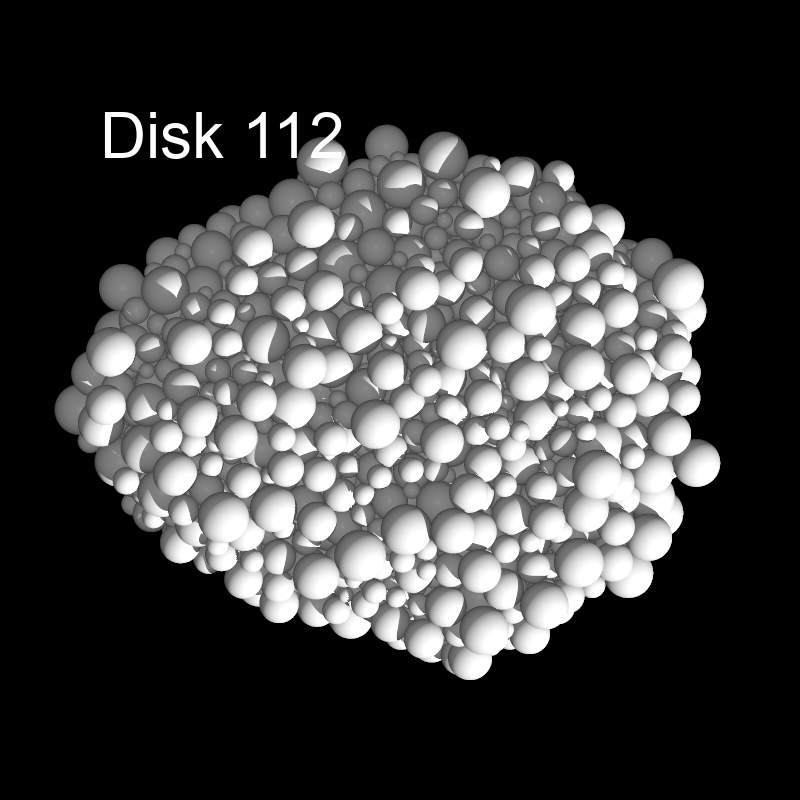}

\newpage
$\phi=35^\circ$\\
\includegraphics[width=0.24\textwidth]{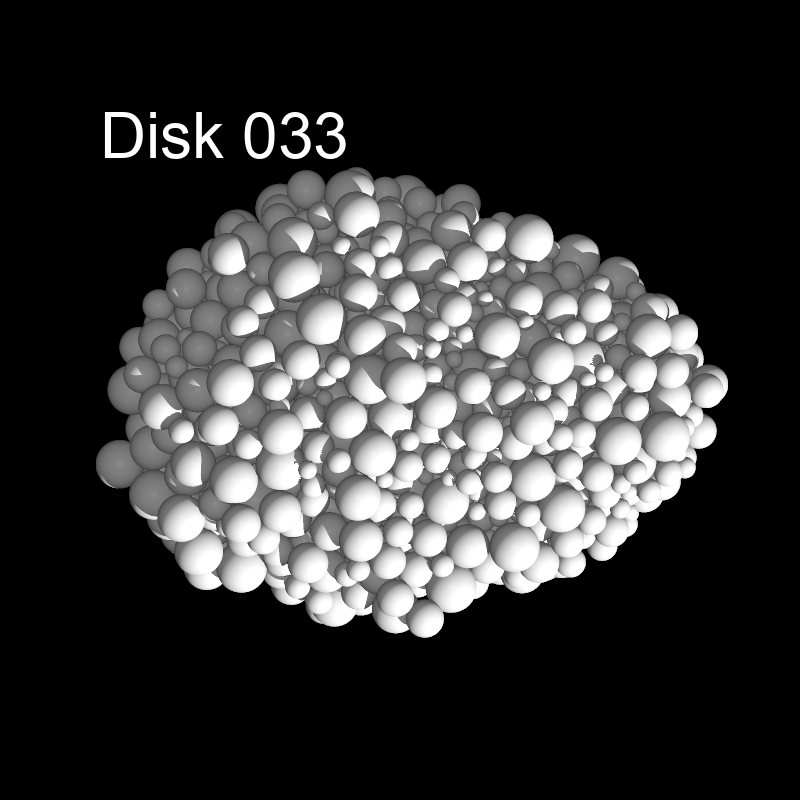}
\includegraphics[width=0.24\textwidth]{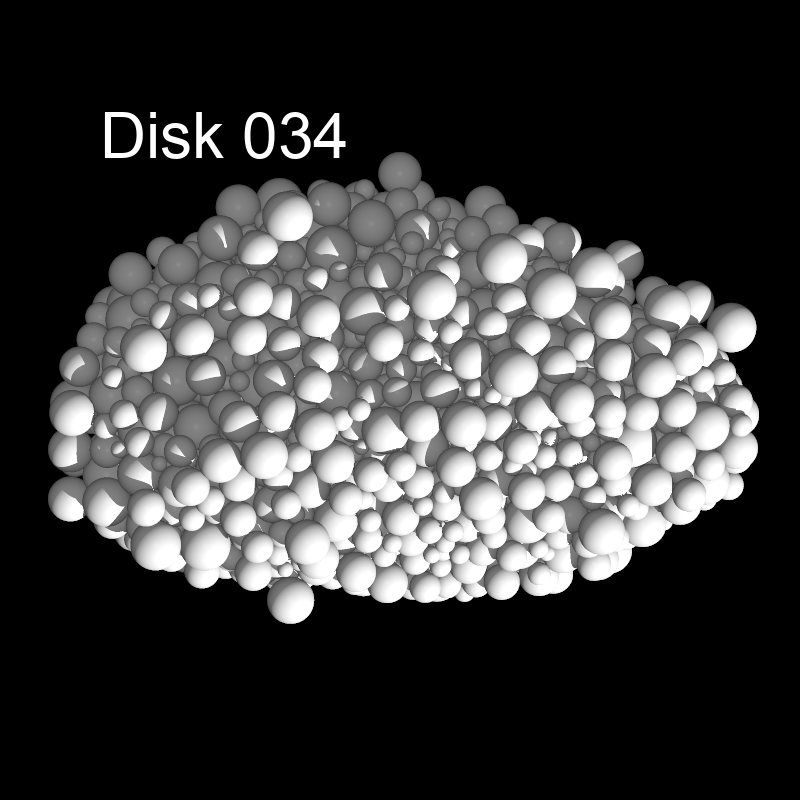}
\includegraphics[width=0.24\textwidth]{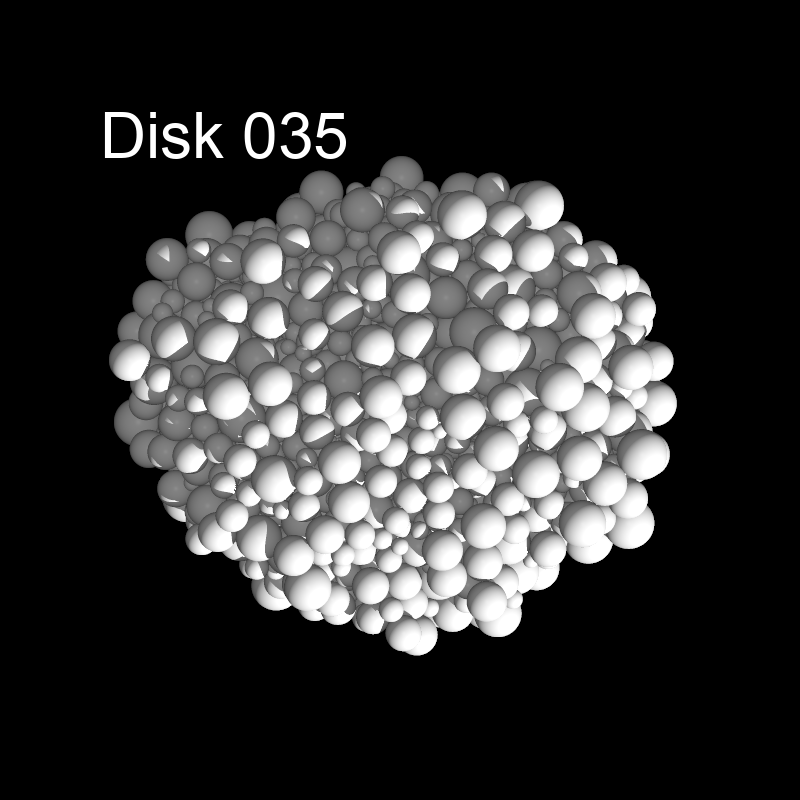}
\includegraphics[width=0.24\textwidth]{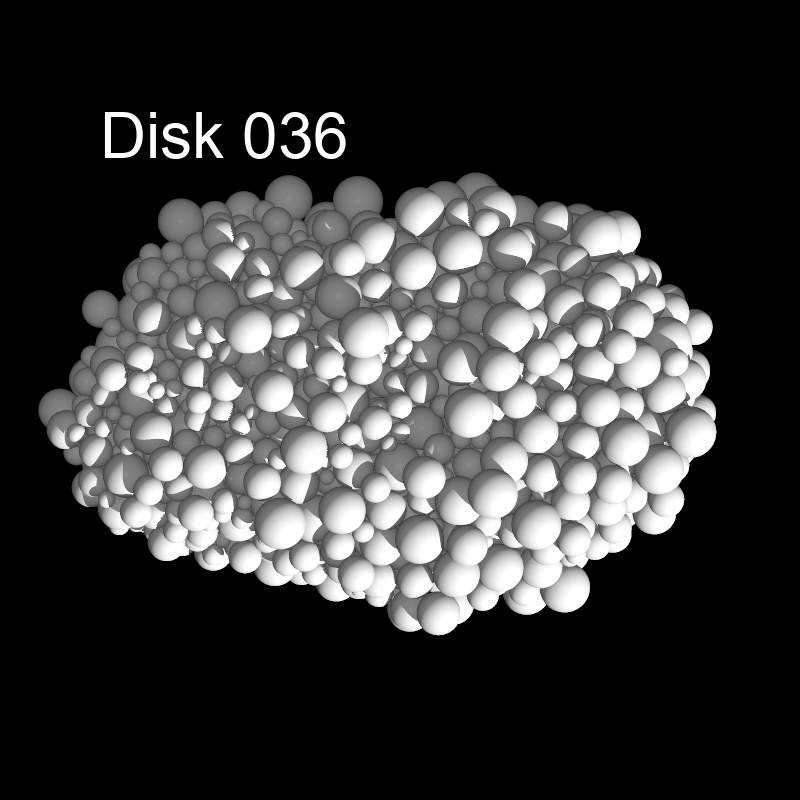}
\includegraphics[width=0.24\textwidth]{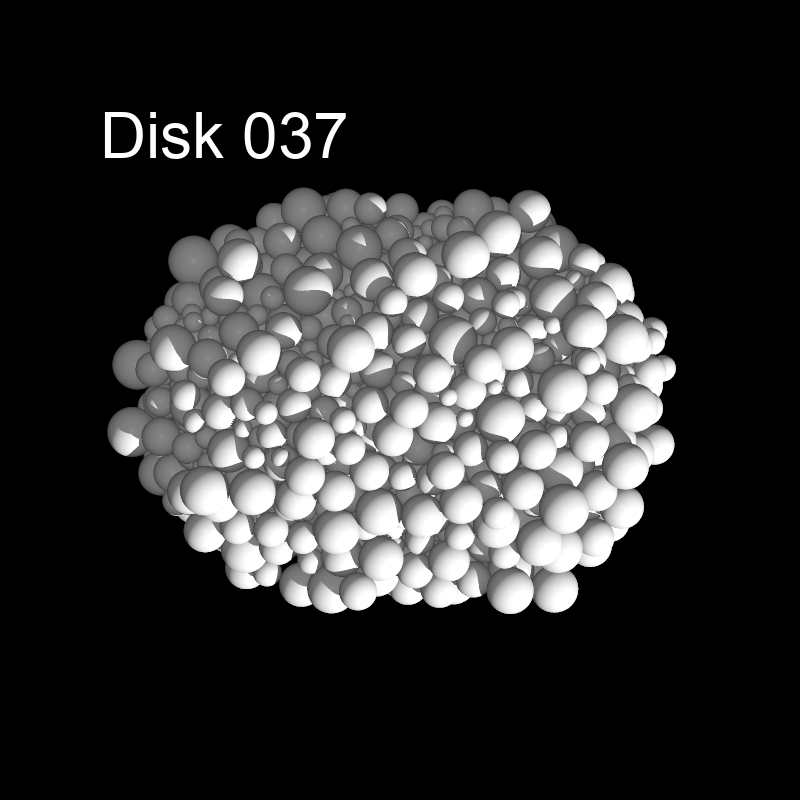}
\includegraphics[width=0.24\textwidth]{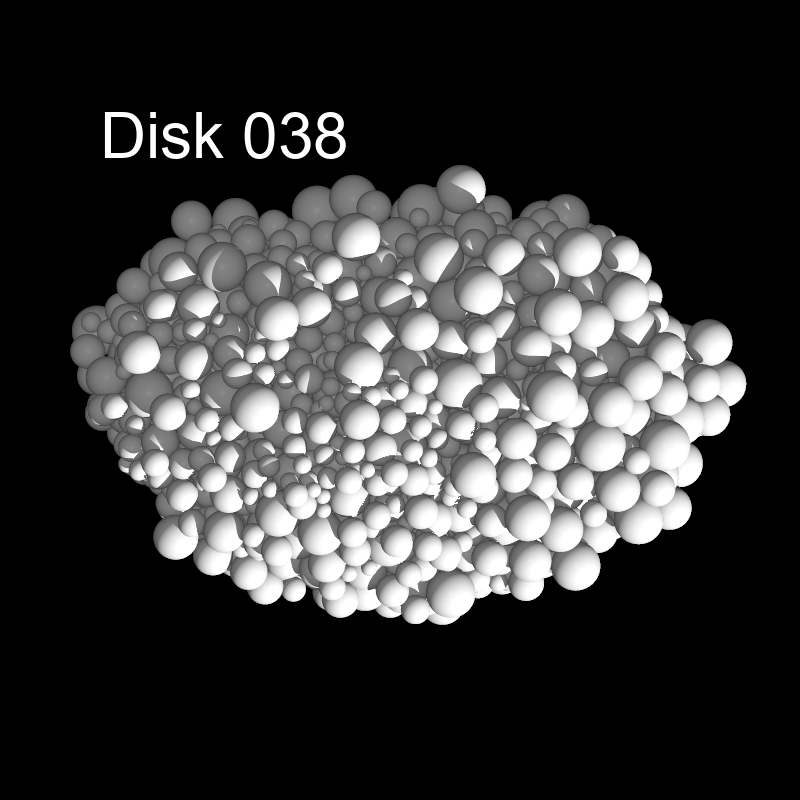}
\includegraphics[width=0.24\textwidth]{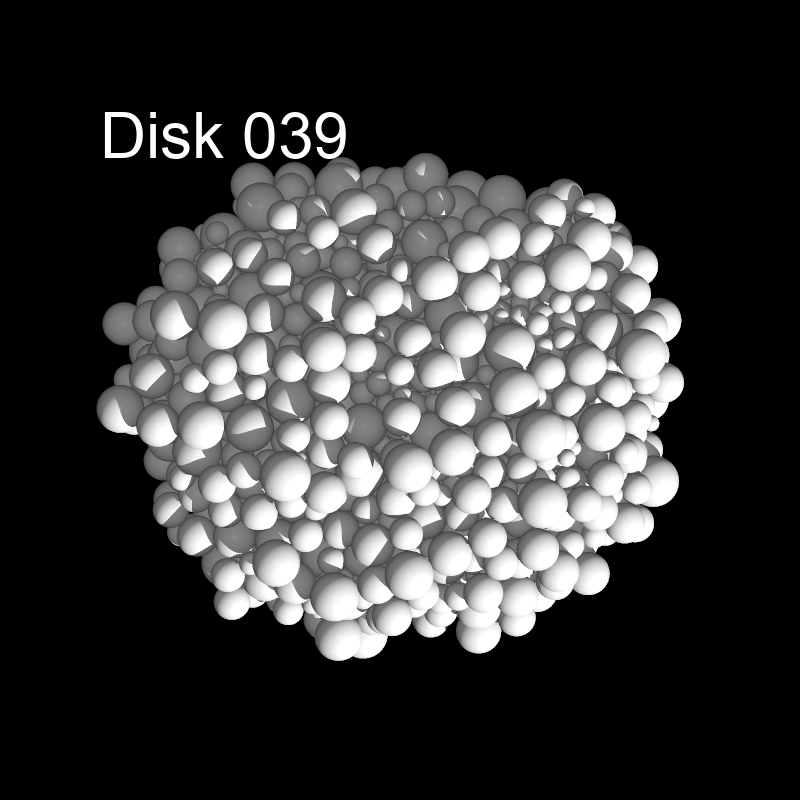}
\includegraphics[width=0.24\textwidth]{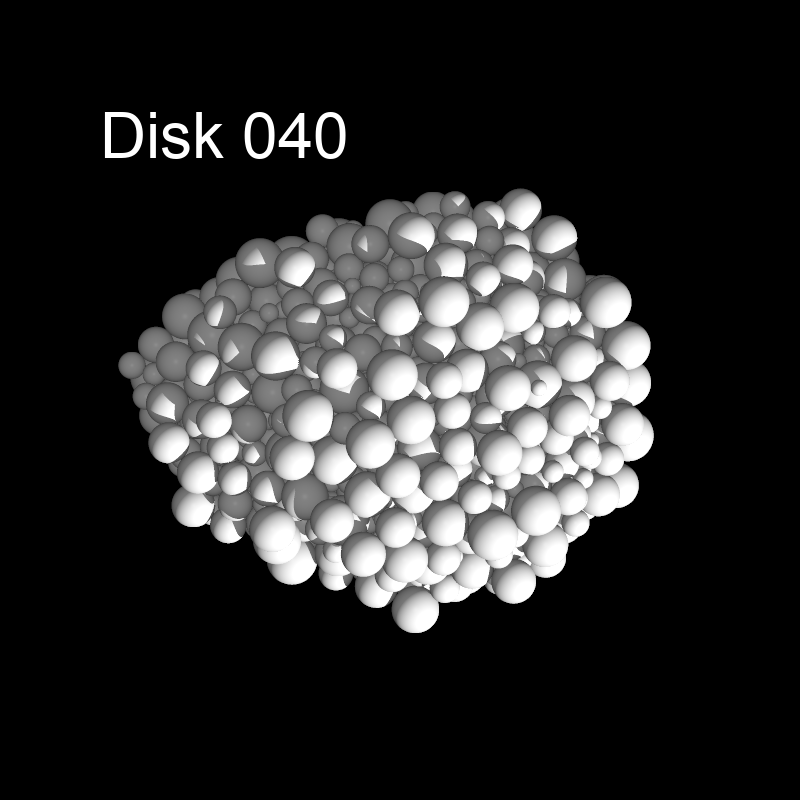}

$\phi=40^\circ$\\
\includegraphics[width=0.24\textwidth]{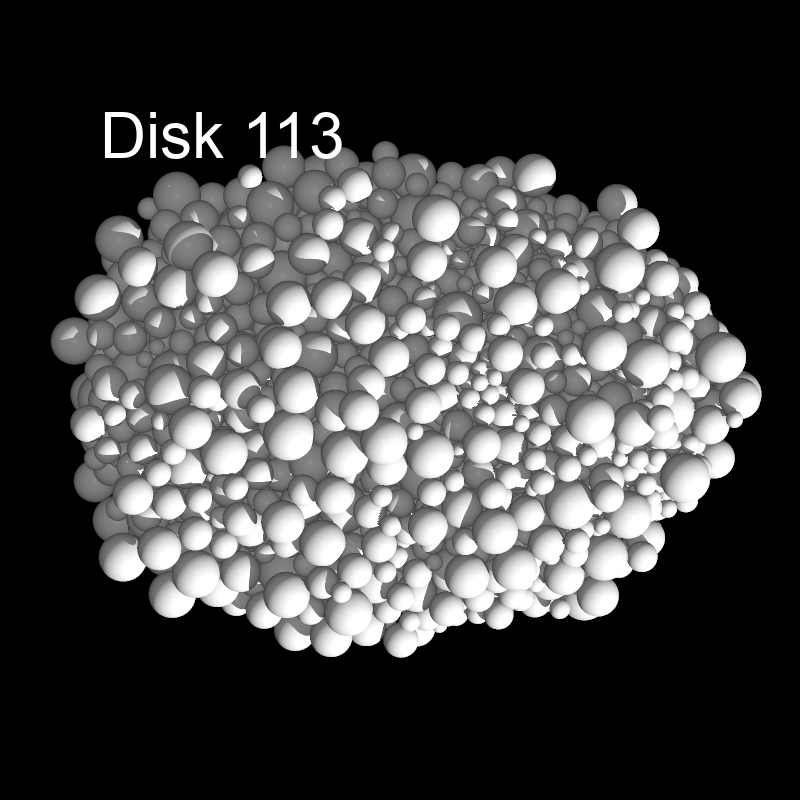}
\includegraphics[width=0.24\textwidth]{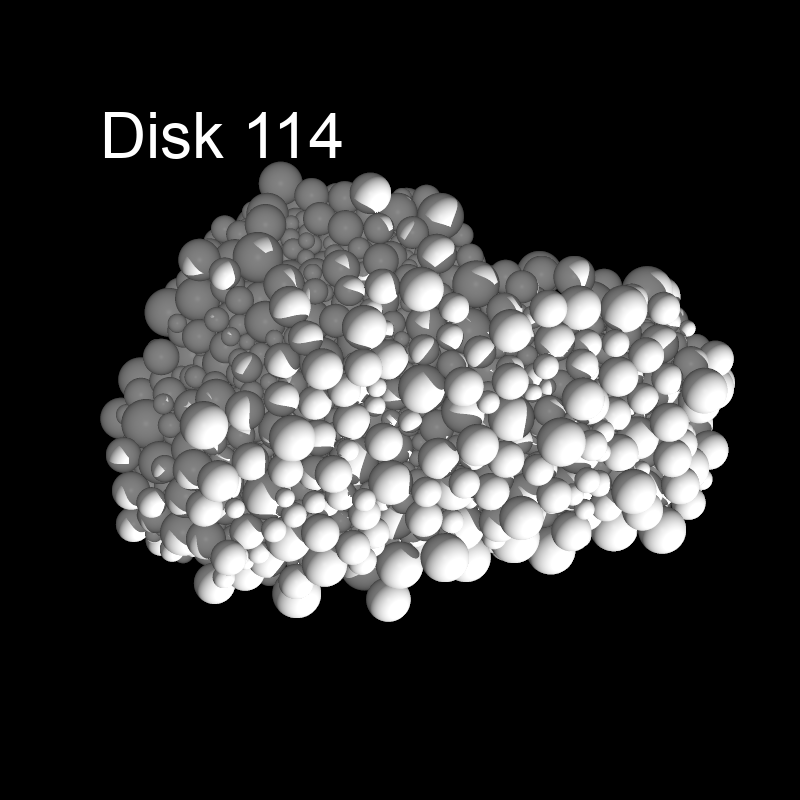}
\includegraphics[width=0.24\textwidth]{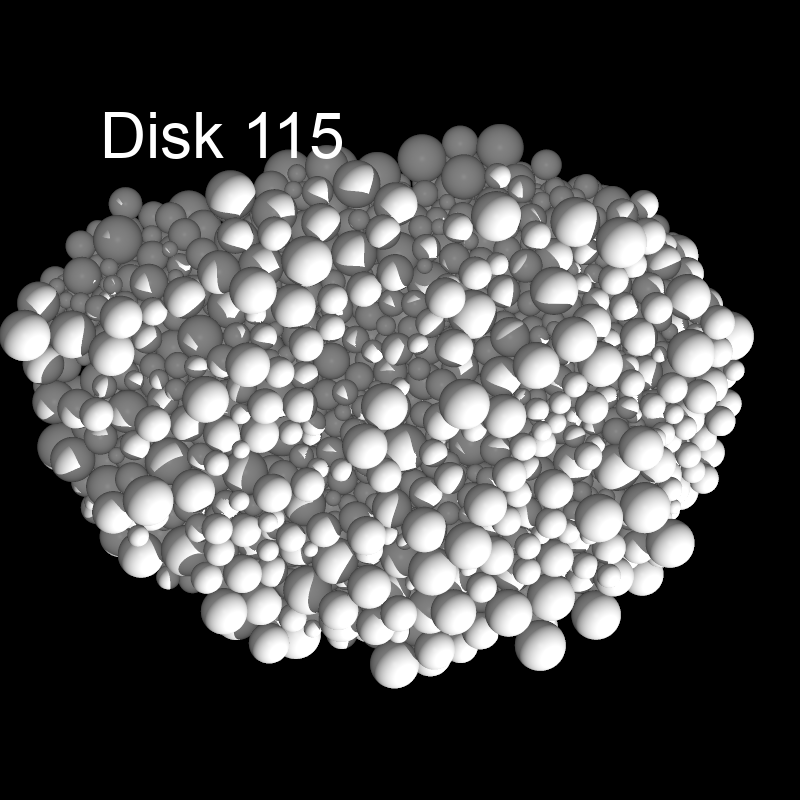}
\includegraphics[width=0.24\textwidth]{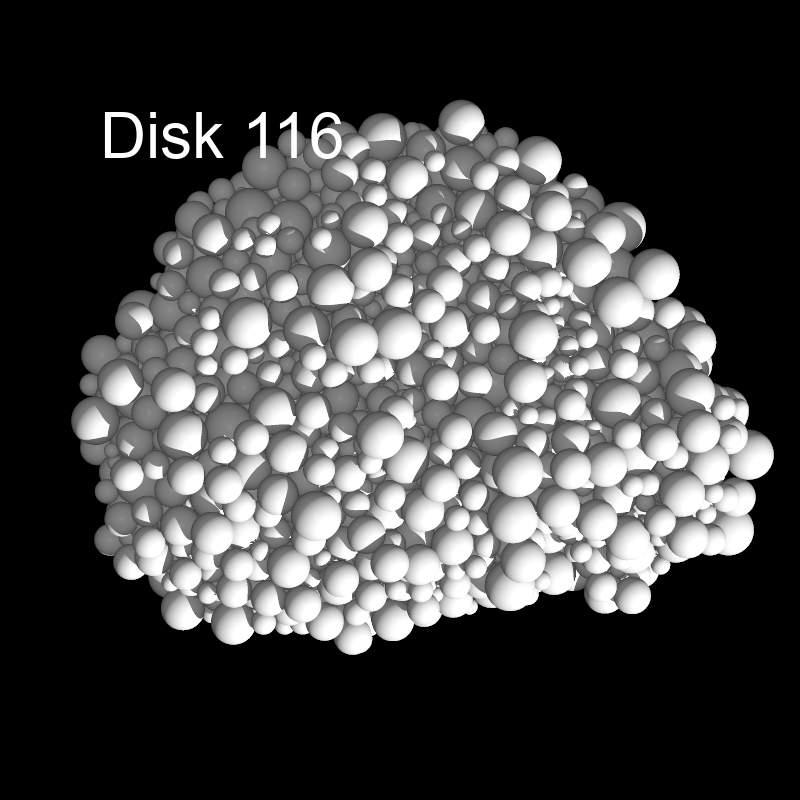}
\includegraphics[width=0.24\textwidth]{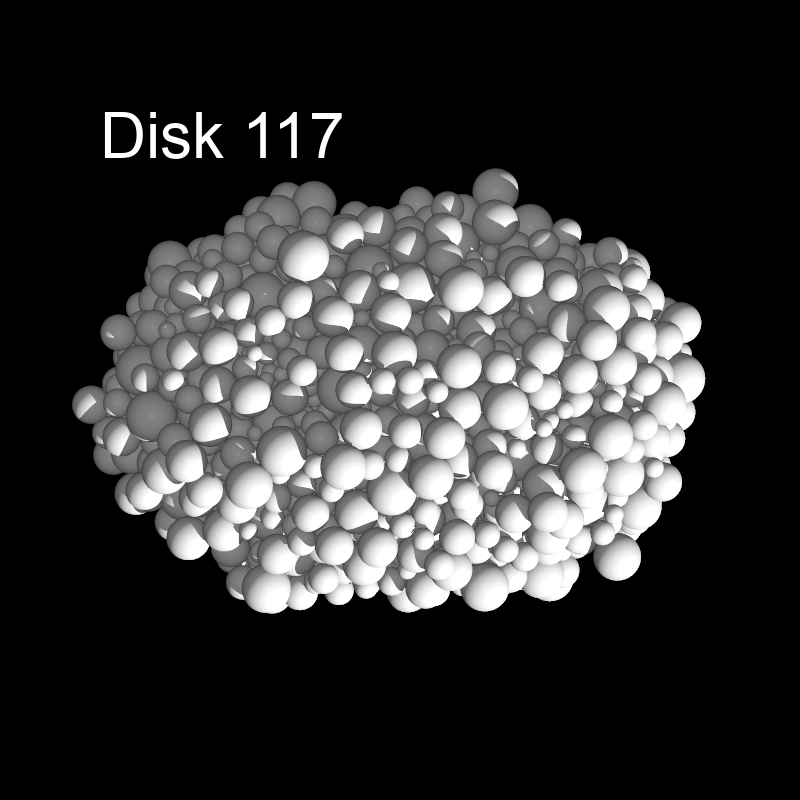}
\includegraphics[width=0.24\textwidth]{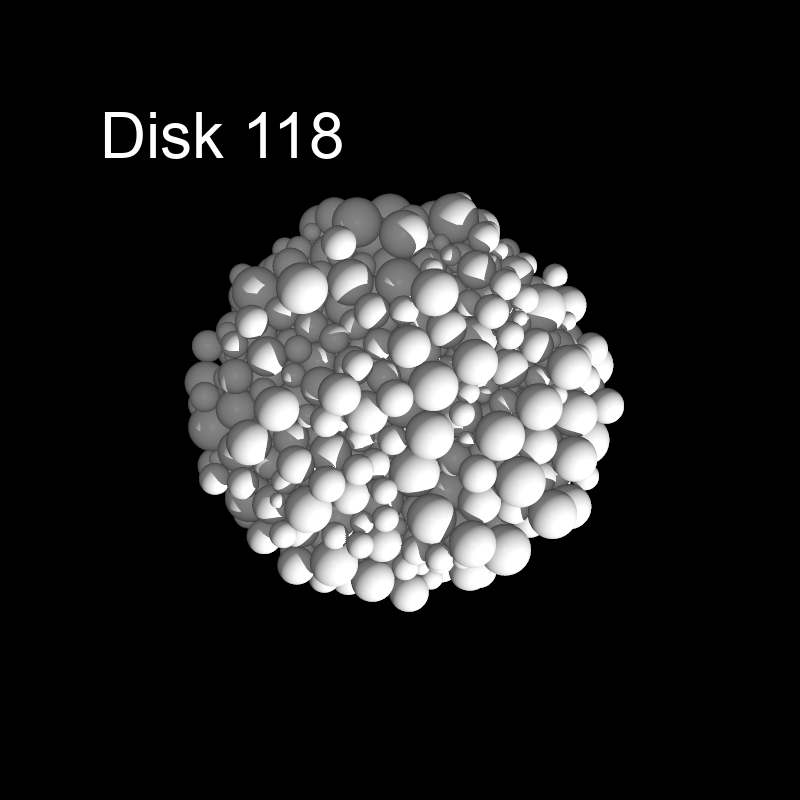}
\includegraphics[width=0.24\textwidth]{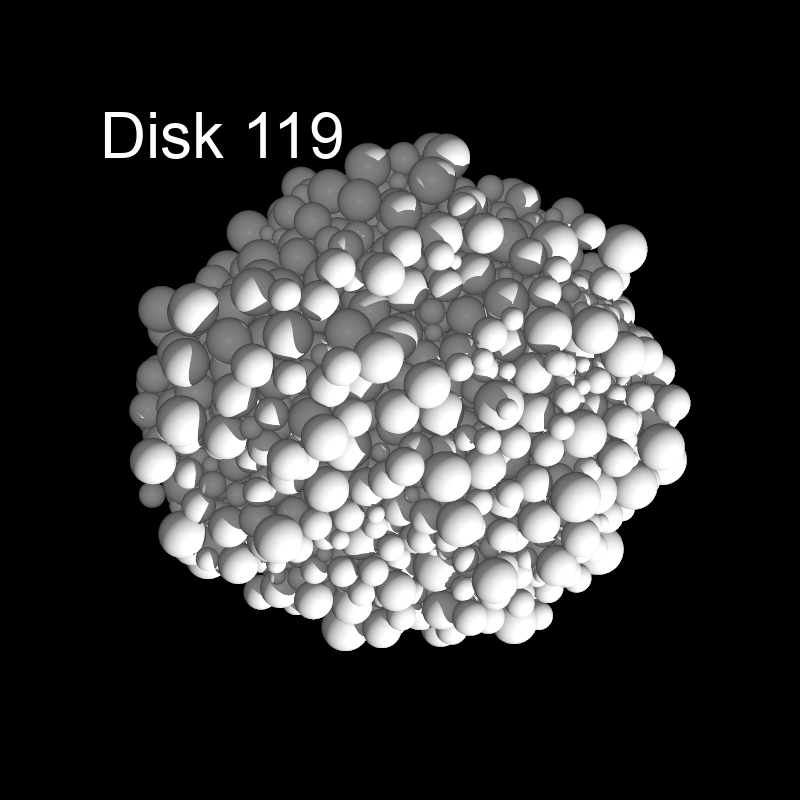}
\includegraphics[width=0.24\textwidth]{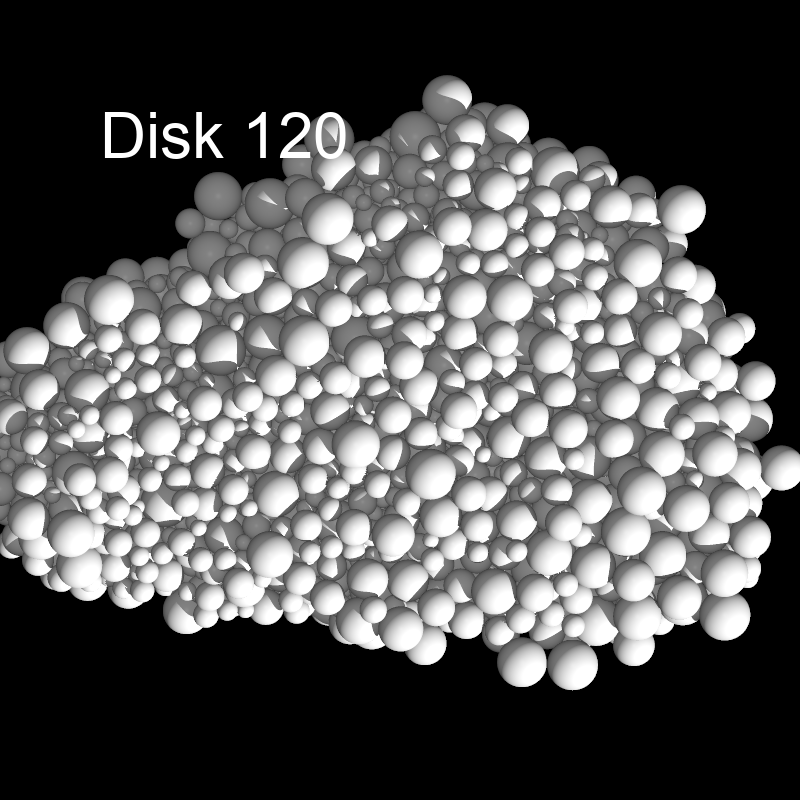}


\bibliography{references}{}
\bibliographystyle{aasjournal}



\end{document}